\documentclass[a4paper,12pt]{article}
\usepackage[a4paper, left=2.2cm, right=2.2cm, top=2.2cm, bottom=2.2cm]{geometry}
\usepackage{amsmath,xfrac}
\usepackage{xr}
\usepackage{bm}
\usepackage{enumitem}  
\usepackage{amsthm,amsbsy}
\usepackage[utf8]{inputenc} 
\usepackage[english]{babel} 
\usepackage{xcolor} 
\usepackage{tikz}
\usepackage{adjustbox} 
\usepackage[onehalfspacing]{setspace}
\usepackage{graphicx}
\usepackage{pgfplots}
\usepackage{xcolor, lipsum} 
\usepackage{caption}
\usepackage{mathcomp}
\usepackage{verbatim} 
\usepackage{enumitem}
\usepackage[overload]{abraces}
\usepackage{array}
\usepackage{pict2e} 
\usepackage{amssymb}
\usepackage{bbm}
\usepackage{epstopdf}
\usepackage{pbox}
\usepackage{mathrsfs}
\usepackage{wrapfig}
\usepackage{multirow}
\usepackage{subcaption}
\usepackage{colortbl}
\usepackage{mathtools}
\usepackage{multicol}
\usepackage{wasysym}
\usepackage{makecell}
\usepackage{eurosym}
\usepackage{booktabs}
\newcommand{\ra}[1]{\renewcommand{\arraystretch}{#1}}
\usepackage{pdflscape}
\usepackage{afterpage}
\usepackage{arydshln}
\usepackage{capt-of}
\usepackage{rotating}
\usepackage{longtable}
\usepackage{pdfpages}
\usepackage[hidelinks,
colorlinks=true,
urlcolor=blue,
linkcolor=blue,
citecolor=blue,
hyperfootnotes=true]{hyperref}
\usepackage{xurl}
\usepackage{natbib}
\usepackage{mathtools}

\DeclarePairedDelimiter\floor{\lfloor}{\rfloor}
\usetikzlibrary{decorations.pathreplacing}
\usetikzlibrary{positioning, shapes, arrows.meta, fit}
\allowdisplaybreaks
\makeatletter
\newcommand*{\addFileDependency}[1]{
	\typeout{(#1)}
	\@addtofilelist{#1}
	\IfFileExists{#1}{}{\typeout{No file #1.}}
}
\makeatother

\newcommand*{\myexternaldocument}[1]{%
	\externaldocument{#1}%
	\addFileDependency{#1.tex}%
	\addFileDependency{#1.aux}%
}

\myexternaldocument{webappendix_v3}

\renewcommand{\ref}[1]{\hyperref[#1]{\ref{#1}}}

\newtheorem{proposition}{Proposition}
\newtheorem{lemma}{Lemma}
\newtheorem{corollary}{Corollary}
\newtheorem{observation}{Observation}

\DeclareMathOperator*{\argmax}{arg\,max}

\renewcommand{\thefootnote}{\fnsymbol{footnote}}
\begin{document}
	\onehalfspacing
	\begin{center}
		{\LARGE \textbf{{\LARGE \textbf{Learning in a Small/Big World}}}\footnote{I am grateful to Tilman Börgers, Jacques Cr\'emer, Matthew Elliott, Renato Gomes, Philippe Jehiel, Hamid Sabourian, Mikhael Safronov, Larry Samuelson, Tak-Yuen Wong, anonymous referees, and the audiences at various seminars and conferences for their insightful discussions and comments.}
		}\\
		\vspace{5mm}
		\textit{Benson Tsz Kin Leung\footnote{Hong Kong Baptist University. Email: \href{mailto:btkleung@hkbu.edu.hk}{\texttt{btkleung@hkbu.edu.hk}}, \href{mailto:ltkbenson@gmail.com}{\texttt{ltkbenson@gmail.com}}. }}\\
		\textit{\today}\\

	\end{center}
	\vspace{4mm} 
	\begin{abstract}
		\fontsize{11}{12}\selectfont
		Complexity and limited ability affect how we learn and make decisions under uncertainty. Using finite automata to model belief formation, this paper studies the characteristics of optimal learning behavior in small and big worlds, where the complexity of the environment is low and high, respectively, relative to the cognitive ability of the decision maker. The optimal long-run average payoff and distribution of actions are well approximated by the Bayesian benchmark in very small worlds, but the gap widens as the worlds get bigger. In big worlds, it can be optimal for the decision maker to ignore some states entirely, concentrating limited cognitive resources on a subset of states. This ``optimal ignorance'' is reminiscent of well-documented non-Bayesian learning patterns such as the use of heuristics, correlation neglect, and inattentive learning. The results establish a relationship between complexity, cognitive ability, and the scope for non-Bayesian learning behavior.

		\vspace{5mm}
		\textbf{Keywords:} Learning, Bounded Memory, Bayesian, Complexity, Cognitive Ability

		\textbf{JEL codes:} D83, D91
		
	\end{abstract}
	
	\renewcommand{\thefootnote}{\arabic{footnote}}
	\setcounter{footnote}{0}

	\newpage
	\section{Introduction}


	\cite{savage1972foundations} introduces the distinction between ``small'' and ``big'' worlds, arguing that Bayesian decision theory applies only to small world problems but not big world problems, where the latter refers to scenarios where it is difficult to form prior beliefs on states and signal structures or construct the state space.\footnote{In Savage's words: ``[A person] is free to \ldots\ look before he leaps \ldots\ when the world is sufficiently small'' \citep[p.~16]{savage1972foundations}.} This paper borrows the terminology but bases the distinction on the complexity of the inference problem relative to the cognitive ability of the individual: ``small'' worlds are inference problems that are simple relative to the individual's cognitive ability. Analyzing a theoretical model with finite automata, I show that optimal learning in big worlds --- but not in small worlds --- exhibits ignorance reminiscent of well-documented non-Bayesian patterns, including the use of heuristics (\cite{kahneman1982judgment}), correlation neglect (\cite{enke2019correlation}), persistent over-confidence (\cite{hoffman2017worker}), and inattentive learning (\cite{graeber2019inattentive}). These ``biased'' learning behaviors are thus more prominent in more complex problems and among individuals with lower cognitive ability.

	More specifically, I consider a decision maker (DM) who tries to learn the true state of the world from a finite state space, where the number of possible states $N$ measures the complexity of the inference problem. In each period $t=1,\cdots,\infty$, the DM guesses the true state of the world and gets a higher utility if he is correct. After making a guess, he receives a signal and updates his belief. As $t\rightarrow\infty$, a Bayesian would learn the true state perfectly. To model limited cognitive ability, I instead confine the DM's belief to an $M$-sized automaton, as in the seminal work of \cite{hellman1970learning} and \cite{wilson2014bounded}.\footnote{See also \cite{compte2012belief}, \cite{monte2014value}, \cite{basu2015interim}, \cite{chauvin2019euclidean}, \cite{chatterjee2020game}, \cite{chatterjee2021learning} in the economic literature that model belief updating and the aversion of complexity with finite automata. See also \cite{oprea2020makes} and \cite{banovetz2020complexity} for experimental evidence.} In each period, the DM makes his guess based on his current memory state (the decision rule), then transits to another memory state given the signal received (the transition rule). The DM thus has a coarser idea of the likelihood of different states than a Bayesian, with the coarseness decreasing in $M$; thus $M$ measures cognitive ability.\footnote{Note that if the individual tracks his belief not with a finite automaton but with a real number statistic, the cardinality of the belief statistics is much larger than $N$, and the model collapses to a Bayesian model.} I define small worlds as decision problems where $\frac{N}{M}$ is small, and big worlds otherwise: whether a problem is a small or big world depends on the complexity of the world relative to the individual's cognitive ability.

\begin{table}
		\begin{center}
			\ra{1.2}
			\begin{tabular}{@{}ccc@{}}
				\toprule
				& \pbox{18cm}{\small Small Worlds:\\\small low complexity relative\\\small to cognitive ability $\frac{N}{M}$}  & \pbox{18cm}{\small Big Worlds:\\\small high complexity relative\\\small to cognitive ability $\frac{N}{M}$} \\
				\midrule
			\small	Are actions close to Bayesian? &\small Yes &\small No \\
			\small	Could ignorance in learning be ``optimal''? &\small No &\small Yes\\
			\small	Could disagreement be persistent? &\small No &\small Yes\\
				\bottomrule
			\end{tabular}
		\end{center}
	\caption{Differences in learning behaviors in small/big worlds}
	\label{tab:results}
	\end{table}


	It is, in general, not tractable to characterize the optimal updating mechanism, i.e., the starting memory state, transition rule, and decision rule that maximize the DM's asymptotic utility: the design space is a product of $M\times\vert S\vert$ stochastic transition matrices and $M$ action assignments, and the objective, while linear in the stationary distributions, is non-convex and discontinuous in the transition matrices. Without an explicit characterization, however, I can compare the characteristics of optimal updating mechanisms in small and big worlds, and hence shed light on how perceived complexity affects learning; Table~\ref{tab:results} summarizes the results. First, I analyze how the DM's actions differ from the Bayesian benchmark --- that is, when the Bayesian model serves as a good approximation of decision-making under uncertainty. In small worlds, the DM's long-run average payoff and distribution of actions approach the benchmark: as $\frac{N}{M}\rightarrow 0$, the DM almost always makes the same guess as a Bayesian. As $\frac{N}{M}$ increases, the DM makes more mistakes and his payoff diverges further from the benchmark.


	The second and main result shows that in big worlds, as the DM faces a trade-off in allocating his scarce memory states, it is optimal in some decision environments to ignore some states and focus learning on the rest; such ``ignorant'' behavior is never optimal in small worlds. This establishes a relationship between complexity and ignorance in learning, and the structure of optimal ignorance is reminiscent of several well-documented learning biases --- including the use of heuristics, correlation neglect, persistent over-confidence, and inattentive learning.


	Before I briefly describe the intuition of the result, to see how ``ignorance'' captures different biases, consider persistent over-confidence (\cite{hoffman2017worker}, \cite{heidhues2018unrealistic}). Let the state of the world comprise the DM's and his teammate's ability, each high or low, with team performances as signals. The DM is observationally indistinguishable from exhibiting persistent over-confidence when he never guesses the states where his own ability is low --- behaving as if he always believes his ability is high and updating only his belief about his teammate --- even after a sequence of bad team performances. Similarly, for correlation neglect (\cite{enke2019correlation}), consider the state comprising a strong or weak stock market and positive, no, or negative correlation among data. The DM is observationally indistinguishable from exhibiting correlation neglect when he ignores the possible correlation and updates only his belief about the market, as if the data were uncorrelated.


	As mentioned above, such ignorant behavior is optimal only when relative complexity is large. In small worlds, the DM faces little trade-off between learning about different states and ignores none. In big worlds, i.e., when relative complexity is high, the trade-off kicks in, and the DM optimally ignores some states to focus learning on others --- especially states that are a priori unlikely or supported by weak information. For example, a DM confident about his own ability focuses on learning his teammate's; if it is difficult to detect correlation, the DM ignores the possibility that data are correlated. More surprisingly, even when the states and information structures are fully symmetric, ignorance is optimal in environments where learning is difficult --- when signals are noisy or the state space is large. This echoes that complexity, or difficulty in learning, drives ignorance, and shows that optimal ignorance is not a knife-edge result.


	Lastly, I analyze whether disagreement persists (with an infinite sequence of i.i.d.\ signals) in small and big worlds. In small worlds, as the asymptotic distribution of actions is close to the Bayesian benchmark, individuals eventually agree and make the same guesses. In big worlds, individuals with different prior beliefs and/or cognitive abilities adopt different optimal learning mechanisms, may ignore different states, and can disagree with each other with probability $1$ even after receiving the same infinite sequence of public information. For example, when two individuals with persistent over-confidence each ignore the states where their own ability is low and attribute bad team performance to the other. Moreover, I show a novel driving force of disagreement: two individuals with the same prior beliefs, observing the same infinite sequence of public signals with no uncertainty in signal structures, can eventually disagree when they have different levels of cognitive ability $M$.


	This paper is organized as follows. Section~\ref{section:literature} discusses the related literature; Section~\ref{section:model} presents the model; Section~\ref{section:example} develops a series of illustrative examples. Section~\ref{section:small} analyzes optimal learning behavior in small and big worlds, including limit results as $N,M\rightarrow\infty$. Section~\ref{section:discussion} concludes with a discussion of the implications. The proofs are presented in the Appendix, and additional material and extensions are presented in the online Appendix.\footnote{The online Appendix could be found on \url{https://sites.google.com/site/ltkbenson/research}.}

	\section{Literature}\label{section:literature}

	In this section, I discuss the existing literature and the contribution of this paper.


	This paper is most closely related to the literature using finite automata to model learning with aversion to complexity (\cite{hellman1970learning}, \cite{borgers2004complexity}, \cite{compte2012belief}, \cite{wilson2014bounded}, \cite{chatterjee2021learning}, etc.).\footnote{Also see \cite{chatterjee2020game} for a review.} These papers study how complexity aversion shapes learning, but not how \textit{different levels} of complexity and cognitive ability affect learning and the prominence of learning biases; moreover, they focus on binary state spaces because of technical difficulties. While this paper does not ``fully'' characterize the optimal automaton, its results shed light on the optimal automaton's characteristics when $N>2$ --- in particular, that ignorance becomes a crucial feature of optimal learning even when states are a priori identical, unlike when $N=2$.


	This paper also contributes to a growing literature that explains behavioral anomalies as optimal/efficient strategies in light of limited cognitive ability: rational inattention explains sticky prices and micro-founds the multinomial logit choice model (\cite{sims2003implications}; \cite{matvejka2015rational}); optimal responses to noise in perceiving lotteries generate the probability weighting of prospect theory (\cite{steiner2016perceiving}); and capacity constraints on belief updating drive confirmation bias (\cite{jehiel2018selective}; \cite{leung2020limited}). In contrast, this paper connects a larger class of biases within a single framework and links their prominence to the perceived complexity of the environment.


	Closest to Section~\ref{sec:ignorance}, \cite{caplin2019rational} present conditions under which the DM ignores some actions in a rational inattention setting. Their conditions depend on the prior belief and the utility matrix, but not on the level of complexity or the informativeness of signals. Moreover, in contrast to \cite{caplin2019rational}, I show that ignoring states can be optimal even in symmetric environments with uniform prior and symmetric utilities.


	Last, this paper's results on asymptotic disagreement contribute to a large literature in which disagreement is driven by differences in signal distributions across states or in learning mechanisms (\cite{morris1994trade}, \cite{mailath2020learning}, \cite*{gilboa2020learning}), the lack of identification or uncertainty in signal distributions (\cite*{acemoglu2016fragility}), confirmation bias (\cite{rabin1999first}), or model misspecification (\cite{freedman1963asymptotic,freedman1965asymptotic}, \cite{berk1966limiting}). This paper instead ties disagreement to limited cognitive ability: it characterizes when asymptotic disagreement arises and when it cannot, depending on the relative complexity of the inference problem, and shows a novel mechanism by which disagreement stems solely from differences in cognitive ability.

	\section{Model}\label{section:model}
	I consider a world with $N\geq 2$ possible true states, i.e., $\omega\in\Omega=\lbrace 1,2,\cdots, N\rbrace$, and a decision-maker (DM). In each period $t=1,\cdots,\infty$, the DM receives a signal and tries to guess what the true state of the world is. Formally, the unknown state of the world $\omega$ is realized before period 1 and is unchanged across all periods. In each period $t$, the DM takes an action $a_{t}\in \mathscr{A}=\Omega$ and gets utility $u(a,\omega)\in\mathscr{R}$ where $a=\omega$ is the unique maximizer of $u(a,\omega)$.\footnote{This assumption rules out ``safe'' actions that are not optimal under complete information, i.e., actions that are not maximizers in any state but yield good payoffs in multiple states. However, actions that are maximizers in some state and also yield good payoffs in other states are not ruled out. The assumption plays two further roles. First, it identifies each state with a unique optimal action, so that ignorance can be defined and interpreted through observable actions (Definitions~1 and~2). Second, it makes all $N$ states payoff-distinct and ties the size of the action space to $N$, so that $N$ alone measures the complexity of the world.} The DM does not observe his utility after taking action; thus, $u(a,\omega)$ is best interpreted as an intrinsic utility of being correct. Otherwise, the problem becomes trivial as the DM will experiment and learn perfectly the true state after observing the utility. The prior belief of the DM is $(p^{\omega})_{\omega=1}^{N}$ where $\sum_{\omega=1}^{N}p^{\omega}=1$ and $p^{\omega}>0$ for all $\omega\in\Omega$.

	\newtheorem{innercustomgeneric}{\customgenericname}
	\providecommand{\customgenericname}{}
	\newcommand{\newcustomtheorem}[2]{%
		\newenvironment{#1}[1]
		{%
			\renewcommand\customgenericname{#2}%
			\renewcommand\theinnercustomgeneric{##1}%
			\innercustomgeneric
		}
		{\endinnercustomgeneric}
	}
	\newcustomtheorem{assumption}{Assumption}
	
	In each period after taking an action, the DM receives a signal $s_{t}$ that is independently drawn from a set $S$ across different periods.\footnote{The order, i.e., whether the DM receives a signal before or after taking an action in each period, does not affect the result.} For the ease of exposition, I assume that $S$ is finite and the probability mass function in state $\omega$ is denoted as~$f^{\omega}$.\footnote{To ease exposition, I assume that signals follow a discrete distribution, but all results hold with more general probability measures including non-absolutely continuous distributions.} I assume that no signal perfectly rules out any state of the world, i.e., $f^\omega(s)>0$ for all $(s,\omega)$, such that asymptotic learning is non-trivial. We denote 
	\begin{equation*}
		\varsigma\equiv\min_{\omega'\neq\omega,s\in S}\frac{f^{\omega}(s)}{f^{\omega'}(s)}.
	\end{equation*}
	As $S$ and $\Omega$ are finite, $\varsigma>0$. Intuitively, $\varsigma$ measures (inverse) signal informativeness: small $\varsigma$ means some signal sharply rules out one state in favor of another, while $\varsigma$ close to $1$ means no signal carries strong information. Moreover, I assume no pairs of signal structures are the same, i.e., there are no $\omega$ and $\omega'\neq\omega$ such that $f^{\omega}(s)=f^{\omega'}(s)$ for all $s\in S$. This implies that states are identifiable, and a Bayesian learns the true state perfectly as $t\rightarrow\infty$.\footnote{More specifically, \cite{doob1949application} shows that Bayesian posterior converges to a mass point at the true state of the world with probability 1.} In contrast, I focus on the bounded memory setting I now describe.
    

	The DM is subject to a memory constraint such that he can only update his belief using an $M$-memory-state automaton. In each period, his belief is represented by a memory state $m_{t}\in\lbrace1,2,\cdots,M\rbrace$. With some abuse of notations, I sometimes use $M$ to denote the set of memory states. An updating mechanism specifies an initial state $m_1\in \bigtriangleup M$, a (stationary) transition function between the $M$ memory states given a signal~$s\in S$, which is denoted as $\mathscr{T}: M\times S\rightarrow \bigtriangleup M$, and a (stationary) decision rule $d: M\rightarrow A$.\footnote{Restricting attention to deterministic decision rules is without loss of generality: given any stochastic decision rule, as the long-run average payoff is linear in the action probabilities at each memory state and additively separable across memory states, one can replace each memory state's randomized action with the action that maximizes the expected per-period payoff at that memory state, weakly increasing the long-run average payoff. Thus, for ease of exposition, I restrict attention to deterministic decision rules unless stated otherwise.} For the ease of exposition, I denote the set of memory states where the DM chooses action $\omega$ as $M^{\omega}$.
	
	In each period $t=1,2,\cdots$, the DM starts with some memory state $m_t$, takes action $a_t= d(m_t)$, receives a signal $s_t$, and transits to memory state $m_{t+1}\sim\mathscr{T}(m_{t},s_{t})$. This paper analyzes the asymptotic learning of the DM, i.e., before period~1, the DM designs and commits to an updating mechanism that maximizes his expected long-run per-period utility:\footnote{Switching between multiple $M$-memory-state automata requires more than $M$ memory states, and the current setting allows switching between smaller automata, as illustrated in Online Appendix~\ref{section:switch}.} 
	\begin{equation*}
		\lim_{T\rightarrow\infty}E_{\omega,m_1,\mathscr{T},d}\left[\frac{1}{T}\sum_{t=1}^{T}u(a_{t},\omega)\right].
	\end{equation*}
	An alternative setup is that the DM maximizes the discounted sum of utility as in \cite{wilson2014bounded}. As shown in Appendix~\ref{section:wilson}, the results in this paper hold qualitatively when the discount factor is close to $1$. 

Given a state $\omega\in\Omega$, the sequence $m_{t}$, together with some specified initial memory state $m_1$, forms a Markov chain. Denote $\mu^{\omega}_{m}$ as the long-run proportion of time that the DM is in memory state $m$ when the true state of the world is $\omega$, and $\mathbf{Q}^{\omega}$ as the matrix of transition probabilities, i.e., $\mathbf{Q}^{\omega}=[\sum_{s\in S}\Pr\lbrace \mathscr{T}(m,s)=m'\rbrace f^\omega(s)]_{mm'}$. By the Birkhoff–Khinchin theorem, the distribution $\bm{\mu}^{\omega}=(\mu_{1}^{\omega},\mu_{2}^{\omega},\cdots,\mu_{M}^{\omega})^{T}$ solves the following equation:
\begin{equation}\label{eq:fixpoint}
	\bm{\mu}^{\omega}=(\bm{\mu}^{\omega})^{T}\mathbf{Q}^{\omega}.
\end{equation}
By the Brouwer fixed-point theorem, a solution always exists. When there is a unique solution, e.g., when the transition matrix is irreducible, $\bm{\mu}^{\omega}$ depends only on the transition function $\mathscr{T}$ but not on the initial memory state; when there are multiple solutions, it implies that the Markov chain is reducible, and the long-run distribution is one of the solutions of Equation~\eqref{eq:fixpoint} and is uniquely pinned down by the initial memory state. For simplicity of exposition, I omit the arguments in $\bm{\mu}^{\omega}$ throughout the paper.


 		The asymptotic utility, or the long-run per-period utility, of an updating mechanism $(m_1,\mathscr{T},d)$ is equal to:
 	\begin{equation}
 		U(m_1,\mathscr{T},d)\equiv\lim_{T\rightarrow\infty}E_{\omega,m_1,\mathscr{T},d}\left[\frac{1}{T}\sum_{t=1}^{T}u(a_{t},\omega)\right]=\sum_{\omega=1}^{N}\left[p^{\omega}\left(\sum_{m=1}^{M}u\left(d(m),\omega\right)\mu^{\omega}_{m}\right)\right]
 	\end{equation}
 	and the asymptotic utility loss is equal to:
 	\begin{equation}
 		\begin{split}
 			L(m_1,\mathscr{T},d)
 			&\equiv\sum_{\omega=1}^{N}\left[p^{\omega}u(\omega,\omega)\right]-U(m_1,\mathscr{T},d).
 		\end{split}
 	\end{equation}
 	The DM maximizes the asymptotic utility or, equivalently, minimizes the asymptotic utility loss. I mostly refer to the optimal design of the updating mechanism as the minimization of $L$. In general, following similar arguments as in \cite{hellman1970learning}, an optimal mechanism may not exist.\footnote{Although the set of transition probabilities is compact, an optimal mechanism may not exist because the long-run distribution $\bm{\mu}^\omega$ is not continuous in the transition probabilities. To illustrate, consider $M=2$ and $\mathscr{T}(m,s)=(1-\delta)\times\lbrace m\rbrace+\delta\times \lbrace m'\rbrace$ for $m'\neq m$ and $m=1,2$. That is, the DM transits from memory state~$1$ to $2$ (and $2$ to $1$) with probability $\delta$. For all $\delta\in(0,1]$, the long-run distribution is $(\frac{1}{2},\frac{1}{2})$; however, at $\delta=0$ the long-run distribution jumps to $(1,0)$ or $(0,1)$ depending on the starting memory state. The optimal loss $L_M^*$ is therefore defined as an infimum that may not be attained. Example~2 in Section~\ref{section:example} illustrates the non-existence.} Therefore, the rest of the paper focuses on $\epsilon$-optimal updating mechanisms (for small $\epsilon$) that are defined as follows. Define
 	\begin{equation*}
 		L_{M}^{*}\equiv\inf_{m_1,\mathscr{T},d} L(m_1,\mathscr{T},d).
 	\end{equation*} 
 	An updating mechanism $(m_1,\mathscr{T},d)$ is \textit{$\epsilon$-optimal} if and only if $L(m_1,\mathscr{T},d)\leq L_{M}^{*}+\epsilon$. 
    Throughout the paper, I focus on the case where
 	\begin{equation}\label{assumption:learningisuseful}
 		L_{M}^{*}<\min_{a}\sum_{\omega=1}^{N}p^{\omega}\left[u(\omega,\omega)-u(a,\omega)\right],
 	\end{equation}
 	where the right-hand side is the minimum long-run utility loss from always taking the same fixed action regardless of signals. This inequality says that the DM can strictly improve his utility by learning, i.e., the memory constraint is not so severe as to make learning worthless. 
 	
  	In Section~\ref{section:small}, I compare the characteristics of the $\epsilon$-optimal updating mechanisms in small and big worlds. Roughly speaking, $N$ represents the complexity of the world, and $M$ represents the cognitive ability of the DM. I call $\frac{N}{M}$ the relative, or the perceived, complexity. This gives a natural definition of small and big worlds: an inference problem is a small world when $\frac{N}{M}$ is small, and is a big world otherwise. Throughout the paper, unless stated otherwise, I assume $M\geq N$, so that the DM has enough memory states to potentially choose every action.\footnote{When $M<N$, the DM is forced to ignore at least $N-M$ states (never take those actions) simply because he lacks enough memory states for them, making ignoring trivially necessary rather than an optimal allocation of memory states.}
	
	\newcustomtheorem{definition}{Definition}

    \section{Illustrative Examples}\label{section:example}
    Before presenting the main results, I go through a series of examples to illustrate how the design of an optimal updating mechanism changes as $N$ and $M$ vary. The examples build on each other and collectively motivate some of the research questions and main results of the paper.

    \subsection*{Example 1: $N=2$, $M=2$}


	First consider $N=2$ states with uniform prior $p^1=p^2=\tfrac{1}{2}$, $u(a,\omega)=1$ if $a=\omega$ and $0$ otherwise, a binary signal space $S=\{s^1,s^2\}$, and $M=2$ memory states. With $N=2$, each signal can be deterministically defined as supporting the state under which it is more likely: signal $s$ supports state~$\omega$ if and only if $f^{\omega}(s)>f^{-\omega}(s)$. Without loss of generality, suppose $s^1$ supports state~1, and hence $s^2$ supports state~2 (as $f^\omega(s^2)=1-f^\omega(s^1)$). For simplicity, in this example I assume the information structure is symmetric, $f^1(s^1)=f^2(s^2)$, and denote $\mathscr{I}\equiv\frac{f^1(s^1)}{f^2(s^1)}=\frac{f^2(s^2)}{f^1(s^2)}$, which measures the informativeness of the signals.

	An optimal updating mechanism is illustrated in Figure~\ref{fig:examplen2m2}: the DM chooses action~$\omega$ in memory state~$\omega$, and switches memory states if and only if he receives a signal supporting the other state. The transition and decision rules are symmetric, i.e., invariant to permutations of states and memory states. Balancing the long-run flows into and out of memory state~1 in each state of the world,
    \begin{equation*}
        \begin{split}
            \mu^1_1 f^{1}(s^2)&=\mu^1_2 f^{1}(s^1)\\
            \mu^2_1 f^{2}(s^2)&=\mu^2_2 f^{2}(s^1),
        \end{split}
    \end{equation*}
    which rearranges to
    \begin{equation*}
        \frac{\mu^1_1}{\mu^1_2}=\frac{\mu^2_2}{\mu^2_1}=\mathscr{I}
        \qquad\text{and}\qquad
        U=\frac{\mathscr{I}}{1+\mathscr{I}},\quad L=\frac{1}{1+\mathscr{I}}.
    \end{equation*}
    Intuitively, the DM transits to memory state~$\omega$ upon receiving $s^\omega$, and receives $s^\omega$ more often in state~$\omega$; he is therefore more likely to be in memory state~$\omega$ when the state is $\omega$, matching his action with the state. Moreover, the more informative the signals, the lower the DM's utility loss. The mechanism in Figure~\ref{fig:examplen2m2} is in fact optimal: it maximizes the ``separability'' of the two memory states, i.e., the distance between the posteriors held in them --- \cite{hellman1970learning} show the likelihood ratios of two memory states can differ by at most a factor of $\mathscr{I}^{2}$, which Figure~\ref{fig:examplen2m2} attains ($\frac{\mu^{1}_{1}}{\mu^{2}_{1}}/\frac{\mu^{1}_{2}}{\mu^{2}_{2}}=\mathscr{I}^{2}$).
    
\tikzstyle{cloud} = [draw, align=left, circle,fill=cyan!20, node distance=3cm,
minimum height=1em]
\tikzstyle{cloud1} = [draw, align=left, circle,fill=black, node distance=1cm,
minimum height=0.01cm]
\tikzstyle{cloud2} = [draw, align=left, circle,fill=red, node distance=1cm,
minimum height=0.01cm]
\tikzset{
	dot/.style = {circle, fill, minimum size=#1,
		inner sep=0pt, outer sep=0pt},
	dot/.default = 3pt 
}
    \begin{figure}
		\centering
		\begin{tikzpicture}[every text node part/.style={align=center}]
			\node[cloud, fill=cyan!60] (m1) {1};
			\node[cloud, right = 3em of m1] (m2) {2};
			\draw[-Latex]
			(m1) edge[bend right] node [below, midway] {$s^{2}$} (m2)
			(m2) edge[bend right] node [above, midway] {$s^{1}$} (m1)
			(m1) edge[loop left,looseness=10] node [left] {$s^{1}$} (m1)
			(m2) edge[loop right,looseness=10] node [right] {$s^{2}$} (m2);
		\end{tikzpicture}
		\caption{Optimal updating mechanism with $N=2$, $M=2$, and $S=\{s^1,s^2\}$, where $f^{1}(s^1)=f^{2}(s^2)>f^{1}(s^2)=f^{2}(s^1)$. Decision rule follows: $d(1)=1$, $d(2)=2$. The DM transits to memory state~$\omega$ upon receiving signal $s^\omega$, and otherwise stays in the current memory state.}
		\label{fig:examplen2m2}
	\end{figure}
    
    \subsection*{Example 2: $N=2$, $M=3$}

    Now suppose the DM gains one extra memory state, so $M=3$, while keeping all other model parameters the same. Consider the updating mechanism illustrated in Figure~\ref{fig:examplen2m3}. Similar to the example with $N=M=2$ illustrated in Figure~\ref{fig:examplen2m2}, the DM transits to a higher memory state only if he receives a signal supporting state~2, and to a lower memory state only if he receives a signal supporting state~1. The decision rule is $d(1)=d(2)=1$ and $d(3)=2$.

	\begin{figure}
		\centering
		\begin{tikzpicture}[every text node part/.style={align=center}]
			\node[cloud, fill=cyan!30] (m1lo) {$2$};
			\node[cloud, fill=cyan!60, left = 3em of m1lo] (m1hi) {$1$};
			\node[cloud, right = 3em of m1lo] (m2) {$3$};
			\draw[-Latex]
			(m1lo) edge[bend right]  node [above, midway] {$s^1$} (m1hi)
			(m1hi) edge[bend right]  node [below, midway] {$s^2\times\delta$} (m1lo)
			(m1lo) edge[bend right] node [below, midway] {$s^2$} (m2)
			(m2)   edge[bend right] node [above, midway] {$s^1\times\delta$} (m1lo)
			(m1hi) edge[loop left, looseness=10] node [left] {$s^1$} (m1hi)
			(m2)   edge[loop right, looseness=10] node [right] {$s^2$} (m2);
		\end{tikzpicture}
		\caption{Optimal updating mechanism with $N=2$, $M=3$, and $S=\{s^1,s^2\}$.
        The DM chooses action~1 in both memory states~$1$ and~$2$, and action~2 in memory state~$3$.
        Memory state~$1$ acts as a ``high-confidence'' state for action~$1$: an $s^2$ signal only demotes confidence to memory state~$2$, not all the way to action~$2$.}
		\label{fig:examplen2m3}
	\end{figure}

	Crucially, although the DM chooses action~1 in both memory states~1 and~2, he is more confident about it in memory state~1. The long-run distribution follows:
    \begin{equation*}
        \begin{split}
            \mu^1_1\delta f^1(s^2)=\mu^1_2 f^1(s^1); \quad \mu^1_2 f^1(s^2)=\mu^1_3\delta f^1(s^1)\\
            \mu^2_1\delta f^2(s^2)=\mu^2_2 f^2(s^1); \quad \mu^2_2 f^2(s^2)=\mu^2_3\delta f^2(s^1),\\
        \end{split}
    \end{equation*}
    which gives
    \begin{equation*}
        \begin{split}
            \mu^1_1=\frac{\mathscr{I}^2}{1+\delta\mathscr{I}+\mathscr{I}^2};\quad \mu^1_2=\frac{\delta\mathscr{I}}{1+\delta\mathscr{I}+\mathscr{I}^2};\quad \mu^1_3=\frac{1}{1+\delta\mathscr{I}+\mathscr{I}^2}\\
            \mu^2_1=\frac{1}{1+\delta\mathscr{I}+\mathscr{I}^2};\quad \mu^2_2=\frac{\delta\mathscr{I}}{1+\delta\mathscr{I}+\mathscr{I}^2};\quad \mu^2_3=\frac{\mathscr{I}^2}{1+\delta\mathscr{I}+\mathscr{I}^2}
        \end{split}
    \end{equation*}
    and the likelihood ratios
    \begin{equation*}
        \frac{\mu^1_1}{\mu^2_1}=\mathscr{I}^2; \quad \frac{\mu^1_2}{\mu^2_2}=1; \quad \frac{\mu^1_3}{\mu^2_3}=\mathscr{I}^{-2}.
    \end{equation*}
    Thus, in memory state~2, the DM is unsure about the true state. An $\epsilon$-optimal mechanism chooses $\delta$ strictly positive but small: a positive $\delta$ keeps the mechanism sensitive to information, while a small $\delta$ ensures enough ``exploitation'' that the DM predominantly occupies the two ``extreme'' memory states~1 and~3. The chain structure maximizes the separability of the two extreme memory states:
    \begin{equation*}
        \frac{\mu^1_1}{\mu^2_1}/\frac{\mu^1_3}{\mu^2_3}=\mathscr{I}^4.
    \end{equation*}
    As with $N=M=2$, the utility loss $L^*=\frac{1}{1+\mathscr{I}^2}$ decreases in the informativeness $\mathscr{I}$.

    Two further remarks on this example. First, by \cite{hellman1970learning}, $\frac{1}{1+\mathscr{I}^{2}}$ is the infimum asymptotic utility loss over \emph{all} updating mechanisms with three memory states, so the family in Figure~\ref{fig:examplen2m3} is optimal in the limit $\delta\rightarrow0$. Second, the family delivers an illustration of why an exactly optimal mechanism need not exist. More specifically, the infimum asymptotic utility loss is achieved in the limit $\delta\rightarrow0$ but not attained with $\delta=0$: $\delta=0$ leads to absorption and therefore waste of memory states.

    \subsection*{Example 3: $N=3$, $M=3$}
%
%
%

    I now increase the state space to $N=3$ with $M=3$ memory states, continuing to assume uniform prior $p^1=p^2=p^3=\frac{1}{3}$, and $u(a,\omega)=1$ if and only if $a=\omega$ and $0$ otherwise. The signal space is $S=\lbrace s^1,s^2,s^3\rbrace$ with a symmetric distribution, i.e., $f^\omega(s^{\omega'})=f^{\omega'}(s^{\omega})$, where signal $s^{\omega}$ naturally supports state~$\omega$. Generalizing to $N>2$ raises a novel question that cannot arise under $N=2$: \emph{which states should the DM learn?}

    \paragraph{Ignoring states?}
    Assume that $\frac{f^\omega(s^\omega)}{f^{\omega'}(s^\omega)}=\mathscr{I}>1$ for all $\omega\neq\omega'$. That is,
    \begin{equation*}
        f^\omega(s^\omega) = \frac{\mathscr{I}}{\mathscr{I}+2}, \qquad f^\omega(s^{\omega'}) = \frac{1}{\mathscr{I}+2} \quad \text{for } \omega'\neq\omega,
    \end{equation*}
    where $\mathscr{I}$ measures the informativeness of the signal as in the examples with $N=2$.


	When $N>2$, the DM additionally has to decide \emph{which} states to learn: with $N=3$, he can either ignore no state and try to distinguish each state from the other two, or ignore one state to focus on distinguishing the remaining pair.
    
    \paragraph{Non-ignorant mechanism.}
    First, consider a non-ignorant mechanism in which the DM allocates one memory state to each action: $d(m)=m$ for $m\in\{1,2,3\}$. Following the symmetric construction with $N=M=2$, the transition rule is $\mathscr{T}(m,s^\omega)=\omega$ for all $m$. The stationary distribution under state $\omega=1$ assigns probability $\mu^1_1 = \tfrac{\mathscr{I}}{\mathscr{I}+2}$ to memory state~1 and $\tfrac{1}{\mathscr{I}+2}$ to each of memory states~2 and~3.
    The asymptotic utility and asymptotic utility loss under this mechanism are
    \begin{equation}
        U^{\mathrm{NI}} = \frac{\mathscr{I}}{\mathscr{I}+2};\quad L^{\mathrm{NI}} = \frac{2}{\mathscr{I}+2}.
        \label{eq:NI_payoff}
    \end{equation}

    \begin{figure}[h]
		\centering
		\begin{tikzpicture}[every text node part/.style={align=center}]
			\node[cloud] (m1) {\footnotesize $1$};
			\node[cloud, right = 8em of m1] (m2) {\footnotesize $2$};
			\node[cloud, below right = 7em and 4em of m1] (m3) {\footnotesize $3$};
			\draw[-Latex]
			(m1) edge[bend right=10] node [below, midway] {\footnotesize$s^{2}$} (m2)
			(m2) edge[bend right=10] node [above, midway] {\footnotesize$s^{1}$} (m1)
			(m1) edge[bend right=10] node [left, midway] {\footnotesize$s^{3}$} (m3)
			(m3) edge[bend right=10] node [right, midway] {\footnotesize$s^{1}$} (m1)
			(m2) edge[bend right=10] node [left, midway] {\footnotesize$s^{3}$} (m3)
			(m3) edge[bend right=10] node [right, midway] {\footnotesize$s^{2}$} (m2)
			(m1) edge[loop left, looseness=10] node [left] {\footnotesize$s^{1}$} (m1)
			(m2) edge[loop right, looseness=10] node [right] {\footnotesize$s^{2}$} (m2)
			(m3) edge[loop below, looseness=10] node [below] {\footnotesize$s^{3}$} (m3);
		\end{tikzpicture}
		\caption{Non-ignorant updating mechanism with $N=M=3$. The number in each node denotes the action the DM takes in that memory state: $d(m)=m$. Upon receiving signal~$s^\omega$, the DM transitions to memory state~$\omega$ regardless of his current state.}
		\label{fig:step4nonignorant}
	\end{figure}

    \paragraph{Ignorant mechanism.}
    Now consider an ignorant mechanism: the DM ignores state~3 entirely (so $d(m)\in\{1,2\}$ for all $m$) and dedicates $M=3$ memory states to learning between states~1 and~2. Using the $N=2$, $M=3$ construction from Example~2, now illustrated in Figure~\ref{fig:step4ignorant}, the utility and utility loss of such an ignorant mechanism when $\delta\rightarrow 0$ satisfy:
    \begin{equation}
        U^{I}=\frac{2}{3}\left(\frac{\mathscr{I}^2}{1+\mathscr{I}^2}\right); \quad L^{I} = 1 - \frac{2}{3}\left(\frac{\mathscr{I}^2}{1+\mathscr{I}^2}\right).
        \label{eq:ignorant_payoff}
    \end{equation}

    \begin{figure}[h]
		\centering
		\begin{tikzpicture}[every text node part/.style={align=center}]
			\node[cloud, fill=cyan!60] (m1hi) {\footnotesize $1$};
			\node[cloud, fill=cyan!30, right = 5em of m1hi] (m1lo) {\footnotesize $2$};
			\node[cloud, right = 5em of m1lo] (m2) {\footnotesize $3$};
			\draw[-Latex]
			(m1lo) edge[bend right]  node [above, midway] {\footnotesize$s^1$} (m1hi)
			(m1hi) edge[bend right]  node [below, midway] {\footnotesize$s^2\times\delta$} (m1lo)
			(m1lo) edge[bend right] node [below, midway] {\footnotesize$s^2$} (m2)
			(m2)   edge[bend right] node [above, midway] {\footnotesize$s^1\times\delta$} (m1lo)
			(m1hi) edge[loop left, looseness=10] node [left] {\footnotesize$s^1$} (m1hi)
			(m2)   edge[loop right, looseness=10] node [right] {\footnotesize$s^2$} (m2);
		\end{tikzpicture}
		\caption{Ignorant updating mechanism with $N=3$, $M=3$, ignoring state~$3$. The DM chooses action~$1$ in memory states~$1$ and~$2$, and action~$2$ in memory state~$3$. Memory state~$1$ serves as a ``high-confidence'' state for action~$1$: receiving signal~$s^2$ only demotes the DM to low-confidence state~$2$, not all the way to action~$2$. Signal~$s^3$ is uninformative about states~$1$ vs.\ $2$ and triggers no transition.}
		\label{fig:step4ignorant}
	\end{figure}

    Comparing the ignorant and non-ignorant mechanisms, the ignorant mechanism obviously yields a lower utility in state~3. However, as the ignorant mechanism allocates three memory states to distinguishing states~1 and~2, utilities in those two states are higher. The comparison of utilities across states is shown in Table~\ref{tab:utilbystates}.

    \begin{table}[h]
    \centering
    \begin{tabular}{c|ccc}
         & Non-Ignorant & & Ignorant \\ \hline
        $\omega=1$ & $\dfrac{\mathscr{I}}{2+\mathscr{I}}$ & $<$ & $\dfrac{\mathscr{I}^2}{1+\mathscr{I}^2}$ \\[10pt]
        $\omega=2$ & $\dfrac{\mathscr{I}}{2+\mathscr{I}}$ & $<$ & $\dfrac{\mathscr{I}^2}{1+\mathscr{I}^2}$ \\[10pt]
        $\omega=3$ & $\dfrac{\mathscr{I}}{2+\mathscr{I}}$ & $>$ & $0$
    \end{tabular}
    \caption{Comparison of asymptotic utility by state under the non-ignorant and ignorant mechanisms with $N=M=3$.}
    \label{tab:utilbystates}
    \end{table}


	Note that utilities in states~1 and~2 are higher under the ignorant mechanism for two reasons. First, ignoring state~3 sharpens the distinction between states~1 and~2: the probability ratio of choosing action~1 vis-\`a-vis action~2 in state~1 is $\mathscr{I}^2$ rather than $\mathscr{I}$, i.e., the $\mathscr{I}^2$ versus $\mathscr{I}$ terms in Table~\ref{tab:utilbystates}. Second, ignoring one of three states reduces the effective probability of receiving (and attending to) a ``wrong'' signal from $\frac{2}{2+\mathscr{I}}$ to $\frac{1}{1+\mathscr{I}}$, i.e., the ``$2$'' versus ``$1$'' terms in Table~\ref{tab:utilbystates}. The following observation compares the utility losses of the two mechanisms.
    
    \begin{observation}
        There exists a threshold $\bar{\mathscr{I}}>1$ such that for all $\mathscr{I}\in(1,\bar{\mathscr{I}})$, the ignorant mechanism in Figure~\ref{fig:step4ignorant} achieves strictly lower asymptotic utility loss than the non-ignorant mechanism in Figure~\ref{fig:step4nonignorant}.
    \end{observation}


	Intuitively, when signals are noisy ($\mathscr{I}$ close to~1), distinguishing any pair of states is difficult. As the probability of receiving ``wrong'' signals is high, the non-ignorant mechanism is too sensitive to signals and thus noisy. By ignoring one state, the DM devotes more memory to distinguishing the other two and improves utility. Proposition~\ref{prop:symmetricignorance} formalizes this logic.

    \subsection*{Example 4: which signal supports which state?}
    The examples so far feature symmetric signal structures, under which signal $s^{\omega}$ naturally supports state~$\omega$. The final example illustrates a mathematical complication that arises for general signal structures with $N>2$: even defining what constitutes a supporting signal for each state becomes subtle. As argued above, when $N=2$, a signal supports state~$\omega$ if and only if $f^{\omega}(s)>f^{-\omega}(s)$. With this definition, the sets of signals supporting each state are non-empty. However, when $N>2$, there might exist no signals supporting some state, which makes the design of a good updating mechanism considerably more subtle.

    To see why, consider the following symmetric signal structure with $S=\{s^1,s^2,s^3\}$:
    \begin{equation*}
        \begin{array}{c|ccc}
             & s^1 & s^2 & s^3 \\ \hline
            f^1 & 0.50 & 0.40 & 0.10 \\
            f^2 & 0.40 & 0.20 & 0.40 \\
            f^3 & 0.10 & 0.40 & 0.50
        \end{array}
    \end{equation*}
    If we extend the definition of supporting signals from the example with $N=2$, a signal $s$ supports state~$\omega$ if and only if $\omega=\argmax_{\omega'} f^{\omega'}(s)$. Under this definition, $s^1$ supports state~1, $s^3$ supports state~3, and $s^2$ supports state~1 or state~3. Thus, no signal supports state~2. In fact, one can show that no deterministic definition of supporting signals can give state~2 any confirmatory signal. Note that the optimal mechanisms when $N=2$, as in \cite{hellman1970learning} and \cite{wilson2014bounded}, depend on the existence of confirmatory signals for both states. Therefore, generalizing the optimal mechanisms from $N=2$ to $N>2$ is non-trivial.

    As I will discuss in the next section, a fix is to stochastically ``label'' each signal $s$ as confirmatory for state~$\omega$ with probability proportional to $f^{\omega}(s)\big/\sqrt{\sum_{s'\in S}[f^{\omega}(s')]^{2}}$ --- the likelihood of $s$ in state~$\omega$, scaled by the Euclidean norm of $f^{\omega}$ --- with a common normalizing constant chosen so that the probabilities sum to at most~1 across states (the residual probability corresponds to the case where the signal $s$ is labeled as non-confirmatory for any state). In the example, the norms are $\sqrt{0.42}\approx 0.65$ for states~1 and~3 and $0.6$ for state~2; taking the common constant to be $0.6$, $s^2$ is labeled as confirmatory for state~1 with probability $0.37$, for state~2 with probability $0.20$, for state~3 with probability $0.37$, and as non-confirmatory with probability $0.06$.
    Under this stochastic labeling, there is a strictly positive probability that the DM receives a signal labeled as supporting~$\omega$ for all~$\omega$. Moreover, the probability of receiving a confirmatory signal for the \emph{correct} state strictly exceeds the probability for any wrong state, in every state of the world:
    \begin{equation*}
        \begin{split}
        \Pr[\text{signals support }\omega \mid \omega] &= c\,\frac{\sum_{s\in S}[f^{\omega}(s)]^{2}}{\sqrt{\sum_{s\in S}[f^{\omega}(s)]^{2}}}= c\sqrt{\sum_{s\in S}[f^{\omega}(s)]^{2}}\\
        &> c\,\frac{\sum_{s\in S}f^{\omega}(s)f^{\omega'}(s)}{\sqrt{\sum_{s\in S}[f^{\omega'}(s)]^{2}}}= \Pr[\text{signals support }\omega' \mid \omega],
        \end{split}
    \end{equation*}
    where $c$ is the common normalizing constant and the inequality follows from the Cauchy--Schwarz inequality (strictly, since $f^{\omega}\neq f^{\omega'}$). As I will discuss, this stochastic definition of supporting signals and its property---that the DM is more likely to receive a signal supporting the true state---is key to generalizing the updating mechanisms from \cite{hellman1970learning} and \cite{wilson2014bounded} to $N>2$.


	\paragraph{How does utility change when both $N,M$ increase?} Finally, extending to $N>2$ also allows us to ask how the DM's utility and optimal updating mechanism change as the world becomes more complex ($N$ increases) while the DM becomes more capable ($M$ increases). The answer is presented in the next section through limit results in which $N$ and $M$ grow at different rates.

	\section{Results}\label{section:small}
	\subsection{Are actions close to Bayesian?}
	In the current setup, a Bayesian individual learns perfectly the true state of the world asymptotically: by \cite{doob1949application}, the Bayesian posterior converges almost surely to a mass point at the true state, so a Bayesian chooses $a=\omega$ with probability~1 as $t\rightarrow \infty$ and achieves asymptotic utility loss equal to $0$. Thus, we interpret $L^*_{M}$ as a measure of the distance between the DM's payoff and distribution of actions and that of a Bayesian individual: the higher $L^*_{M}$ is, the more likely the DM chooses $a\neq\omega$, differing from the decision of a Bayesian individual.\footnote{More precisely, $L^*_M$ measures the gap in long-run average payoff between the DM and a Bayesian. Under the additional condition $u(\omega,\omega)-u(a,\omega)=u(a,a)-u(\omega,a)$ for all $a,\omega$ --- which holds for example for $u(a,\omega)=\mathbbm{1}[a=\omega]$ --- $L^*_M$ equals the expected total variation distance between the DM's action and the Bayesian's action $\omega$.} The following Proposition presents the formal results on how relative complexity affects $L^*_{M}$. For ease of exposition, I denote $\bm{u}^{\omega}\equiv ( u(a,\omega))_{a=1}^N$.

	\begin{proposition}
		We have the following results regarding $L^*_{M}$:
		\begin{enumerate}[label=(\roman*)]
			\item $L_{M}^{*}$ is strictly decreasing in $M$ and converges to $0$ as $M\rightarrow \infty$. Moreover, convergence to the Bayesian benchmark is exponential in $\frac{M}{N}$: for each $N$ and $(\bm{u}^{\omega}, p^{\omega}, f^{\omega})_{\omega=1}^{N}$, there exist constants $K>0$ and $r<1$ such that $L^*_M< K r^{\floor{\frac{M-1}{N}}}$;
			\item For each $N$ and $ (f^{\omega})_{\omega=1}^{N}$, there exists a sequence $(m_1,\mathscr{T}_{M},d_M)_{M=1}^{\infty}$ of ``prior-free'' updating mechanisms --- depending only on the signal structures $(f^{\omega})_{\omega=1}^{N}$ but not on the prior beliefs $(p^{\omega})_{\omega=1}^{N}$ or the utility function $(\bm{u}^{\omega})_{\omega=1}^{N}$ --- such that $\lim_{M\rightarrow\infty}L(m_1,\mathscr{T}_{M},d_M)=0$ for all $(\bm{u}^{\omega}, p^{\omega})_{\omega=1}^{N}$. That is, the same mechanism achieves a distribution of actions that approximates the Bayesian benchmark regardless of the prior and payoffs.
		\end{enumerate}
		\label{prop_smallworld}
	\end{proposition}

		To briefly sum up, since the utility loss $L$ is a weighted average of the probabilities of taking incorrect actions across states, $L^*_M\rightarrow 0$ implies that the long-run fraction of each action converges to the Bayesian benchmark. Thus, the Proposition shows that the DM's distribution of actions is closer to, and converges to, Bayesian as the DM's cognitive ability increases. Moreover, the convergence speed depends on the relative complexity of the world, i.e., $\frac{N}{M}$. Proposition~\ref{prop_smallworld} presents results of comparative statics w.r.t.~$M$, thus one natural question is whether $\frac{N}{M}$ truly affects the payoff and distribution of actions. When $N$ changes, naturally, the optimal updating mechanism depends hugely on how the prior belief $p^\omega$, utility matrix $u$ and signals structure $f^\omega$ change. Later in this subsection, we analyze the case where both $N,M\rightarrow \infty$ and impose extra assumptions at the limit, and show the optimal mechanism depends crucially on the ratio $\frac{N}{M}$ in the limit.


	It is also worth contrasting the optimally designed automaton with a natural alternative model of limited memory: a ``naive'' or truncated-Bayesian memory, under which the DM recalls only the last $T$ signals and updates as a Bayesian on that truncated history. Such a memory is itself an automaton whose memory states are the $|S|^{T}$ possible signal strings. A budget of $M$ memory states therefore supports the recall horizon $T=\frac{\log M}{\log\vert S\vert}$. The performance of the naive memory is thus influenced by the size of the \emph{signal space}: as $\vert S\vert$ grows, the recall horizon that a fixed $M$ can support collapses, because the naive memory treats every signal string as distinct regardless of its decision relevance. In contrast, the bound in Proposition~\ref{prop_smallworld}(i), $L^{*}_{M}<Kr^{\floor{(M-1)/N}}$, is free of $\vert S\vert$: the designed updating mechanism takes into account how strongly signals favor each state, and its performance is governed by the relative complexity $\frac{N}{M}$ alone. In particular, along sequences where $\vert S\vert$ grows quickly with $M$, the naive memory's recall horizon stalls and its utility loss remains bounded away from the Bayesian benchmark even as $M\rightarrow\infty$, while the mechanisms of Proposition~\ref{prop_smallworld} still drive the loss to $0$. Even when $\vert S\vert$ is fixed, truncated-Bayesian memory learns at a much slower rate than that characterized in Proposition~\ref{prop_smallworld}(i). Online Appendix~\ref{section:naivememory} presents a numerical illustration.

	In the following, we briefly present the proof of Proposition~\ref{prop_smallworld}. 
	Proving the Proposition is tricky for two reasons. First, characterizing the optimal updating mechanism is intractable: the space of feasible mechanisms is large, and the asymptotic utility loss is neither continuous nor convex in the transition probabilities. Second, as illustrated in Example~4 in Section~\ref{section:example}, it is non-trivial to define the set of confirmatory signals for each state~$\omega$ when $N>2$. As illustrated in Example~4 in Section~\ref{section:example}, a fix is to define whether a signal~$s$ is confirmatory for state~$\omega$ stochastically: a signal~$s$ confirms state~$\omega'$ with probability proportional to $f^{\omega'}(s)$, and confirms no state with the complementary probability, which ensures that in every state~$\omega$ the DM receives confirmatory signals for state~$\omega$ more often than for any other state. In the following, when I say the DM transits from $m$ to $m'$ upon receiving a signal supporting state~$\omega$, I mean that the transition probability from $m$ to $m'$ given signal~$s$ is proportional to $f^\omega(s)$.

	 Now I briefly describe the proof that $L^*_M$ strictly decreases in $M$. The strategy is to show that any mechanism with $M$ memory states can be strictly improved by a mechanism with $M+1$ memory states, and that the improvement is bounded away from~$0$. Take an $\epsilon$-optimal mechanism with $M$ memory states and long-run distributions $(\mu^{\omega}_{m})$, and let $U_{\omega}$ denote its long-run utility in state~$\omega$. Because at least two actions are taken and each action is uniquely optimal in its own state, there is a memory state $\tilde{m}$ whose action $d(\tilde{m})$ is better than the mechanism's average performance in some states of the world and worse in others: the vector with entries $p^{\omega}\mu^{\omega}_{\tilde{m}}\left[U_{\omega}-u(d(\tilde{m}),\omega)\right]$, $\omega=1,\dots,N$, has entries of both signs. I attach a new memory state $M+1$ --- a ``waiting room'' --- to $\tilde{m}$, in three steps:
	 \begin{itemize}
	 	\item \emph{Decision rule.} $d(M+1)=d(\tilde{m})$: in the waiting room the DM takes action $d(\tilde{m})$. All other decisions are unchanged.
	 	\item \emph{Entry.} At $\tilde{m}$, upon receiving a signal $s$, the DM moves to $M+1$ with a small probability $\varepsilon\varphi(s)$; all other transitions are scaled by $1-\varepsilon$ and the DM otherwise stays put. This preserves the \emph{relative} long-run frequencies of the original $M$ memory states: their long-run distribution is only scaled down by a common factor.
	 	\item \emph{Exit.} At $M+1$, upon receiving a signal $s$, the DM returns to $\tilde{m}$ with probability $\psi(s)$, and otherwise stays.
	 \end{itemize}
	 Figure~\ref{fig:Mplus1construction} illustrates the construction. Note that because the asymptotic utility earned while in the original memory states is unchanged (their relative frequencies are preserved), the change in asymptotic utility comes entirely from the time spent in the waiting room, i.e., memory state~$M+1$. In state~$\omega$, the long-run probability of the waiting room is proportional to $\varepsilon\sum_{s}f^{\omega}(s)\varphi(s)\big/\sum_{s}f^{\omega}(s)\psi(s)$, the ratio of the expected entry to the expected exit probability, which differs across states. The change in asymptotic utility is thus a sum over states of how much better or worse $d(\tilde{m})$ is than the mechanism's average performance, weighted by $\varepsilon\sum_{s}f^{\omega}(s)\varphi(s)\big/\sum_{s}f^{\omega}(s)\psi(s)$. The DM takes action $d(\tilde{m})$ more often in every state, but crucially the proof shows that $(\varphi,\psi)$ can always be chosen so that he does so disproportionately in the states where the action is relatively good --- making the change strictly positive and, importantly, bounded away from~$0$ along the $\epsilon$-optimal sequence. Typically, entering after a single signal that is on balance more likely in those states, and leaving after one period, suffices. When no single signal separates the relevant states (e.g., binary signals with $N\geq 3$), a signal-dependent exit rule --- in the spirit of the stochastic labeling above --- makes the time in the waiting room depend on the run of signals between entry and exit; the extra memory state then aggregates evidence across periods, and identifiability guarantees that some finite string of signals does the job.

\begin{figure}[t]
	\centering
	\begin{tikzpicture}[
		box/.style = {draw, rounded corners=8pt, minimum width=4.2cm, minimum height=3.6cm, align=center},
		newstate/.style = {draw, circle, fill=cyan!20, minimum size=2.4em, align=center},
		>=latex
	]
	\node[box, fill=blue!8] (Mbox) {};
	\node[above] at (Mbox.north) {original $M$ memory states};
	\node[circle, fill=cyan!20, draw, minimum size=1.2em, inner sep=0pt] at ([yshift=14pt, xshift=-14pt]Mbox.center) (ma) {};
	\node[circle, fill=cyan!20, draw, minimum size=1.2em, inner sep=0pt] at ([yshift=14pt, xshift=14pt]Mbox.center) (mb) {};
	\node[circle, fill=cyan!20, draw, minimum size=1.2em, inner sep=0pt] at ([yshift=-12pt, xshift=-14pt]Mbox.center) (mc) {};
	\node[circle, fill=cyan!60, draw, thick, minimum size=1.5em, inner sep=0pt] at ([yshift=-12pt, xshift=18pt]Mbox.center) (mt) {\footnotesize $\tilde{m}$};
	\node[below=0.1em of mt, font=\footnotesize] {action $d(\tilde{m})$};
	\draw[->, gray, thick] ([xshift=6pt]Mbox.south) arc[start angle=0, end angle=-270, radius=6pt];
	\node[below=0.3em of Mbox.south, gray, font=\footnotesize] {original transitions, scaled by $1-\varepsilon$};

	\node[newstate, right=3.6cm of Mbox] (Mplus1) {\footnotesize $M\!+\!1$};
	\node[right=0.5em of Mplus1, align=left, font=\footnotesize] {waiting room:\\[-2pt] action $d(\tilde{m})$};
	\draw[->, thick, gray] (Mplus1) edge[loop above, looseness=8] node[above, font=\footnotesize] {otherwise stay} (Mplus1);

	\draw[->, thick, blue!70!black, bend left=20] (mt.east) to node[above, pos=0.62, align=center, font=\footnotesize] {signal $s$: enter\\[-2pt] with prob.\ $\varepsilon\varphi(s)$} (Mplus1.west);
	\draw[->, thick, red!70!black, bend left=20] (Mplus1.south west) to node[below, pos=0.4, align=center, font=\footnotesize] {signal $s$: return\\[-2pt] with prob.\ $\psi(s)$} (mt.south east);
	\end{tikzpicture}
	\caption{Construction of the $(M+1)$-state mechanism from an $M$-state mechanism. The new memory state $M+1$ is a waiting room attached to a single memory state $\tilde{m}$ and assigned $\tilde{m}$'s action. The DM enters it from $\tilde{m}$ upon signal $s$ with probability $\varepsilon\varphi(s)$ and exits back to $\tilde{m}$ upon signal $s$ with probability $\psi(s)$. Time spent in the waiting room differs across states of the world, and the entry and exit rules are chosen so that the DM lingers on action $d(\tilde{m})$ disproportionately in the states where it is relatively good.}
	\label{fig:Mplus1construction}
\end{figure}

	 \tikzstyle{cloud} = [draw, align=left, circle,fill=cyan!20, node distance=3cm,
	 minimum height=1em]
	 \tikzstyle{cloud1} = [draw, align=left, circle,fill=black, node distance=1cm,
	 minimum height=0.01cm]
	 \tikzstyle{cloud2} = [draw, align=left, circle,fill=red, node distance=1cm,
	 minimum height=0.01cm]
	 \tikzset{
	 	dot/.style = {circle, fill, minimum size=#1,
	 		inner sep=0pt, outer sep=0pt},
	 	dot/.default = 3pt 
	 }

	Now to prove the rest of Proposition~\ref{prop_smallworld}, I construct a simple updating mechanism, illustrated in Figures~\ref{fig:simplelearningalgosmallworld} and~\ref{fig:simplelearningalgosmallworldzoomin}. The $M$ memory states are segmented into $N$ branches of equal size and a central memory state, reminiscent of a star network. For ease of exposition, the remaining memory states are discarded, and in the following, I rename the memory states as $\lbrace0\rbrace\cup\lbrace k1,\cdots,k\floor{\frac{M-1}{N}}\rbrace_{k=1}^{N}$ such that memory state $0$ is the central memory state, and $(k1,\cdots,k\floor{\frac{M-1}{N}})$ are memory states in the $k$-th branch. In all memory states in the $k$-th branch, the DM chooses action~$k$; otherwise, e.g., when the DM is in the central memory state, he takes action~$1$.\footnote{The decision rules in memory states not in $\lbrace k1,\cdots,k\floor{\frac{M-1}{N}}\rbrace_{k=1}^{N}$ do not affect the proof and result.} The transition rule is described as follows.


	\begin{figure}
		\centering
		\begin{tikzpicture}
		\node[cloud2] (0lambda) {\footnotesize $0$};
		
		\node[cloud, above left = 2em and 10em of 0lambda] (1lambda-1) {};
		\node[cloud, above = 1em of 1lambda-1] (1lambda-2) {};
		\node[above = 1em of 1lambda-2] (1lambda-3) {$\vdots$};
		\node[cloud, above = 1em of 1lambda-3] (1lambda-4) {\footnotesize};
		\node[cloud, above = 1em of 1lambda-4] (1lambda-5) {\footnotesize};
		\node[above left = 1em and 1em of 1lambda-5] (1topleft) {};
		\node[above right = 1em and 1em of 1lambda-5] (1topright) {};
		\draw [decorate,decoration={brace,amplitude=5pt},xshift=0pt,yshift=0pt]
		(1topleft) -- (1topright) node [black,midway,yshift=0.6cm]
		{\footnotesize action~$1$};
		
		\node[below left = 0.5em and 1.5em of 1lambda-1] (arrowbottom) {};
		\node[above left = 0.5em and  1.5em of 1lambda-5] (arrowtop) {};
		
		\node[cloud, above left = 2em and 5em of 0lambda] (2lambda-1) {};
		\node[cloud, above = 1em of 2lambda-1] (2lambda-2) {};
		\node[above = 1em of 2lambda-2] (2lambda-3) {$\vdots$};
		\node[cloud, above = 1em of 2lambda-3] (2lambda-4) {};
		\node[cloud, above = 1em of 2lambda-4] (2lambda-5) {};
		\node[above left = 1em and 1em of 2lambda-5] (2topleft) {};
		\node[above right = 1em and 1em of 2lambda-5] (2topright) {};
		\draw [decorate,decoration={brace,amplitude=5pt},xshift=0pt,yshift=0pt]
		(2topleft) -- (2topright) node [black,midway,yshift=0.6cm]
		{\footnotesize action~$2$};
		
		\node[above = 2em of 0lambda] (3lambda-1) {$\cdots$};
		\node[above = 2em of 3lambda-1] (3lambda-2) {$\vdots$};
		\node[above = 2em of 3lambda-2] (3lambda-3) {$\cdots$};
		
		\node[cloud, above right = 2em and 5em of 0lambda] (4lambda-1) {};
		\node[cloud, above = 1em of 4lambda-1] (4lambda-2) {};
		\node[above = 1em of 4lambda-2] (4lambda-3) {$\vdots$};
		\node[cloud, above = 1em of 4lambda-3] (4lambda-4) {};
		\node[cloud, above = 1em of 4lambda-4] (4lambda-5) {};
		\node[above left = 1em and 1em of 4lambda-5] (4topleft) {};
		\node[above right = 1em and 1em of 4lambda-5] (4topright) {};
		\draw [decorate,decoration={brace,amplitude=5pt},xshift=0pt,yshift=0pt]
		(4topleft) -- (4topright) node [black,midway,yshift=0.6cm]
		{\footnotesize action~$N-1$};	
		
		\node[cloud, above right = 2em and 10em of 0lambda] (5lambda-1) {};
		\node[cloud, above = 1em of 5lambda-1] (5lambda-2) {};
		\node[above = 1em of 5lambda-2] (5lambda-3) {$\vdots$};
		\node[cloud, above = 1em of 5lambda-3] (5lambda-4) {};
		\node[cloud, above = 1em of 5lambda-4] (5lambda-5) {};
		\node[above left = 1em and 1em of 5lambda-5] (5topleft) {};
		\node[above right = 1em and 1em of 5lambda-5] (5topright) {};
		\draw [decorate,decoration={brace,amplitude=5pt},xshift=0pt,yshift=0pt]
		(5topleft) -- (5topright) node [black,midway,yshift=0.6cm]
		{\footnotesize action~$N$};
		\draw [decorate,decoration={brace,mirror, amplitude=5pt},xshift=2cm,yshift=0pt]
		([xshift=.5cm]5lambda-1.east) -- ([xshift=.5cm]5lambda-5.east) node [align=center,black,midway,xshift=1.2cm,yshift=0cm]
		{\footnotesize with length \\\footnotesize$\floor{\frac{M-1}{N}}$};
		
		\draw[-Latex]
		(0lambda) edge[loop below,looseness=10] node [left] {} (0lambda)
		(1lambda-1) edge[loop left,looseness=10] node [left] {} (1lambda-1)
		(1lambda-2) edge[loop left,looseness=10] node [right] {} (1lambda-2)
		(1lambda-4) edge[loop left,looseness=10] node [left] {} (1lambda-4)
		(1lambda-5) edge[loop left,looseness=10] node [right] {} (1lambda-5)		
		(2lambda-1) edge[loop left,looseness=10] node [left] {} (2lambda-1)
		(2lambda-2) edge[loop left,looseness=10] node [right] {} (2lambda-2)
		(2lambda-4) edge[loop left,looseness=10] node [left] {} (2lambda-4)
		(2lambda-5) edge[loop left,looseness=10] node [right] {} (2lambda-5)
		(4lambda-1) edge[loop right,looseness=10] node [left] {} (4lambda-1)
		(4lambda-2) edge[loop right,looseness=10] node [right] {} (4lambda-2)
		(4lambda-4) edge[loop right,looseness=10] node [left] {} (4lambda-4)
		(4lambda-5) edge[loop right,looseness=10] node [right] {} (4lambda-5)
		(5lambda-1) edge[loop right,looseness=10] node [left] {} (5lambda-1)
		(5lambda-2) edge[loop right,looseness=10] node [right] {} (5lambda-2)
		(5lambda-4) edge[loop right,looseness=10] node [left] {} (5lambda-4)
		(5lambda-5) edge[loop right,looseness=10] node [right] {} (5lambda-5);
		
		\draw[-Latex]
		(1lambda-1) edge[bend right] node [below, midway] {} (1lambda-2)
		(1lambda-2) edge[bend right] node [below, midway] {} (1lambda-1)
		(1lambda-3) edge[bend right] node [below, midway] {} (1lambda-2)
		(1lambda-2) edge[bend right] node [below, midway] {} (1lambda-3)		
		(1lambda-3) edge[bend right] node [below, midway] {} (1lambda-4)
		(1lambda-4) edge[bend right] node [below, midway] {} (1lambda-3)
		(1lambda-4) edge[bend right] node [below, midway] {} (1lambda-5)
		(1lambda-5) edge[bend right] node [below, midway] {} (1lambda-4);
		
		\draw[-Latex]
		(2lambda-1) edge[bend right] node [below, midway] {} (2lambda-2)
		(2lambda-2) edge[bend right] node [below, midway] {} (2lambda-1)
		(2lambda-3) edge[bend right] node [below, midway] {} (2lambda-2)
		(2lambda-2) edge[bend right] node [below, midway] {} (2lambda-3)		
		(2lambda-3) edge[bend right] node [below, midway] {} (2lambda-4)
		(2lambda-4) edge[bend right] node [below, midway] {} (2lambda-3)
		(2lambda-4) edge[bend right] node [below, midway] {} (2lambda-5)
		(2lambda-5) edge[bend right] node [below, midway] {} (2lambda-4);
		
		\draw[-Latex]
		(4lambda-1) edge[bend right] node [below, midway] {} (4lambda-2)
		(4lambda-2) edge[bend right] node [below, midway] {} (4lambda-1)
		(4lambda-3) edge[bend right] node [below, midway] {} (4lambda-2)
		(4lambda-2) edge[bend right] node [below, midway] {} (4lambda-3)		
		(4lambda-3) edge[bend right] node [below, midway] {} (4lambda-4)
		(4lambda-4) edge[bend right] node [below, midway] {} (4lambda-3)
		(4lambda-4) edge[bend right] node [below, midway] {} (4lambda-5)
		(4lambda-5) edge[bend right] node [below, midway] {} (4lambda-4);
		
		\draw[-Latex]
		(5lambda-1) edge[bend right] node [below, midway] {} (5lambda-2)
		(5lambda-2) edge[bend right] node [below, midway] {} (5lambda-1)
		(5lambda-3) edge[bend right] node [below, midway] {} (5lambda-2)
		(5lambda-2) edge[bend right] node [below, midway] {} (5lambda-3)		
		(5lambda-3) edge[bend right] node [below, midway] {} (5lambda-4)
		(5lambda-4) edge[bend right] node [below, midway] {} (5lambda-3)
		(5lambda-4) edge[bend right] node [below, midway] {} (5lambda-5)
		(5lambda-5) edge[bend right] node [below, midway] {} (5lambda-4);
		
		\draw[-Latex]
		(0lambda) edge[bend left=40] node [below, midway] {} (1lambda-1)
		(1lambda-1) edge[bend right=70] node [below, midway] {} (0lambda)
		(0lambda) edge[bend right] node [below, midway] {} (2lambda-1)
		(2lambda-1) edge[bend right] node [below, midway] {} (0lambda)
		(0lambda) edge[bend right] node [below, midway] {} (3lambda-1)
		(3lambda-1) edge[bend right] node [below, midway] {} (0lambda)
		(0lambda) edge[bend right] node [below, midway] {} (4lambda-1)
		(4lambda-1) edge[bend right] node [below, midway] {} (0lambda)
		(0lambda) edge[bend right=40] node [below, midway] {} (5lambda-1)
		(5lambda-1) edge[bend left=70] node [below, midway] {} (0lambda);
		\end{tikzpicture}
		\caption{A simple updating mechanism that achieves perfect learning in small worlds for all $N$, $(\bm{u}^{\omega},p^\omega)_{\omega=1}^{N}$, and with convergence rate $\frac{1}{N}$.}
		\label{fig:simplelearningalgosmallworld}
	\end{figure}

\begin{figure}
		\centering
		\begin{tikzpicture}[every text node part/.style={align=center}, >=Latex]
		\node[cloud2] (0) at (0,0) {\footnotesize $0$};


		\node[cloud] (11) at (-3.46, 2.0) {\footnotesize $11$};
		\node[cloud] (12) at (-5.46, 3.15) {\footnotesize $12$};
		\node (1dots) at (-7.06, 4.1) {$\vdots$};

		\node[cloud] (21) at (3.46, 2.0) {\footnotesize $21$};
		\node[cloud] (22) at (5.46, 3.15) {\footnotesize $22$};
		\node (2dots) at (7.06, 4.1) {$\vdots$};

		\node[cloud] (w1) at (0, -4.0) {\footnotesize $\omega 1$};
		\node[cloud] (w2) at (0, -5.8) {\footnotesize $\omega 2$};
		\node (wdots) at (0, -7.3) {$\vdots$};

		\draw[->] (0) edge[bend left=30] node[below left=-2pt, font=\footnotesize] {$f^1(s)$} (11);
		\draw[->] (0) edge[bend right=30] node[below right=-2pt, font=\footnotesize] {$f^2(s)$} (21);
		\draw[->] (0) edge[bend left=30] node[right, font=\footnotesize] {$f^\omega(s)$} (w1);

		\draw[->] (11) edge[bend left=30] node[above=5pt, font=\footnotesize] {$\tfrac{1}{\delta}\textstyle\sum_{\omega'\neq 1}f^{\omega'}(s)$} (0);
		\draw[->] (21) edge[bend right=30] node[above=5pt, font=\footnotesize] {$\tfrac{1}{\delta}\textstyle\sum_{\omega'\neq 2}f^{\omega'}(s)$} (0);
		\draw[->] (w1) edge[bend left=30] node[left, font=\footnotesize] {$\tfrac{1}{\delta}\textstyle\sum_{\omega'\neq \omega}f^{\omega'}(s)$} (0);

		\draw[->] (11) edge[bend left=30] node[below left=-2pt, font=\footnotesize] {$f^1(s)$} (12);
		\draw[->] (12) edge[bend left=30] node[above right=-3pt, font=\footnotesize] {$\tfrac{1}{\delta}\textstyle\sum_{\omega'\neq 1}f^{\omega'}(s)$} (11);

		\draw[->] (12) edge[bend left=30] node[below, font=\footnotesize] {} (1dots);
		\draw[->] (1dots) edge[bend left=30] node[above, font=\footnotesize] {} (12);

		\draw[->] (21) edge[bend right=30] node[below right=-3pt, font=\footnotesize] {$f^2(s)$} (22);
		\draw[->] (22) edge[bend right=30] node[above left=-3pt, font=\footnotesize] {$\tfrac{1}{\delta}\textstyle\sum_{\omega'\neq 2}f^{\omega'}(s)$} (21);

		\draw[->] (22) edge[bend right=30] node[below, font=\footnotesize] {} (2dots);
		\draw[->] (2dots) edge[bend right=30] node[above, font=\footnotesize] {} (22);

		\draw[->] (w1) edge[bend left=30] node[right, font=\footnotesize] {$f^\omega(s)$} (w2);
		\draw[->] (w2) edge[bend left=30] node[left, font=\footnotesize] {$\tfrac{1}{\delta}\textstyle\sum_{\omega'\neq \omega}f^{\omega'}(s)$} (w1);

		\draw[->] (w2) edge[bend left=30] node[right, font=\footnotesize] {} (wdots);
		\draw[->] (wdots) edge[bend left=30] node[left, font=\footnotesize] {} (w2);


		\end{tikzpicture}
		\caption{Zoom-in of Figure~\ref{fig:simplelearningalgosmallworld}, showing the transitions between memory state~$0$ and the first two memory states of three of the $N$ branches. Memory states are labeled $\omega k$ ($k$-th state in branch~$\omega$). Upon receiving signal~$s$, the DM in state~$\omega k$ moves one state further from the center to $\omega(k{+}1)$ with probability $f^\omega(s)$ (a confirmatory signal for state~$\omega$), one state closer to the center to $\omega(k{-}1)$ with probability $\frac{1}{\delta}\sum_{\omega'\neq\omega}f^{\omega'}(s)$ (a disconfirmatory signal, underweighted by $\frac{1}{\delta}$), and stays with complementary probability. At the extreme ($k=\lfloor\frac{M-1}{N}\rfloor$), the DM stays instead of moving further; at the bottom ($\omega 1$), a disconfirmatory signal sends the DM back to the central state~$0$.}
		\label{fig:simplelearningalgosmallworldzoomin}
	\end{figure}

    First, the DM starts in the central memory state~$0$.\footnote{The starting memory state has no impact on the long-run distribution over the memory states and does not affect the asymptotic payoff.} If he receives a confirmatory signal for state~$\omega$, he transits to memory state~$\omega 1$, the first state in branch~$\omega$; if he receives a signal that does not confirm any state, he stays in memory state~$0$. Second, suppose the DM is in memory state~$\omega k$, the $k$-th state in branch~$\omega$. If he receives a confirmatory signal for state~$\omega$, he moves one state further from the center to~$\omega(k{+}1)$, unless he is already at the extreme of the branch ($k=\floor{\frac{M-1}{N}}$), in which case he stays. Third, if he receives a confirmatory signal for some other state~$\omega'\neq\omega$, then with probability $\frac{1}{\delta}<1$ he moves one state closer to the center: to~$\omega(k{-}1)$ if $k\geq 2$, or back to the central state~$0$ if $k=1$; with probability $1-\frac{1}{\delta}$ he stays in his current memory state. Lastly, if he receives a signal that does not confirm any state, he stays in his current memory state. Intuitively, this simple updating mechanism tracks only one action at a time --- the action associated with the branch the DM currently occupies --- and a more extreme position within the branch reflects greater certainty that this action is optimal. The non-transition upon receiving disconfirmatory signals (captured by $\frac{1}{\delta}<1$) ensures that the DM does not switch between branches too readily.

The proof involves choosing a sufficiently large $\delta$ such that the DM is more likely to move towards the extreme of a branch than towards the center, as in Example~2. This ensures enough ``exploitation'' that the DM does not switch between actions too often. Crucially, when $\frac{M}{N}$ increases, the maximum number of confirmatory signals the DM can ``store'' in each branch increases, and the DM is more likely to be in the correct branch choosing the correct action, i.e., matching the state of the world. When $\frac{M}{N}\rightarrow\infty$, the DM almost surely learns perfectly the true state as $t\rightarrow\infty$. Lastly, when $N$ increases, the DM needs more memory states to record confirmatory information for every state, and thus an increase in $M$ induces a smaller improvement in the DM's decisions, explaining why a larger $N$ slows down the convergence rate of $L^*_M$.

\paragraph{Discussion}
	Here I point out several noteworthy implications of the analysis. Several features of the simple updating mechanism reflect a fundamental difficulty that arises when $N\geq 3$: each signal carries information about multiple pairwise comparisons simultaneously, and the mechanism must use this information efficiently within the memory constraint.

	First, as argued previously, unlike \cite{cover1969hypothesis} and \cite{wilson2014bounded} with $N=2$, when $N>2$, the definition of confirmatory signals for each state must in general be stochastic in order for the simple updating mechanism to achieve near-optimal performance. This means that upon receiving the same signal, the individual will sometimes regard it as supporting one state, and sometimes regard it as supporting a different state. With $N=2$, there is a natural deterministic partition of signals into those supporting each state. With $N\geq 3$, however, a signal that is most likely under state~$\omega$ may also be informative for distinguishing other pairs of states; a deterministic labeling would discard this information, while stochastic labeling allows the mechanism to exploit it. This resembles empirical evidence that documents heterogeneous interpretations of the same pieces of information, especially when the piece of information is imprecise (\cite{gaines2007same}).

	Second, the same difficulty underlies the na\"ive-memory (truncated-Bayesian) comparison presented after Proposition~\ref{prop_smallworld}: the na\"ive memory stores raw signal strings, so its memory requirement grows with $\vert S\vert$, whereas the simple updating mechanism compresses signals into decision-relevant statistics --- how strongly the evidence favors each state --- so its requirement depends on $N$ alone. This distinction matters in particular when the DM observes high-dimensional data but the underlying uncertainty is low-dimensional. In such situations, the simple updating mechanism is much more efficient than the na\"ive memory, as the numerical illustration in Online Appendix~\ref{section:naivememory} shows.

	Third, as shown in Proposition~\ref{prop_smallworld}(ii), the simple updating mechanism described in this section approximates perfect learning for all $N$ and $(\bm{u}^{\omega}, p^{\omega})_{\omega=1}^{N}$ in small worlds when $M$ is large. Therefore, no knowledge of prior belief $(p^{\omega})_{\omega=1}^{N}$ and utility matrices $(\bm{u}^{\omega})_{\omega=1}^{N}$ is required. This is particularly consistent with the result that the distribution of actions is close to Bayesian in small worlds, as the simple updating mechanism is parsimonious and easy to implement.

	Fourth, perfect learning with the simple updating mechanism in very small worlds is robust to ``implementation errors'', as shown in the online Appendix~\ref{section:robustness}. Roughly speaking, I assume that the DM mistakenly transits to a neighboring memory state with some probability $\gamma$ in each period regardless of the signal realization $s$. Such local mistakes could be induced by mistakes in the perception of signals or imperfect tracking (local fluctuation) of memory states. Online Appendix~\ref{section:robustness} shows that Proposition~\ref{prop_smallworld}(ii) holds for all $\gamma\in [0,1)$, further strengthening the result of perfect learning in very small worlds.

    Lastly, the Proposition is a comparative statics w.r.t.~$M$. Therefore, it remains to see whether the ratio $N/M$ is important. However, as $N$ changes, the decision problem would be highly sensitive to the parameters related to the newly introduced states. For example, adding a state with $p^\omega$ close to $0$ would be very different from adding a state with a larger prior likelihood. Similarly, adding a state where there are signals easily ruling out that state relative to other states would be different from the situation where signals are close to uninformative. In the following, to side-step these issues, I analyze the limit where $N,M$ both go to $\infty$ and show that the limit payoff and distributional results depend on the ratio $\frac{N}{M}$ at the limit.

    \subsubsection{$N,M\rightarrow \infty$: Assumptions and main result}\label{sec:NMlimit_small}

    In this subsection, I present the analysis for $N,M\rightarrow \infty$, showing that the qualitative distinction between small and big worlds is preserved at the limit. Formally, I consider a sequence of decision problems $\left\lbrace(\bm{u}^{\omega}_{N},p^{\omega}_{N},f^{\omega}_{N})_{\omega=1}^{N},M_{N}\right\rbrace_{N\in\mathcal{N}}$, indexed by the number of states~$N$, where $\mathcal{N}$ is an infinite set of positive integers, $f^{\omega}_{N}$ is a probability mass function on a finite signal space $S_{N}$, $M_{N}\rightarrow\infty$, and $N=\Theta(M_{N}^{h})$.\footnote{$N=\Theta(M_{N}^{h})$ means that there exist constants $0<c\leq C$ such that $cM_{N}^{h}\leq N\leq CM_{N}^{h}$ for all $N\in\mathcal{N}$. Similarly, $x=O(y)$ means $x\leq Cy$ for a constant $C>0$. Indexing the decision problems by the (strictly increasing) number of states is without loss of generality.} With $h<1$, $\frac{N}{M_{N}}\rightarrow 0$, corresponding to small worlds; with $h\geq 1$, $\liminf_{N\rightarrow\infty}\frac{N}{M_{N}}> 0$, corresponding to big worlds. Unless stated otherwise, all limits below are taken along $N\in\mathcal{N}$. I impose the following assumptions on the limit environment.

\begin{assumption}{(Limit-FS)}\label{assumption_fullsupportinfinite}
	For any infinite subset $\mathcal{N}'\subseteq\mathcal{N}$ and any sequence of subsets of states $(\Omega_{N})_{N\in\mathcal{N}'}$ with $\Omega_{N}\subseteq \lbrace 1,\cdots, N\rbrace$, for which the two limits below exist as $N\rightarrow\infty$ along $\mathcal{N}'$,\footnote{Quantifying over subsequences avoids presupposing that any limit exists along the full sequence $\mathcal{N}$: the two limits may fail to exist when the environments vary with $N$, and the assumption then constrains the sequence through every subsequence along which they do exist.}
	\begin{equation}
		\lim_{N\rightarrow \infty}\sum_{\omega\in\Omega_{N}}p_{N}^{\omega}>0\text{ \quad if $\lim_{N\rightarrow \infty}\frac{\vert\Omega_{N}\vert}{N}>0$.}
		\label{eq:fullsupportinfinite}
	\end{equation}
\end{assumption}

That is, any sequence of subsets of states that constitutes a non-negligible fraction of the state space at the limit must also carry non-negligible prior probability. This is the natural analogue of the full-support assumption $p^\omega>0$ in the baseline model. For example, the uniform prior $p^{\omega}_{N}=\frac{1}{N}$ satisfies Assumption~\ref{assumption_fullsupportinfinite}. Define
\begin{equation}
	\bar{p}_{0}\coloneqq \sup\left\lbrace\lim_{N\rightarrow \infty}\sum_{\omega\in\Omega_{N}}p_{N}^{\omega}:\; \Omega_{N}\subseteq \lbrace 1,\cdots, N\rbrace,\;\lim_{N\rightarrow\infty}\frac{\vert\Omega_{N}\vert}{M_{N}}=0\right\rbrace,
	\label{eq:p0bar}
\end{equation}
the maximum prior probability that can concentrate on a sequence of sets of states whose cardinality is vanishing relative to the memory size $M_{N}$. Note that when $h<1$ (small worlds), $M_{N}$ grows faster than $N$, so even $\Omega_{N}=\lbrace 1,\ldots, N\rbrace$ satisfies $\vert\Omega_{N}\vert/M_{N}\rightarrow 0$; hence $\bar{p}_{0}=1$. When $h\geq 1$ (big worlds), $M_{N}\leq O(N)$, so the constraint $\vert\Omega_{N}\vert/M_{N}\rightarrow 0$ is binding and $\bar{p}_{0}<1$ in general.

%
%

A stronger version of \ref{assumption_fullsupportinfinite} replaces the ``if'' with ``if and only if'':

\begin{assumption}{(Limit-FS$^+$)}\label{assumption_fullsupportinfiniteplus}
	For any infinite subset $\mathcal{N}'\subseteq\mathcal{N}$ and any sequence of subsets of states $(\Omega_{N})_{N\in\mathcal{N}'}$ with $\Omega_{N}\subseteq \lbrace 1,\cdots, N\rbrace$, for which the two limits below exist as $N\rightarrow\infty$ along $\mathcal{N}'$,
	\begin{equation}
		\lim_{N\rightarrow \infty}\sum_{\omega\in\Omega_{N}}p_{N}^{\omega}>0\text{ \quad if and only if $\lim_{N\rightarrow \infty}\frac{\vert\Omega_{N}\vert}{N}>0$.}
		\label{eq:fullsupportinfiniteplus}
	\end{equation}
\end{assumption}

Assumption~\ref{assumption_fullsupportinfiniteplus} strengthens \ref{assumption_fullsupportinfinite} by additionally requiring that any sequence of subsets constituting a \emph{negligible} fraction of the state space must also carry negligible prior probability. This implies $\bar{p}_{0}=0$ in big worlds. In other words, at the limit, no state is infinitely more likely than any other state: prior probability cannot concentrate on a vanishing fraction of the state space. For example, priors with an atom --- $p^{1}_{N}=\frac{1}{2}$ and $p^{\omega}_{N}=\frac{1}{2(N-1)}$ for $\omega\geq 2$ --- satisfy Assumption~\ref{assumption_fullsupportinfinite} but violate Assumption~\ref{assumption_fullsupportinfiniteplus}; in big worlds they yield $\bar{p}_{0}=\frac{1}{2}$, as prior probability concentrates on the atom. I also impose the following two assumptions on the information structures.

%
%

\begin{assumption}{(Limit-NRO)}\label{assumption_noperfectsignalinfinite}
	There exists $\varsigma>0$ such that for all $N\in\mathcal{N}$ and all $\omega,\omega'\in\lbrace 1,\cdots,N\rbrace$,
	\begin{equation*}
		\min_{s\in S_{N}}\frac{f_{N}^{\omega}(s)}{f_{N}^{\omega'}(s)}\geq\varsigma.
	\end{equation*}
\end{assumption}

That is, no signal is infinitely more likely under one state than under another, uniformly along the sequence --- the analogue of the baseline assumption that no signal perfectly rules out any state, with the implied bound $\varsigma>0$ holding uniformly in~$N$.

\begin{assumption}{(Limit-ID)}\label{assumption_identifiableinfinte}
	For every $\varepsilon>0$, there exists a sequence of subsets $\widehat{\Omega}_{N}\subseteq\lbrace 1,\cdots, N\rbrace$ such that $	\liminf_{N\rightarrow\infty}\sum_{\omega\in\widehat{\Omega}_{N}}p_{N}^{\omega}\geq 1-\varepsilon$
	\begin{equation*}
		\kappa:=\liminf_{N\rightarrow\infty}\,\min_{\substack{\omega,\omega'\in\widehat{\Omega}_{N}\\ \omega'\neq\omega}}\left\lbrace-\log\dfrac{\sum_{s\in S_{N}} f_{N}^{\omega}(s)f_{N}^{\omega'}(s)}{\sqrt{\sum_{s\in S_{N}} (f_{N}^{\omega}(s))^{2}}\sqrt{\sum_{s\in S_{N}} (f_{N}^{\omega'}(s))^{2}}}\right\rbrace>0.
	\end{equation*}
\end{assumption}

That is, there is a sequence of sets of states, carrying almost all prior probability, within which states remain pairwise identifiable at the limit: the pairwise Cauchy--Schwarz distances of their signal structures stay uniformly bounded away from~$0$.\footnote{Online Appendix~\ref{section:signalstructure} presents an information structure that satisfies Assumptions~\ref{assumption_noperfectsignalinfinite} and~\ref{assumption_identifiableinfinte}.} This is the analogue of the baseline assumption that no two signal structures coincide.\footnote{Note that Assumption~\ref{assumption_identifiableinfinte} is a genuine restriction: for instance, the symmetric signal structure of Section~\ref{sec:symmetry} violates it for any fixed $\mathscr{I}$, as its Cauchy--Schwarz ratio $\frac{2\mathscr{I}+N-2}{\mathscr{I}^{2}+N-1}$ converges to $1$, so identifiability degrades as $N$ grows --- one of the driving forces behind Proposition~\ref{prop:symmetricignorance}.} For ease of exposition, I also assume $u_{N}(\omega,\omega)=1$ and $u_{N}(\omega',\omega)=0$ for all $\omega\neq\omega'$ and all $N$. Denote $L^*_N$ as the infimum asymptotic utility loss with $N$ states and $M_{N}$ memory states.


    \begin{proposition}\label{prop_smallworldinfiniteN}
        Fix a sequence of decision problems $\left\lbrace (\bm{u}^{\omega}_{N},p^{\omega}_{N},f^{\omega}_{N})_{\omega=1}^{N},M_{N}\right\rbrace_{N\in\mathcal{N}}$ satisfying Assumptions~\ref{assumption_fullsupportinfinite}, \ref{assumption_noperfectsignalinfinite}, and \ref{assumption_identifiableinfinte}, where $N=\Theta(M_{N}^{h})$.
        \begin{enumerate}[label=(\roman*)]
            \item \emph{(Small worlds.)} If $h<1$, then $\lim_{N\rightarrow\infty}L^{*}_{N}=0$.
            \item \emph{(Big worlds.)} If $h\geq 1$, then for every $\xi'>0$ and every sequence of updating mechanisms $\lbrace m_{1N},\mathscr{T}_{N},d_{N}\rbrace_{N\in\mathcal{N}}$,
            \begin{equation}\label{eq:limitignorancedisplay}
                \lim_{N\rightarrow\infty}\frac{\left\vert\left\lbrace \omega:\, \lim_{T\rightarrow\infty}\tfrac{1}{T}\sum_{t=1}^{T}\Pr\nolimits^{\omega'}(a_{t}=\omega)\geq\xi'\text{ for some }\omega'\right\rbrace\right\vert}{M_{N}}=0,
            \end{equation}
            where $\Pr^{\omega'}$ denotes the probability distribution over action sequences induced by $(m_{1N},\mathscr{T}_{N},d_{N})$ conditional on the true state being~$\omega'$. Consequently, whenever $L^{*}_{N}$ converges, $\lim_{N\rightarrow\infty}L^{*}_{N}=1-\overline{p}_{0}$; under Assumption~\ref{assumption_fullsupportinfiniteplus}, $L^{*}_{N}$ converges and $\lim_{N\rightarrow\infty}L^{*}_{N}=1$.
        \end{enumerate}
        Moreover, (i) and (ii) imply that, as $N,M_{N}$ grow, in small worlds the loss eventually falls strictly below its level at any fixed point of the sequence, while in big worlds, under Assumption~\ref{assumption_fullsupportinfiniteplus}, it eventually rises strictly above it: for any $N_{0}$ with $L^{*}_{N_{0}}>0$ (resp.\ $L^{*}_{N_{0}}<1$), we have $L^{*}_{N}<L^{*}_{N_{0}}$ (resp.\ $L^{*}_{N}>L^{*}_{N_{0}}$) for all large enough $N$.\footnote{Term-by-term monotonicity of $L^{*}_{N}$ cannot hold at this level of generality: the assumptions do not restrict how the environment varies along the sequence, so the ``difficulty'' of the decision problem may oscillate to any degree.}
    \end{proposition}

    The threshold $h=1$ has a clean economic interpretation: it separates environments in which cognitive resources scale faster than complexity ($h<1$) from those in which they scale at most as fast ($h\geq 1$). When $h<1$ (small worlds), the proof constructs the same simple updating mechanism as in Proposition~\ref{prop_smallworld}: as $\frac{M_{N}}{N}\rightarrow\infty$, the DM can allocate arbitrarily many memory states to each action, and learning approaches the Bayesian benchmark. When $h\geq 1$ (big worlds), statement (ii) shows that, under \emph{any} updating mechanism, almost all actions are chosen only negligibly often in the long run, whatever the true state --- the DM effectively ignores almost all states, a result that we elaborate in Section~\ref{sec:ignorance}. The utility loss is then bounded below by the prior probability of the ignored states. Under \ref{assumption_fullsupportinfinite} alone, the prior can put non-vanishing mass on a sequence of subsets $\Omega_{N}$ with $\vert\Omega_{N}\vert/M_{N}\rightarrow 0$, i.e., $\bar{p}_{0}$ can be strictly positive. In this case, the DM optimally focuses his memory on such a subset and recovers utility close to $\bar{p}_{0}$, yielding a limit loss of $1-\bar{p}_{0}<1$ (whenever $L^{*}_{N}$ converges). Under \ref{assumption_fullsupportinfiniteplus}, in contrast, every subset with vanishing fraction in $N$ --- and \emph{a fortiori} every subset with vanishing fraction in $M_{N}$, since $M_{N}\leq O(N)$ in big worlds --- carries vanishing prior. Hence $\bar{p}_{0}=0$ and the loss converges to $1$: the DM learns essentially nothing at the limit.

    The final statement of the proposition expresses the same small-world/big-world contrast in terms of how the loss eventually changes along the sequence: as $N$ grows, optimal performance in small worlds eventually improves upon that at any fixed point of the sequence, whereas in big worlds (under Assumption~\ref{assumption_fullsupportinfiniteplus}) it eventually deteriorates below it. In big worlds, the DM's memory per state fails to grow, and the learning problem becomes harder despite the larger absolute resources $M_{N}$.

	\subsection{Is ignorance optimal?}\label{sec:ignorance}
	\subsubsection{General Environments}\label{sec:ignorancegeneral}
	In this subsection, I present the main results on whether ignorant behavior is optimal in small and big worlds.

\begin{definition}{1}
An updating mechanism $(m_1,\mathscr{T},d)$ \emph{ignores} state~$\omega$ if there does not exist any memory state $m$ such that $d(m)=\omega$. An updating mechanism is \emph{ignorant} if it ignores some state, and \emph{non-ignorant} otherwise.
\end{definition}

As argued in the introduction, the structure of optimal ignorance is reminiscent of a large set of behavioral biases that depart from the Bayesian model. Note that given the assumption that information strictly improves utility, i.e., $L^*_M<\min_a\sum_{\omega=1}^Np^\omega[u(\omega,\omega)-u(a,\omega)]$, when $N=2$, no ignorant updating mechanism is $\epsilon$-optimal when $\epsilon$ is sufficiently small. I present the formal results on ignorance in the following Corollary when $N>2$, and offer interpretations in the next paragraph. 


	\begin{corollary}\label{prop_bigworldignorance2}
		Ignorance could be optimal in big worlds but not in small worlds:
		\begin{enumerate}[label=(\roman*)]
			\item For all $N>2$ and $M\geq N$, there exists some environment $(\bm{u}^{\omega'},p^{\omega'},f^{\omega'})_{\omega'=1}^{N}$ and a state~$\omega$ such that:\footnote{The restriction $M\geq N$ is the paper's maintained assumption (Section~\ref{section:model}); it rules out trivial ignorance for lack of memory states. The other trivial route --- states with identical payoff vectors, whose neglect is costless --- is excluded throughout by the maintained assumption that each state has a unique optimal action.}
			\begin{enumerate}[label=(\alph*)]
				\item every mechanism with optimal decision rule, i.e.,
				\begin{equation*}
					d(m)\in\argmax_{a}\sum_{\omega''}p^{\omega''}\mu^{\omega''}_{m}u(a,\omega'') \text{ for all } m,
				\end{equation*}
				ignores state~$\omega$; and
				\item more generally, without any restriction on the decision rule, every $\epsilon$-optimal mechanism picks action~$\omega$ in any state~$\omega'$ with probability at most $O(\epsilon)$, which vanishes as $\epsilon\rightarrow 0$.
			\end{enumerate}
			\item In contrast, fix $N>2$ and some $(\bm{u}^{\omega'},p^{\omega'},f^{\omega'})_{\omega'=1}^{N}$; when $M$ is large enough, for all $\epsilon$ smaller than some threshold $\bar{\bar{\epsilon}}$, every $\epsilon$-optimal updating mechanism ignores no state.
		\end{enumerate}
	\end{corollary}
	

	First, (i) shows that generally, there exist decision environments where ignorance is optimal in two related senses. By (i)(a), every mechanism with the optimal decision rule strictly ignores some state~$\omega$; by (i)(b), every $\epsilon$-optimal mechanism picks $\omega$ with vanishing probability as $\epsilon\rightarrow 0$. With $\frac{M}{N}$ finite, the DM cannot allocate unlimited cognitive resources to every state of the world. He is bound to make mistakes and thus faces a trade-off between mistakes across states. (i) implies that in some environments the DM optimally ignores some states altogether to improve learning in others. On the other hand, (ii) shows that the set of environments where ignorance is optimal vanishes as $M$ grows, or equivalently as the world gets smaller. As $\frac{N}{M}\rightarrow 0$, the DM can allocate many more memory states to each action. The DM learns almost perfectly in every state, the trade-off between states loses force, and he has no incentive to ignore states. Corollary~\ref{prop_bigworldignorance2} also implies that when $\frac{N}{M}$ is small, the DM will not choose actions that are ``safe'' but sub-optimal in every state of the world.
	
	A natural follow-up question is: \emph{which} states does the DM ignore? The following Proposition characterizes three properties of states that make them candidates for optimal ignorance.


	\begin{proposition}\label{coro_bigworldignorance2example}
		Fix any $N>2$, $M$, and $(f^{\omega})_{\omega=1}^{N}$, and suppose $u(\omega',\omega)=0$ for all $\omega'\neq\omega$. There exist thresholds $\xi_p,\xi_u>0$ and $\overline{F}>1$ --- depending only on the model primitives --- such that if
		\begin{enumerate}[label=(\roman*)]
			\item $p^{\omega}<\xi_p$, i.e., state~$\omega$ is a priori unlikely, or
			\item $\frac{u(\omega,\omega)}{\min_{\omega''\neq\omega}u(\omega'',\omega'')}<\xi_u$, i.e., guessing state~$\omega$ correctly is unimportant, or
			\item $\sup_{s}\frac{f^{\omega}(s)}{f^{\omega'}(s)}\times\sup_{s}\frac{f^{\omega'}(s)}{f^{\omega}(s)}\leq\overline{F}$ for some $\omega'\neq\omega$, i.e., distinguishing states~$\omega$ and~$\omega'$ is difficult,
		\end{enumerate}
		then:
		\begin{enumerate}[label=(\alph*)]
			\item every mechanism with optimal decision rule, i.e.,
			\begin{equation*}
				d(m)\in\argmax_{a}\sum_{\omega''}p^{\omega''}\mu^{\omega''}_{m}u(a,\omega'') \text{ for all }m,
			\end{equation*}
			ignores state~$\omega$ in cases~(i) and~(ii), and ignores at least one of $\omega$ and $\omega'$ in case~(iii);
			\item more generally, without any restriction on the decision rule, every $\epsilon$-optimal mechanism picks action~$\omega$ in any state~$\omega''$ with probability at most $O(\epsilon)$ in cases~(i) and~(ii), and in case~(iii), either action~$\omega$ or action~$\omega'$ is picked with probability at most $O(\epsilon)$ in any state~$\omega''$; these probabilities vanish as $\epsilon\rightarrow 0$.
		\end{enumerate}
	\end{proposition}
	
As in Corollary~\ref{prop_bigworldignorance2}(i), part~(a) of Proposition~\ref{coro_bigworldignorance2example} delivers ignorance in the exact sense of Definition~1 for mechanisms with optimal decision rules, whereas for arbitrary $\epsilon$-optimal mechanisms only the vanishing-frequency form of part~(b) can hold: such a mechanism may waste a rarely visited memory state on action~$\omega$ at $O(\epsilon)$ cost.

	In summary, the DM ignores states that are a priori unlikely, relatively unimportant, or hard to distinguish from other states. By (i), when a state is a priori unlikely, the DM would rather allocate his limited memory to learning other states: an individual confident about his own ability updates only his beliefs about his teammates', resembling persistent overconfidence, whereas a Bayesian observing a series of bad team performances would eventually assign positive probability to his own ability being low. By (ii), the DM likewise ignores states whose correct guess yields relatively little utility. Lastly, by (iii), when two states $\omega$ and $\omega'$ are difficult to distinguish ($f^{\omega}\approx f^{\omega'}$), the DM ignores one of them: he would need many memory states to accumulate enough evidence favoring $\omega$ over $\omega'$ and would remain almost unsure --- an inefficient use of memory --- so he redeploys those resources to learn the remaining states, where they improve his utility. Condition (iii) is particularly relevant for correlation neglect: when positive and no correlation are hard to distinguish and positive correlation is slightly less likely, the DM ignores the possibility of correlation and focuses his learning elsewhere.\footnote{For example, consider an investor who wants to learn the state of the stock market --- strong, moderately strong, moderately weak, or weak --- and whether analyst forecasts are positively correlated or independent. A signal is a profile of predictions from market analysts. If the predictions are predominantly favorable, this could indicate either that the market is genuinely strong with independent forecasts, or that the market is only moderately strong but the analysts are drawing from the same source and echoing each other (positive correlation). Because these two scenarios produce similar signal profiles, they are difficult to distinguish. According to condition~(iii), the DM would optimally ignore one of them --- say, the possibility of positive correlation --- and focus his limited cognitive resources on learning whether the market is strong, moderately strong, moderately weak, or weak, behaving as if analyst forecasts are always independent.}

    \subsubsection{$N=M$, symmetric environments and symmetric mechanisms}\label{sec:symmetry}
	
     Corollary~\ref{prop_bigworldignorance2} and Proposition~\ref{coro_bigworldignorance2example} show that ignorance is optimal in some asymmetric environments when $\frac{M}{N}$ is small. In the following, I show that ignorance is also optimal in symmetric environments where learning is ``difficult''. It implies that optimal ignorance is not a knife-edge result, but rather a general phenomenon when the individual has limited resources for learning. Unlike in Proposition~\ref{coro_bigworldignorance2example}, where the DM ignores states that are unlikely, unimportant, or difficult to distinguish from another state, here the primitives treat every state identically, so ignorance is driven purely by the scarcity of memory.
    
    I consider the following simple symmetric environment where states and actions are ex-ante identical: There are $N$ possible signal realizations, i.e., $S=\lbrace s^1,\cdots, s^N\rbrace$, and $p^{\omega}$, $u(\omega,\omega)$, $u(\omega',\omega)$, $\frac{f^{\omega}(s^\omega)}{f^{\omega}(s^{\omega'})}>1$ are invariant across all $\omega$ and $\omega'\neq\omega$. Moreover, I focus on settings where $M=N$ and is an even number. 

    As mentioned earlier, it is generally intractable to determine the optimal updating mechanism when $N>2$. In the following, I restrict attention to the class of \emph{symmetric} mechanisms and ask whether the optimal mechanism \emph{within this class} is ignorant. Three considerations motivate the restriction. First, it makes the comparison tractable. Second, in the well-understood case $N=2$ \citep{hellman1970learning}, the infimum loss in a symmetric environment is the same whether or not one restricts to symmetric mechanisms: for every $M$, one of the optimal mechanisms is symmetric, as in Examples~1 and~2, so nothing is lost by the restriction there.  Third, the class is rich in the relevant dimension: it contains both non-ignorant mechanisms and mechanisms ignoring any number of states (symmetry is only required on the non-ignored states), so restricting to it does not predetermine the answer to the ignorance question. Formally, denote the set of states that are not ignored by an updating mechanism as $\Omega^{NI}=\lbrace\omega: \exists\, m  \text{ s.t. } d(m)=\omega\rbrace$. An updating mechanism is \emph{symmetric} if for any permutation $\sigma:\Omega^{NI}\rightarrow \Omega^{NI}$, there exists a one-to-one mapping $\sigma^M: M\rightarrow M$ such that for all $\omega\in\Omega^{NI}$,
    \begin{equation*}
        \begin{split}
            d(\sigma^M(m))&=\sigma(d(m))\\
            \Pr\lbrace\mathscr{T}(\sigma^M(m),s^{\sigma(\omega)})=\sigma^M(m')\rbrace&=\Pr\lbrace\mathscr{T}(m,s^{\omega})=m'\rbrace.
        \end{split}
    \end{equation*}
    That is, after any permutation of the non-ignored states, there exists a relabeling of memory states such that the updating mechanism remains unchanged.
    

    \begin{proposition}\label{prop:symmetricignorance}
        In symmetric environments, consider optimality within the class of symmetric mechanisms, and let $\mathscr{I}=\frac{f^{\omega}(s^\omega)}{f^{\omega}(s^{\omega'})}$ (with $\omega\neq\omega'$) denote the informativeness of the signals. There is a threshold $\bar{\mathscr{I}}(N)>1$, increasing in $N$ with $\bar{\mathscr{I}}(N)\rightarrow\infty$, such that whenever $\mathscr{I}\leq\bar{\mathscr{I}}(N)$, for all $\epsilon$ smaller than some threshold, every $\epsilon$-optimal symmetric mechanism ignores some state.\footnote{More precisely, $\bar{\mathscr{I}}(N)=\frac{1}{4}\left(\sqrt{9N^{2}-32N+32}+N\right)$.} In particular, all $\epsilon$-optimal symmetric updating mechanisms (with small enough $\epsilon$) are ignorant (i) when the state space is large enough for any fixed informativeness $\mathscr{I}$, and (ii) when signals are noisy enough for any fixed $N$.
    \end{proposition}

    Proposition~\ref{prop:symmetricignorance} also has an implication that does not restrict the class of mechanisms: since its proof shows that a symmetric ignorant mechanism strictly outperforms the best symmetric non-ignorant mechanism, no symmetric non-ignorant mechanism is $\epsilon$-optimal among \emph{all} mechanisms for small $\epsilon$. Every $\epsilon$-optimal mechanism must therefore break the symmetry of the environment --- by ignoring some states, or by treating ex-ante identical states asymmetrically.
    
	In the proof (in Online Appendix~\ref{section:onlineproofs}), I first show that among all symmetric non-ignorant mechanisms, the one illustrated in Figure~\ref{fig:symmetricallstates1} is optimal. In this mechanism, the decision rule $d(m)$ is a one-to-one mapping from $m$ to $\omega$, i.e., the DM chooses each action in one and only one unique memory state, and the DM transits to the memory state corresponding to action~$\omega$ if and only if he receives signal $s^{\omega}$ (the signal that supports state~$\omega$), as in the non-ignorant mechanism in Example~3. Second, I compare the asymptotic utility of this best symmetric non-ignorant mechanism with that of a simple symmetric ignorant mechanism that ignores half of the states, as illustrated in Figure~\ref{fig:symmetricignorestates1}.
	
	The result is illustrated in Figure~\ref{fig:ignorantoutperformnonignorant}; and Figure~\ref{fig:compareignoreornonignorant} shows the comparison of the non-ignorant and ignorant mechanisms where $N=M\in\lbrace 4,6,8\rbrace$. The y-axis is the asymptotic utility of the updating mechanism that ignores half of the states minus that of the best non-ignorant updating mechanism. Thus, if the y-axis is above $0$, all $\epsilon$-optimal mechanisms must be ignorant for small enough $\epsilon$. The x-axis is the informativeness of the signal structure, i.e., $\frac{f^{\omega}(s^\omega)}{f^{\omega}(s^{\omega'})}$ where $\omega'\neq\omega$.  Both Figure~\ref{fig:ignorantoutperformnonignorant} and Figure~\ref{fig:compareignoreornonignorant} show that ignorance is optimal when the informativeness is smaller than some threshold, or when the state space $N$ is large enough. In other words, ignorance is optimal when it is difficult to learn, which echoes the results in Corollary~\ref{prop_bigworldignorance2}.\footnote{The magnitude of the thresholds also suggests that ignorance is not an extreme event. For example, when $N=4$, ignorance is better when the likelihood ratio of the signals is smaller than (around) $3$. } The result also contrasts with the case $N=2$, in which symmetric non-ignorant updating mechanisms approach the infimum utility loss in such symmetric environments.

		\begin{figure}
		\centering
		\begin{tikzpicture}[every text node part/.style={align=center}]
			\node[cloud] (m1) {\footnotesize 1};
			\node[cloud, right = 10em of m1] (m2) {\footnotesize 2};
			\node[cloud, below = 10em of m1] (m3) {\footnotesize 3};
			\node[cloud, below = 10em of m2] (m4) {\footnotesize 4};
			\draw[-Latex]
			(m1) edge[bend right=10] node [left, midway] {\footnotesize$
            \times s^{3}$} (m3)
			(m3) edge[bend right=10] node [right, midway] {\footnotesize$
            s^{1}$} (m1)
			(m1) edge[bend right=10] node [below, midway] {\footnotesize$
            s^{2}$} (m2)
			(m2) edge[bend right=10] node [above, midway] {\footnotesize$
            s^{1}$} (m1)
			(m2) edge[bend right=10] node [left, midway] {\footnotesize$
            s^{4}$} (m4)
			(m4) edge[bend right=10] node [right, midway] {\footnotesize$
            s^{2}$} (m2)
			(m3) edge[bend right=10] node [below, midway] {\footnotesize$
            s^{4}$} (m4)
			(m4) edge[bend right=10] node [above, midway] {\footnotesize$
            s^{3}$} (m3)
			(m2) edge[bend right=90, looseness=2] node [above left, midway] {\footnotesize$
            s^{3}$} (m3)
			(m3) edge[bend right=90, looseness=2] node [below right, midway] {\footnotesize$
            s^{2}$} (m2)
			(m1) edge[bend right=90, looseness=2] node [below left, midway] {\footnotesize$
            s^{4}$} (m4)
			(m4) edge[bend right=90, looseness=2] node [above right, midway] {\footnotesize $
            s^{1}$} (m1);
		\end{tikzpicture}
		\caption{The optimal non-ignorant updating mechanism that considers all states, with $N=M=4$. The number in the node denotes the action that the DM takes when he is in this memory state. Moreover, in memory state $\omega$, and upon receiving a signal that supports state~$\omega'\neq\omega$, the DM transits to memory state~$\omega'$ with probability $
        1$ and stays in his current memory state otherwise.}
		\label{fig:symmetricallstates1}
	\end{figure}
	
	\begin{figure}
		\centering
		\begin{tikzpicture}[every text node part/.style={align=center}]
			\node[cloud] (m1) {\footnotesize 1};
			\node[cloud, right = 5em of m1] (m2) {\footnotesize 2};
			\node[cloud, right = 5em of m2] (m3) {\footnotesize 3};
			\node[cloud, right = 5em of m3] (m4) {\footnotesize 4};
			\draw[-Latex]
			(m1) edge[bend right=10] node [below, midway] {\footnotesize$\delta\times s^{2}$} (m2)
			(m2) edge[bend right=10] node [above, midway] {\footnotesize$s^{1}$} (m1)
			(m2) edge[bend right=10] node [below, midway] {\footnotesize$\delta\times s^{2}$} (m3)
			(m3) edge[bend right=10] node [above, midway] {\footnotesize$\delta\times s^{1}$} (m2)
			(m3) edge[bend right=10] node [below, midway] {\footnotesize$s^{2}$} (m4)
			(m4) edge[bend right=10] node [above, midway] {\footnotesize$\delta\times s^{1}$} (m3);
		\end{tikzpicture}
		\caption{An example of an updating mechanism that ignores two states, with $N=M=4$. The DM takes action $1$ in memory states $1$ and $2$, and takes action~$2$ in memory states $3$ and $4$. In memory state~$3$, if the DM receives a signal supporting state~$2$, he transits to memory state~$4$; if the DM receives a signal supporting state~$1$, he transits to memory state~$2$ with probability $\delta$ and stays in his current memory state otherwise. In memory state~$4$, if the DM receives a signal supporting state~$1$, he transits to memory state~$3$ with probability $\delta$  and stays in his current memory state otherwise. The transition functions in memory state~$1$ and $2$ are defined accordingly. $\delta$ is chosen to be close to $0$ to maximize the asymptotic utility.}
		\label{fig:symmetricignorestates1}
	\end{figure}
	
	\begin{figure}
		\centering
		\begin{tikzpicture}[yscale=0.5]
			\draw[->] (0.8,1)--(9,1) node[right] {$\overline{l}^{\omega\omega'}$};
			\draw[->] (1,0.8)--(1,12) node[left] {$N$};
			\draw[black, domain=1.01:8.5] plot (\x, {sqrt((4+\x*(2*\x-4)*(\x+1))/(\x-1)+(\x/2-2)*(\x/2-2))-\x/2+2});
			\node[align=center] at (3.5,8) {\footnotesize Ignorant mechanism\\\footnotesize outperforms \\\footnotesize non-ignorant mechanism};
			\draw[dashed] (5,6.27)--(5,1) node[below] {$5$};
			\draw[dashed] (5,6.27)--(1,6.27) node[left] {$6.30$};
			\draw[dashed] (3,4.27)--(3,1) node[below] {$2.73$};
			\draw[dashed] (3,4.27)--(1,4.27) node[left] {$4$};
		\end{tikzpicture}
		\caption{The ignorant mechanism outperforms the non-ignorant mechanism when information is noisy enough or when $N$ is big enough.} 
		\label{fig:ignorantoutperformnonignorant}
	\end{figure}
	
	\begin{figure}
		\centering
		\includegraphics[width=20em]{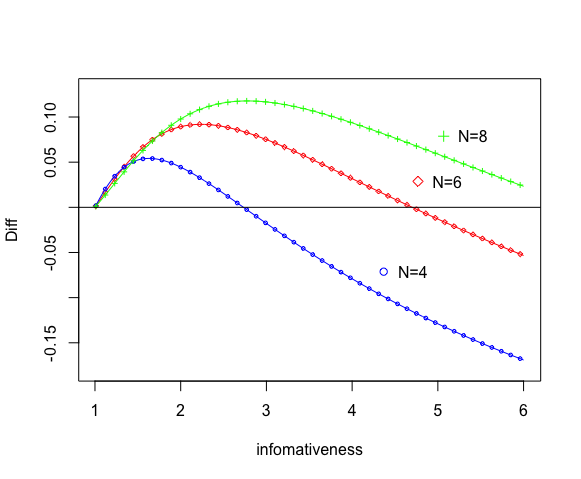}
		\caption{The y-axis is the (estimated) asymptotic utility of the ignorant mechanism illustrated in Figure~\ref{fig:symmetricignorestates1} minus that of the non-ignorant mechanism illustrated in Figure~\ref{fig:symmetricallstates1}, while the x-axis is $\frac{f^{\omega}(s^\omega)}{f^{\omega}(s^{\omega'})}$. The ignorant mechanism outperforms the non-ignorant mechanism when $\frac{f^{\omega}(s^\omega)}{f^{\omega}(s^{\omega'})}$  is smaller than some threshold. Moreover, the threshold is larger as $N$ increases from 4 to 6, and from 6 to 8.} 
		\label{fig:compareignoreornonignorant}
	\end{figure}

	
  	The intuition of the result is as follows. To consider all states, the DM allocates one memory state to each action. It is thus easy for the DM to alternate between different actions and unavoidably make mistakes. Put differently, the updating mechanism is ``noisy''. This is especially true when signals are very noisy or when $N$ is large: even when the likelihood ratio $\frac{f^{\omega}(s^\omega)}{f^{\omega}(s^{\omega'})}$ is fixed, $f^\omega(s^\omega)$ vanishes as $N\rightarrow\infty$ because probability mass is spread across $N-1$ other signal realizations. Formally, note that in the symmetric environment,
	\begin{equation*}
		f^\omega(s^\omega) = \frac{f^{\omega}(s^\omega)/f^{\omega}(s^{\omega'})}{f^{\omega}(s^\omega)/f^{\omega}(s^{\omega'}) + N - 1},
	\end{equation*}
	which vanishes as $N$ grows. Each signal becomes almost uninformative, yet with only one memory state per action, the DM cannot accumulate enough evidence for any single state.

	If, in contrast, the DM ignores half of the actions, he allocates two memory states to each of the actions that he considers and can therefore ``records'' two consecutive signals supporting the same state. As it is exponentially unlikely to receive consecutive signals supporting the wrong state, the ignorant mechanism improves the individual's decision-making among the smaller set of states that he considers. When $N$ is large or when $\frac{f^{\omega}(s^\omega)}{f^{\omega}(s^{\omega'})}$ is small, the improvement outweighs the loss he incurs among the states that he ignores, because the loss is small to begin with, i.e., the asymptotic utility of a ``noisy'' updating mechanism that considers all states is small. In particular, Online Appendix~\ref{section:symmetricsupp} further extends this intuition by analyzing an ignorant mechanism that ignores $2/3$ of the states of the world: when the signals are very noisy, ignoring $2/3$ of the states outperforms ignoring half of them, which in turn outperforms the non-ignorant mechanism.

	\subsubsection{$N,M\rightarrow\infty$}\label{sec:NMlimit_ignorance}

	The results of Section~\ref{sec:ignorancegeneral} and~\ref{sec:symmetry} are results for fixed finite $N$. The following results show that their qualitative conclusions extend to the limit $N,M\rightarrow\infty$, further confirming that the ratio $\frac{N}{M}$ is the key determinant of ignorance.

	When $M\rightarrow\infty$, the finite-case definition of ignorance (Definition~1) is no longer adequate: a mechanism could assign a single memory state to action~$\omega$ without affecting long-run utility, since the cost of ``wasting'' one memory state out of infinitely many vanishes in the limit. Accordingly, I adopt the following definition, based on the long-run frequencies of actions.


	\begin{definition}{2}\label{def:limitignorance}
		Fix $\xi\in(0,1)$. An updating mechanism $(m_{1N},\mathscr{T}_{N},d_{N})$ \emph{$\xi$-ignores} state~$\omega$ if, for every state~$\omega'$,
		\begin{equation*}
			\lim_{T\rightarrow\infty}\frac{1}{T}\sum_{t=1}^{T}\text{\emph{Pr}}^{\omega'}(a_{t}=\omega)<\xi,
		\end{equation*}
		where $\Pr^{\omega'}$ denotes the probability distribution over action sequences induced by $(m_{1N},\mathscr{T}_{N},d_{N})$ conditional on the true state being~$\omega'$. An updating mechanism is \emph{$\xi$-ignorant} if it $\xi$-ignores some state.
	\end{definition}
	That is, a mechanism $\xi$-ignores state~$\omega$ if action~$\omega$ is chosen with long-run frequency below $\xi$ regardless of the true state. Definition~\ref{def:limitignorance} thus replaces the previous definition of ignorance --- ``state~$\omega$ is not in the range of $d$'' --- with one defined in terms of observable actions; the former implies the latter, as a state outside the range of $d_{N}$ is chosen with frequency $0$ and is therefore $\xi$-ignored for every $\xi$. The definition has two advantages under the setting $N,M\rightarrow\infty$. First, as noted above, it is robust to redundant memory allocation that does not affect the actions taken --- a key concern when $M\rightarrow\infty$. Second, it involves no limit along the sequence: it is a property of the $N$-th mechanism alone, so it remains meaningful even when the identity of the neglected states varies along the sequence. In the following, I examine to what extent the $N$-th mechanism $\xi$-ignores states as $N$ grows, in small and in big worlds.

%

	In small worlds ($h<1$), Proposition~\ref{prop_smallworldinfiniteN}(i) established that $L^*_N\rightarrow 0$. Call a sequence of updating mechanisms $\lbrace m_{1N},\mathscr{T}_{N},d_{N}\rbrace_{N\in\mathcal{N}}$ \emph{$\epsilon$-optimal at the limit} $N\rightarrow\infty$ if its loss asymptotically exceeds the optimum by at most $\epsilon$: $\limsup_{N\rightarrow\infty}\,[\,L(m_{1N},\mathscr{T}_{N},d_{N})-L^{*}_{N}\,]\leq\epsilon$. The following corollary shows that such sequences can $\xi$-ignore only states that carry little prior probability for any $\xi\in(0,1)$:

\begin{corollary}\label{coro_ignorancesmallworldinfinteN}
    	Under the same conditions as Proposition~\ref{prop_smallworldinfiniteN} with $h<1$, fix any $\xi\in(0,1)$ and any sequence of updating mechanisms $\lbrace m_{1N},\mathscr{T}_{N},d_{N}\rbrace_{N\in\mathcal{N}}$ that is $\epsilon$-optimal at the limit $N\rightarrow \infty$. Then the states $\xi$-ignored by the $N$-th mechanism carry total prior probability at most $\frac{\epsilon}{1-\xi}$ in the limit:
    	\begin{equation*}
    		\limsup_{N\rightarrow\infty}\,\sum_{\omega\text{ $\xi$-ignored by the $N$-th mechanism}}p^{\omega}_{N}\,\leq\,\frac{\epsilon}{1-\xi},
    	\end{equation*}
    	which vanishes as $\epsilon\rightarrow 0$.
    \end{corollary}

	For big worlds ($h\geq 1$), the counterpart result is part (ii) of Proposition~\ref{prop_smallworldinfiniteN}, which is considerably stronger than Corollary~\ref{prop_bigworldignorance2}: it holds for \emph{all} environments (not just some) and for \emph{all} sequences of updating mechanisms (not only optimal ones). In the language of Definition~\ref{def:limitignorance}, Equation~\eqref{eq:limitignorancedisplay} states that, for any $\xi'\in(0,1)$, all but a vanishing fraction of states are $\xi'$-ignored by the $N$-th mechanism for large $N$: any DM in a big world ignores almost all states.

	This is the structural reason behind the big-world utility loss in Proposition~\ref{prop_smallworldinfiniteN}(ii): by the memory constraint alone, the DM is forced to focus on a vanishing fraction of states regardless of the environment, and therefore captures at most the $\bar{p}_0$-measure of states he attends to --- yielding the loss $1-\bar{p}_0$. The result sharpens the contrast between small and big worlds: in the former the DM attends to essentially all states, while in the latter he attends to vanishingly few. Under \ref{assumption_fullsupportinfiniteplus}, $\bar{p}_0=0$, so the DM attends to a set of states with vanishing prior probability and the loss converges to~$1$.

	The intuition is that to ever choose action~$\omega$ in the long run, the DM must accumulate enough evidence to make $\omega$ stand out among the alternatives, which requires devoting a chain of memory states to tracking signals favoring $\omega$. As we have established in Section~\ref{sec:symmetry}, when $N$ increases, signals favoring $\omega$ become less likely even when state~$\omega$ is true, since there are many alternative states. Tracking signals for many states simultaneously consumes memory roughly linearly in the number of states. With memory $M_{N}$ scaling no faster than $N$ in big worlds, the DM's budget cannot accommodate tracking for more than a vanishing fraction of states, forcing him to ignore almost all states regardless of the prior.

	\subsection{Is disagreement persistent?}

	Lastly, I turn to the question of whether disagreement can persist asymptotically in small and big worlds. Consider two individuals $A$ and $B$ who may differ in their utility matrices, prior beliefs, and/or signal structures, and who each learn with their own updating mechanism. Roughly speaking, sample one long-run action from each individual: their \emph{disagreement} in state~$\omega$ is the probability that the two sampled actions differ, and the two individuals \emph{almost always disagree} (resp.\ \emph{agree}) for a given $\epsilon$ if the disagreement between their $\epsilon$-optimal updating mechanisms equals $1$ (resp.\ $0$) in every state. Online Appendix~\ref{section:disagreementonline} contains the formal definitions and results; I summarize their content here. In big worlds, disagreement can persist: for all $N>2$ and memory sizes $M_{A},M_{B}$, there exist environments in which the two individuals almost always disagree as $\epsilon\rightarrow 0$. In small worlds, by contrast, disagreement is bounded above by a term that vanishes as $\frac{N}{\min\lbrace M_{A},M_{B}\rbrace}\rightarrow 0$ and $\epsilon\rightarrow 0$: individuals with different prior beliefs and different information are nonetheless bound to agree with certainty. The same conclusions extend to the limit $N,M\rightarrow\infty$: under a same-limit-support condition, small-world individuals must almost always agree at the limit, while in big worlds there exist sequences of decision problems --- even with identical priors --- under which $\epsilon$-optimal individuals disagree in all states.

	The intuition is as follows: as different individuals with different prior beliefs and utilities would adopt different updating mechanisms, they could focus their learning on different subsets of states of the world when they have limited memory. In particular, consider an example with $N=4$. If individual~$A$ ignores states~$1$ and~$2$ and individual~$B$ ignores states~$3$ and~$4$, they will never choose the same action and thus disagree with certainty.\footnote{Note that the result of disagreement continues to hold in a framework where individuals observe each other's actions. More specifically, one could re-define the signal structures in the current setting to incorporate the information conveyed by the actions taken by the two individuals.} However, as $\frac{N}{M}$ goes to $0$, decisions get closer to Bayesian, and the two individuals almost always choose the same actions.
	
	The following result further shows that disagreement could be driven solely by differences in cognitive ability. For simplicity, I allow $M$ to be smaller than $N$ in the following result.
	\begin{corollary}\label{prop_disagreementbasedonM}
		Disagreement can be driven solely by differences in cognitive ability even when individuals have the same prior beliefs and preferences. There exists $M_A\neq M_B$ and $(\bm{u}^{\omega}_{A},p^{\omega}_{A},f^{\omega}_{A})_{\omega=1}^{N}=(\bm{u}^{\omega}_{B},p^{\omega}_{B},f^{\omega}_{B})_{\omega=1}^{N}$  such that individual~A and~B almost always disagree when both adopt $\epsilon$-optimal updating mechanisms as $\epsilon\rightarrow 0$.
	\end{corollary}
	
	I prove the Corollary (shown in Online Appendix~\ref{section:onlineproofs}) with an example where $M_A=1<M_B=2$, $N=3$, and $u(\omega,\omega)=1>u(\omega,\omega')=0$ for all $\omega$ and $\omega'\neq\omega$. Importantly, state~$1$ is a priori more likely than state~$2$ and~$3$, while it is easier to distinguish state~$2$ and~$3$ than to distinguish state~$1$ and state~$2$, or state~$1$ and state~$3$.\footnote{Mathematically, $\sup_{s}\frac{f^{3}(s)}{f^{2}(s)}\sup_{s}\frac{f^{2}(s)}{f^{3}(s)}>\max\left\lbrace\sup_{s}\frac{f^{1}(s)}{f^{2}(s)}\sup_{s}\frac{f^{2}(s)}{f^{1}(s)},\sup_{s}\frac{f^{1}(s)}{f^{3}(s)}\sup_{s}\frac{f^{3}(s)}{f^{1}(s)}\right\rbrace$.} In this example, individual $A$ always chooses action~$1$ as he does not have sufficient cognitive resources to learn. On the other hand, if $p^{1}$ is not too large compared to $p^2$ and $p^3$, individual~$B$ ignores state~$1$. He takes advantage of the informative information structure to distinguish state~$2$ and~$3$, such that he can be confident that he doesn't take action~$2$ in state~$3$ and vice versa. Thus, as long as state~$1$ is not a priori too likely, ignoring it would yield a higher utility.\footnote{One may argue that after seeing individual $B$ choosing action~$2$ or~$3$, individual $A$ should change his action. However, this is not possible, as he has only one unit of memory capacity ($M=1$) and thus has to effectively commit to one action. In particular, one can generalize this framework so that the two individuals also see each other's actions as signals, and Proposition~$\ref{prop_disagreementbasedonM}$ would still hold.} To sum up, as individual $A$ has a lower cognitive ability, his learning and actions would depend a lot on the prior belief, while individual $B$ has the ability to take advantage of the information structure.\footnote{Note that although this example imposes strong assumptions in particular on the size of the automaton of individual $A$, it generates a strong form of disagreement in which the two individuals disagree asymptotically with certainty. Similar intuition implies that even when the assumption is relaxed, the difference in $M$ would lead to asymptotic disagreement at least probabilistically.} As a result, they adopt different updating mechanisms, which leads to disagreement.

	\section{Discussion and Conclusion}\label{section:discussion}
	This paper shows that optimal ignorance under limited cognitive resources produces behavior that is observationally indistinguishable from a wide range of behavioral anomalies, through a single mechanism: efficient allocation of limited cognitive resources in light of complexity. Importantly, the comparison of small and big worlds illustrates a link between the degree of (relative) complexity of the inference problem and the aforementioned non-Bayesian learning behaviors, which is supported by the experimental results in \cite{enke2019correlation} and \cite{graeber2019inattentive}. \cite{enke2019correlation} shows that correlation neglect negatively correlates with the cognitive ability of subjects and ``an extreme reduction in the environment's complexity eliminates the bias'', while \cite{graeber2019inattentive} shows that a reduction in the complexity of the problem by removing a deciphering stage of signals reduces inattentive learning behavior. This paper theoretically demonstrates that a relationship exists between biases, complexity, and cognitive ability, extending to other behavioral biases. It illustrates that heterogeneity across both individual and decision environments leads to varying prominence of behavioral biases.

	\newpage
	\bibliographystyle{econ} 
	\bibliography{reference_big}

	\newpage
	\appendix
	\renewcommand\theequation{\thesection.\arabic{equation}}
	\numberwithin{equation}{section}
	\setcounter{equation}{0}
	\renewcommand\theproposition{\thesection.\arabic{proposition}}
	\numberwithin{proposition}{section}
	\setcounter{equation}{0}
	\renewcommand\thelemma{\thesection.\arabic{lemma}}
	\numberwithin{lemma}{section}
	\renewcommand\thecorollary{\thesection.\arabic{corollary}}
	\numberwithin{corollary}{section}

\section{Proofs}\label{section:proof}
Before I present the proofs of the results in the main text, I first prove the following Lemma that is used throughout the proof of the different results.

\begin{lemma}\label{lemmma:appendixiff}
    Fix $N, (f^\omega)_{\omega=1}^N$ and any updating mechanism (irreducible or reducible), for all $\omega,\omega'$,
    \begin{equation}\label{eq:profignorance}
        \min_{m}\frac{\mu^{\omega'}_{m}}{\mu^{\omega}_{m}}\geq \varsigma^{2(M-1)}.
    \end{equation}
    It implies that $\mu^\omega_m=0$ if and only if $\mu^{\omega'}_m=0$ for all $\omega,\omega',m$; in other words, $\mu^\omega_m>0$ if and only if $\mu^{\omega'}_m>0$ for all $\omega,\omega',m$.
\end{lemma}
\begin{proof}
    I first focus on irreducible updating mechanisms. Consider state $\omega$ and $\omega'$, recall that the long-run distribution in state~$\omega$ and $\omega'$ are the solution of the following fixed point equations:
	\begin{equation}
		\begin{split}
			\bm{\mu}^{\omega}=(\bm{\mu}^{\omega})^{T}\mathbf{Q}^{\omega};\\
			\bm{\mu}^{\omega'}=(\bm{\mu}^{\omega'})^{T}\mathbf{Q}^{\omega'}.
		\end{split}
		\label{eq:proofstationarymatrix}
	\end{equation}
	Equation~\eqref{eq:proofstationarymatrix} shows that the long-run distributions $\bm{\mu}^{\omega}$ and $\bm{\mu}^{\omega'}$, and hence the ratio $(\frac{\mu^{\omega}_{m}}{\mu^{\omega'}_{m}})_{m=1}^{M}$, depend only on the updating mechanism $(m_1,\mathscr{T},d)$ and on the conditional signal distributions $f^{\omega}$ and $f^{\omega'}$: the mechanism is defined on memory states and signals without reference to the number of states, and under state~$\omega$ the memory process is a Markov chain governed by $\mathscr{T}$ and $f^{\omega}$ alone. They therefore do not depend on $N$, on the prior $(p^{\omega''})_{\omega''=1}^{N}$, or on the other states' signal distributions $(f^{\omega''})_{\omega''\neq\omega,\omega'}$, and coincide with the long-run distributions of the same mechanism in the two-state decision problem with states $\omega$ and $\omega'$. Theorem 2 of \cite{hellman1970learning}, stated for $N=2$, therefore applies:
	\begin{equation}\label{eq:proofignorance}
		\begin{split}
			\frac{\max_{m}\frac{\mu^{\omega'}_{m}}{\mu^{\omega}_{m}}}{\min_{m}\frac{\mu^{\omega'}_{m}}{\mu^{\omega}_{m}}}\leq (\overline{l}^{\omega\omega'}\overline{l}^{\omega'\omega})^{(M-1)}.
		\end{split}
	\end{equation}
	where $\overline{l}^{\omega\omega'}=\sup_{s}\frac{f^{\omega}(s)}{f^{\omega'}(s)}$. First, $\min_{m}\frac{\mu^{\omega'}_{m}}{\mu^{\omega}_{m}}>0$ for all $\omega'\neq\omega$. Suppose to the contrary that for some $\omega'$, $\min_{m}\frac{\mu^{\omega'}_{m}}{\mu^{\omega}_{m}}=0$, Equation~\eqref{eq:proofignorance} implies that $\max_{m}\frac{\mu^{\omega'}_{m}}{\mu^{\omega}_{m}}=0$ and $\mu^{\omega'}_{m}=0$ for all $m$. It in turn implies that $\max_{m}\frac{\mu^{\omega}_{m}}{\mu^{\omega'}_{m}}$ is unbounded and $\min_{m}\frac{\mu^{\omega}_{m}}{\mu^{\omega'}_{m}}=0$ and it contradicts
	\begin{equation*}
		\frac{\max_{m}\frac{\mu^{\omega}_{m}}{\mu^{\omega'}_{m}}}{\min_{m}\frac{\mu^{\omega}_{m}}{\mu^{\omega'}_{m}}}\leq (\overline{l}^{\omega'\omega}\overline{l}^{\omega\omega'})^{(M-1)}.
	\end{equation*}
	Now since $\min_{m}\frac{\mu^{\omega'}_{m}}{\mu^{\omega}_{m}}>0$ for all $\omega'$, we have
	\begin{align*}
		\min_{m}\frac{\mu^{\omega'}_{m}}{\mu^{\omega}_{m}}&\geq (\overline{l}^{\omega\omega'}\overline{l}^{\omega'\omega})^{-(M-1)}\max_{m}\frac{\mu^{\omega'}_{m}}{\mu^{\omega}_{m}} \nonumber\\
		\min_{m}\frac{\mu^{\omega'}_{m}}{\mu^{\omega}_{m}}&\geq \varsigma^{2(M-1)}\max_{m}\frac{\mu^{\omega'}_{m}}{\mu^{\omega}_{m}}\nonumber\\
		\min_{m}\frac{\mu^{\omega'}_{m}}{\mu^{\omega}_{m}}&\geq \varsigma^{2(M-1)}
	\end{align*}
	The second inequality follows $\varsigma\equiv\min_{\omega'\neq\omega,s\in S}\frac{f^{\omega}(s)}{f^{\omega'}(s)}$ while the third inequality is implied by the fact that $\sum_{m}\mu^{\omega'}_{m}=\sum_{m}\mu^{\omega}_{m}=1$. Now, if $\mu^{\omega'}_m=0$ for some $m$, we must have $\mu^\omega_m=0$ as otherwise it violates Inequality~\eqref{eq:profignorance}.

	The above argument handles the case where the updating mechanism induces an irreducible Markov chain, so that the long-run distribution $\bm{\mu}^{\omega}$ does not depend on the initial memory state $m_{1}$. For a reducible chain, denote the recurrent communicating classes as $\mathscr{R}_{1},\ldots,\mathscr{R}_{r}$ and the set of transient memory states as $\mathscr{R}_{0}$. If $r=1$ or $m_{1}\in\mathscr{R}_{j}$ for some $j\geq 1$, the chain is effectively irreducible on $\mathscr{R}_{j}$ (with $|\mathscr{R}_{j}|\leq M$ states), and the bound above applies within $\mathscr{R}_{j}$, yielding $\min_{m}\mu^{\omega'}_{m}/\mu^{\omega}_{m}\geq\varsigma^{2(|\mathscr{R}_{j}|-1)}\geq\varsigma^{2(M-1)}$.

	In the remaining case, $r\geq 2$ and $m_{1}\in\mathscr{R}_{0}$. Let $\mathscr{P}^{\omega}(\mathscr{R}_{j})$ denote the probability of absorption into $\mathscr{R}_{j}$ under state $\omega$, and let $\mu^{0,\omega}$ be the stationary distribution on $\mathscr{R}_{0}$ for the modified transition rule that re-routes transitions out of $\mathscr{R}_{0}$ back to $m_{1}$. As is standard (see Appendix~2 of \cite{hellman1969learning}),
	\begin{equation*}
		\mathscr{P}^{\omega}(\mathscr{R}_{j})=\sum_{m\in\mathscr{R}_{0}}\mu^{0,\omega}_{m}\sum_{m'\in\mathscr{R}_{j}}q^{\omega}_{mm'},
	\end{equation*}
	where $q^{\omega}_{mm'}=\sum_{s}f^{\omega}(s)\Pr[\mathscr{T}(m,s)=m']$. Let $\mu^{j,\omega}_{m}=\mu^{\omega}_{m}/\sum_{m'\in\mathscr{R}_{j}}\mu^{\omega}_{m'}$ for $m\in\mathscr{R}_{j}$ (within-class stationary distribution). For $m\in\mathscr{R}_{j}$, $\mu^{\omega}_{m}=\mathscr{P}^{\omega}(\mathscr{R}_{j})\cdot\mu^{j,\omega}_{m}$, so
	\begin{equation*}
		\frac{\mu^{\omega'}_{m}}{\mu^{\omega}_{m}}=\frac{\mathscr{P}^{\omega'}(\mathscr{R}_{j})}{\mathscr{P}^{\omega}(\mathscr{R}_{j})}\cdot\frac{\mu^{j,\omega'}_{m}}{\mu^{j,\omega}_{m}}.
	\end{equation*}
	The within-class factor satisfies $\mu^{j,\omega'}_{m}/\mu^{j,\omega}_{m}\geq\varsigma^{2(|\mathscr{R}_{j}|-1)}$ by the irreducible-case bound applied to $\mathscr{R}_{j}$. The absorption-probability ratio satisfies
	\begin{equation*}
		\frac{\mathscr{P}^{\omega'}(\mathscr{R}_{j})}{\mathscr{P}^{\omega}(\mathscr{R}_{j})}\geq \varsigma^{2}\cdot\min_{m\in\mathscr{R}_{0}}\frac{\mu^{0,\omega'}_{m}}{\mu^{0,\omega}_{m}}\geq \varsigma^{2}\cdot\varsigma^{2(|\mathscr{R}_{0}|-1)}=\varsigma^{2|\mathscr{R}_{0}|},
	\end{equation*}
	where the first inequality uses $q^{\omega'}_{mm'}\geq\varsigma\cdot q^{\omega}_{mm'}$ pointwise and the second applies the irreducible-case bound to the transient sub-chain $\mathscr{R}_{0}$. Combining,
	\begin{equation*}
		\frac{\mu^{\omega'}_{m}}{\mu^{\omega}_{m}}\geq\varsigma^{2|\mathscr{R}_{0}|+2(|\mathscr{R}_{j}|-1)}\geq\varsigma^{2(M-1)},
	\end{equation*}
	where the last inequality uses $|\mathscr{R}_{0}|+|\mathscr{R}_{j}|\leq M$. For $m\in\mathscr{R}_{0}$, $\mu^{\omega}_{m}=\mu^{\omega'}_{m}=0$ in the long run, so the bound holds vacuously.

	The Lemma follows as Inequality~\eqref{eq:profignorance} holds for all $\omega,\omega'$.
\end{proof}

\subsection{Proof of Proposition~\ref{prop_smallworld}}

\allowdisplaybreaks

\begin{proof}[Proof of Proposition~\ref{prop_smallworld} (i, $L^*_M$ strictly decreases in $M$)]
	Take the sequence of $\epsilon$-optimal mechanisms $(m_{1k},\mathscr{T}_k,d_k)_{k=1}^{\infty}$ with common decision rule $d$ and its limit long-run distributions $(\mu^{*\omega}_{m})_{\omega,m}$, and denote the limit utility in state~$\omega$ by
	\begin{equation*}
		U^{*}_{\omega}=\sum_{m=1}^{M}\mu^{*\omega}_{m}\,u(d(m),\omega).
	\end{equation*}

	The proof proceeds in five steps. Step~1 identifies a memory state $\tilde{m}$ at which the $M+1$-states mechanism can be built on. Step~2 describes the $M+1$-states mechanism, attaching the new memory state $M+1$ to $\tilde{m}$. Step~3 computes the resulting long-run distribution exactly and reduces the improvement in utility to a sign condition. Step~4 shows that the sign condition can always be met. Step~5 shows that the improvement is uniform along the sequence.

	\emph{Step 1: a ``suitable'' attachment state.}
	Call a memory state $m$ \emph{used} if $\mu^{*\omega}_{m}>0$ for some $\omega$; by Lemma~\ref{lemmma:appendixiff}, $\mu^{*\omega}_{m}>0$ then holds for \emph{every} $\omega$. Note that $\sum_{m\,\mathrm{used}}\mu^{*\omega}_{m}=1$ for every $\omega$, and that at least two actions are taken across used memory states by Equation~\eqref{assumption:learningisuseful}. For each used $m$, define the weight vector $\bm{z}(m)\in\mathbb{R}^{N}$ by
	\begin{equation*}
		z_{\omega}(m)=p^{\omega}\,\mu^{*\omega}_{m}\left[U^{*}_{\omega}-u(d(m),\omega)\right],\qquad \omega=1,\dots,N.
	\end{equation*}
	I claim that some used $\tilde{m}$ has $\bm{z}(\tilde{m})\neq \bm{0}$. Suppose to the contrary; since $p^{\omega}\mu^{*\omega}_{m}>0$, this requires $U^{*}_{\omega}=u(d(m),\omega)$ for all $\omega$ at every used $m$. Taking two used memory states with distinct actions $a_{1}\neq a_{2}$ yields $u(a_{1},\omega)=U^{*}_{\omega}=u(a_{2},\omega)$ for all $\omega$; evaluating at $\omega=a_{1}$ contradicts the assumption that $a=\omega$ is the unique maximizer of $u(a,\omega)$.

	\emph{Step 2: the construction.} Fix a used memory state $\tilde{m}$, two functions $\varphi,\psi:S\rightarrow[0,1]$ with $\psi(s)\geq\frac{1}{4}$ for all $s$, and a scale $\varepsilon\in(0,1)$; the choices are made in Step~4. Define $(m'_{1},\mathscr{T}',d')$ on $M+1$ memory states as follows:
	\begin{itemize}
		\item every original transition probability is multiplied by $(1-\varepsilon)$, with the freed probability assigned to staying at the current memory state; this modification alone leaves every long-run distribution unchanged;
		\item at $\tilde{m}$, upon receiving signal $s$, the DM additionally transits to memory state $M+1$ with probability $\varepsilon\varphi(s)$ (valid, since the scaled transition probabilities from $\tilde{m}$ sum to at most $1-\varepsilon$);
		\item at $M+1$, upon receiving signal $s$, the DM returns to $\tilde{m}$ with probability $\psi(s)$ and stays at $M+1$ otherwise;
		\item $d'(m)=d(m)$ for $m\leq M$, $d'(M+1)=d(\tilde{m})$, and $m'_{1}=m_{1}$.
	\end{itemize}
	Averaging over the signal, the per-period probability of entering $M+1$ from $\tilde{m}$ in state $\omega$ is $\varepsilon q_{\omega}$, and that of exiting (from $M+1$ to $\tilde{m}$) is $e_{\omega}$, where
	\begin{equation*}
		q_{\omega}=\sum_{s\in S}f^{\omega}(s)\varphi(s),\qquad e_{\omega}=\sum_{s\in S}f^{\omega}(s)\psi(s)\geq\frac{1}{4}.
	\end{equation*}

	\emph{Step 3: exact long-run distribution.} I claim that in state $\omega$ the long-run distribution of $(m'_{1},\mathscr{T}',d')$ is exactly
	\begin{equation}\label{eq:newproofstationary}
		\mu'^{\omega}_{m}=A_{\omega}\,\mu^{*\omega}_{m}\ \ (m\leq M),\qquad \mu'^{\omega}_{M+1}=A_{\omega}\,\mu^{*\omega}_{\tilde{m}}\,\frac{\varepsilon q_{\omega}}{e_{\omega}},\qquad A_{\omega}=\frac{1}{1+\mu^{*\omega}_{\tilde{m}}\,\varepsilon q_{\omega}/e_{\omega}}.
	\end{equation}
	To verify, let $Q^{\omega}_{mm'}=\sum_{s\in S}f^{\omega}(s)\Pr[\mathscr{T}(m,s)=m']$ denote the entries of the original transition matrix $\mathbf{Q}^{\omega}$ (Section~\ref{section:model}), so that $\sum_{m'}\mu^{*\omega}_{m'}Q^{\omega}_{m'm}=\mu^{*\omega}_{m}$. The new transition probabilities in state~$\omega$ are: from $m\leq M$ with $m\neq\tilde{m}$, to $m'\leq M$: $(1-\varepsilon)Q^{\omega}_{mm'}+\varepsilon\mathbbm{1}\lbrace m'=m\rbrace$; from $\tilde{m}$: to $m'\leq M$: $(1-\varepsilon)Q^{\omega}_{\tilde{m}m'}+\varepsilon(1-q_{\omega})\mathbbm{1}\lbrace m'=\tilde{m}\rbrace$, and to $M+1$: $\varepsilon q_{\omega}$; from $M+1$ to $\tilde{m}$: $e_{\omega}$, staying at $M+1$ with probability $1-e_{\omega}$. Three balance checks confirm Equation~\eqref{eq:newproofstationary}:
	\begin{itemize}
		\item at $m\leq M$, $m\neq\tilde{m}$:
		\begin{equation*}
			\text{inflow}=\sum_{m'\leq M}A_{\omega}\mu^{*\omega}_{m'}\left[(1-\varepsilon)Q^{\omega}_{m'm}+\varepsilon\mathbbm{1}\lbrace m'=m\rbrace\right]=A_{\omega}\left[(1-\varepsilon)\mu^{*\omega}_{m}+\varepsilon\mu^{*\omega}_{m}\right]=\mu'^{\omega}_{m};
		\end{equation*}
		\item at $M+1$: inflow $=\mu'^{\omega}_{\tilde{m}}\,\varepsilon q_{\omega}+\mu'^{\omega}_{M+1}(1-e_{\omega})$, which equals $\mu'^{\omega}_{M+1}$ as $\mu'^{\omega}_{\tilde{m}}\,\varepsilon q_{\omega}=\mu'^{\omega}_{M+1}\,e_{\omega}$ stated by Equation~\eqref{eq:newproofstationary};
		\item at $\tilde{m}$:
		\begin{align*}
			\text{inflow}&=\sum_{m'\leq M}A_{\omega}\mu^{*\omega}_{m'}(1-\varepsilon)Q^{\omega}_{m'\tilde{m}}+A_{\omega}\mu^{*\omega}_{\tilde{m}}\,\varepsilon(1-q_{\omega})+\mu'^{\omega}_{M+1}e_{\omega}\\
			&=A_{\omega}\left[(1-\varepsilon)\mu^{*\omega}_{\tilde{m}}+\varepsilon\mu^{*\omega}_{\tilde{m}}(1-q_{\omega})+\varepsilon\mu^{*\omega}_{\tilde{m}}q_{\omega}\right]=\mu'^{\omega}_{\tilde{m}}.
		\end{align*}
	\end{itemize}
	The normalization $\sum_{m\leq M}\mu'^{\omega}_{m}+\mu'^{\omega}_{M+1}=A_{\omega}\left(1+\mu^{*\omega}_{\tilde{m}}\varepsilon q_{\omega}/e_{\omega}\right)=1$ pins down $A_{\omega}$. Since $d'(M+1)=d(\tilde{m})$, the new utility in state $\omega$ is $U'_{\omega}=A_{\omega}\left[U^{*}_{\omega}+\mu^{*\omega}_{\tilde{m}}(\varepsilon q_{\omega}/e_{\omega})u(d(\tilde{m}),\omega)\right]$, and using $A_{\omega}-1=-A_{\omega}\mu^{*\omega}_{\tilde{m}}\varepsilon q_{\omega}/e_{\omega}$, 
the change in asymptotic utility --- equivalently, the reduction in asymptotic utility loss --- is
	\begin{equation}\label{eq:newproofmaster}
		\Delta U:=U(m'_{1},\mathscr{T}',d')-\sum_{\omega=1}^{N}p^{\omega}U^{*}_{\omega}=\sum_{\omega=1}^{N}p^{\omega}\left(U'_{\omega}-U^{*}_{\omega}\right)=-\varepsilon\sum_{\omega=1}^{N}A_{\omega}\,z_{\omega}(\tilde{m})\,\frac{q_{\omega}}{e_{\omega}}.
	\end{equation}
	Moreover, since $\mu^{*\omega}_{\tilde{m}}\leq 1$, $q_{\omega}\leq 1$ and $e_{\omega}\geq\frac{1}{4}$, we have $0\leq 1-A_{\omega}\leq 4\varepsilon$ and $q_{\omega}/e_{\omega}\leq 4$; hence, if
	\begin{equation}\label{eq:newproofcriterion}
		\gamma:=-\sum_{\omega=1}^{N}z_{\omega}(\tilde{m})\,\frac{q_{\omega}}{e_{\omega}}>0,
	\end{equation}
	then $\Delta U=\varepsilon\left[\gamma+\sum_{\omega}(1-A_{\omega})z_{\omega}(\tilde{m})q_{\omega}/e_{\omega}\right]\geq\varepsilon\left[\gamma-16\varepsilon\sum_{\omega}\vert z_{\omega}(\tilde{m})\vert\right]\geq\frac{\varepsilon\gamma}{2}>0$ whenever $\varepsilon\leq\gamma/\left(32\sum_{\omega}\vert z_{\omega}(\tilde{m})\vert\right)$. It therefore remains to choose $\tilde{m},\varphi,\psi$ satisfying Equation~\eqref{eq:newproofcriterion}.

	\emph{Step 4: achieving Equation~\eqref{eq:newproofcriterion}.} For a used $m$ and signal realizations $s_{1},\dots,s_{r}$, define the \emph{string sum}
	\begin{equation*}
		\langle s_{1}\cdots s_{r}\rangle_{m}:=\sum_{\omega=1}^{N}z_{\omega}(m)\,f^{\omega}(s_{1})\cdots f^{\omega}(s_{r}).
	\end{equation*}
	Length-one string sums drive Case~A below. Longer strings arise in Case~B
	. Since $\sum_{m\,\mathrm{used}}\mu^{*\omega}_{m}=1$ for all $\omega$, for every signal $s$,
	\begin{equation}\label{eq:newproofidentity}
		\sum_{m\,\mathrm{used}}\langle s\rangle_{m}=\sum_{\omega=1}^{N}p^{\omega}f^{\omega}(s)\left[U^{*}_{\omega}-\sum_{m\,\mathrm{used}}\mu^{*\omega}_{m}u(d(m),\omega)\right]=\sum_{\omega=1}^{N}p^{\omega}f^{\omega}(s)\left[U^{*}_{\omega}-U^{*}_{\omega}\right]=0.
	\end{equation}

	\emph{Case A: there exist a used $m'$ and a signal $s^{*}$ with $\langle s^{*}\rangle_{m'}<0$.} Take $\tilde{m}=m'$, $\varphi=\mathbbm{1}\lbrace s=s^{*}\rbrace$ and $\psi\equiv 1$. Then $q_{\omega}=f^{\omega}(s^{*})$, $e_{\omega}=1$, and $\gamma=-\langle s^{*}\rangle_{m'}>0$.

	\emph{Case B: $\langle s\rangle_{m}\geq 0$ for every used $m$ and every $s\in S$.} By Equation~\eqref{eq:newproofidentity} the pointwise sum over used $m$ is zero, so in fact $\langle s\rangle_{m}=0$ for every used $m$ and every $s$. Take the $\tilde{m}$ from Step~1, and write $\bm{z}=\bm{z}(\tilde{m})\neq \bm{0}$ (by step 1). Suppose, toward a contradiction, that no admissible $(\varphi,\psi)$ satisfies Equation~\eqref{eq:newproofcriterion}. Restrict attention to $\varphi=\mathbbm{1}\lbrace s\rbrace$ for a signal $s$, and, for distinct signals $s_{1},\dots,s_{J}$ and $\vert t_{j}\vert\leq\frac{1}{4J}$,
	\begin{equation*}
		\psi=\frac{1}{2}-\sum_{j=1}^{J}t_{j}\,\mathbbm{1}\lbrace s_{j}\rbrace \in\left[\tfrac{1}{4},\tfrac{3}{4}\right],
	\end{equation*}
	which is admissible for either sign of each $t_{j}$. For this choice of $(\varphi,\psi)$, $q_{\omega}=f^{\omega}(s)$; write $\gamma_{s}(\bm{t})$ for the resulting value of $\gamma$ in Equation~\eqref{eq:newproofcriterion}. Case B then implies, for every $s$ and all such $\bm{t}=(t_{1},\dots,t_{J})$,
	\begin{equation}\label{eq:newproofnonneg}
		-\gamma_{s}(\bm{t})=\sum_{\omega=1}^{N}z_{\omega}\,\frac{f^{\omega}(s)}{e_{\omega}(\bm{t})}\geq 0,\qquad \text{ where } e_{\omega}(\bm{t})=\frac{1}{2}-\sum_{j=1}^{J}t_{j}f^{\omega}(s_{j}).
	\end{equation}

	I claim that that \eqref{eq:newproofnonneg} implies all string sums of $\tilde{m}$ vanish, by induction on the length $r$. The base case $r=1$ is the premise of Case-B. Suppose all string sums of length at most $R$ vanish. Expanding the geometric series $1/e_{\omega}(\bm{t})=2\sum_{n\geq 0}\left(\sum_{j}2t_{j}f^{\omega}(s_{j})\right)^{n}$ for small $\vert \bm{t}\vert$,
	\begin{equation*}
		-\gamma_{s}(\bm{t})=\sum_{n\geq 0}2^{n+1}\sum_{\omega=1}^{N}z_{\omega}f^{\omega}(s)\Bigl(\sum_{j=1}^{J}t_{j}f^{\omega}(s_{j})\Bigr)^{n},
	\end{equation*}
	whose coefficient on the monomial $t_{1}^{\alpha_{1}}\cdots t_{J}^{\alpha_{J}}$ (with $\alpha_{1}+\dots+\alpha_{J}=n$) equals, by the multinomial theorem, the string sum $\langle s\,s_{1}^{\alpha_{1}}\cdots s_{J}^{\alpha_{J}}\rangle_{\tilde{m}}$ of length $n+1$ (signal $s$ followed by $s_{j}$ repeated $\alpha_{j}$ times) multiplied by the nonzero constant $2^{n+1}n!/(\alpha_{1}!\cdots\alpha_{J}!)$; these vanish for $n\leq R-1$ by the induction hypothesis. Hence $-\gamma_{s}(\bm{t})=2^{R+1}P_{s}(\bm{t})+O(\vert \bm{t}\vert^{R+1})$ with $P_{s}(\bm{t})=\sum_{\omega}z_{\omega}f^{\omega}(s)\left(\sum_{j}t_{j}f^{\omega}(s_{j})\right)^{R}$, a homogeneous polynomial of degree $R$. Equation~\eqref{eq:newproofnonneg} forces $P_{s}(\bm{t})\geq 0$ for all $\bm{t}\in\mathbb{R}^{J}$: for any $\bm{t}$ and all small enough $\tau>0$, $\tau\bm{t}$ is admissible, so $0\leq-\gamma_{s}(\tau\bm{t})=2^{R+1}\tau^{R}P_{s}(\bm{t})+O(\tau^{R+1})$; dividing by $\tau^{R}$ and letting $\tau\downarrow 0$ gives $P_{s}(\bm{t})\geq 0$. If $R$ is odd, then $P_{s}(-\bm{t})=-P_{s}(\bm{t})$ together with $P_{s}\geq 0$ forces $P_{s}\equiv 0$. If $R$ is even, then summing over the entry signal $s$ and using $\sum_{s}f^{\omega}(s)=1$,
	\begin{equation*}
		\sum_{s\in S}P_{s}(\bm{t})=\sum_{\omega=1}^{N}z_{\omega}\Bigl(\sum_{j=1}^{J}t_{j}f^{\omega}(s_{j})\Bigr)^{R},
	\end{equation*}
	a polynomial whose coefficients are string sums of length $R$ — identically zero by the induction hypothesis; since each $P_{s}\geq 0$ and they sum to the zero polynomial, again $P_{s}\equiv 0$ for every $s$. In both cases all coefficients of every $P_{s}$ vanish; as $s$, $J$, the signals $s_{j}$ and the multiplicities are arbitrary, all string sums of length $R+1$ vanish. This completes the induction.

	Finally, I show that vanishing of all string sums forces $\bm{z}=\bm{0}$, which contradicts step 1. For a signal $s$, let $\bm{f}(s)=\left(f^{1}(s),\dots,f^{N}(s)\right)\in\mathbb{R}^{N}$ denote the vector of its likelihoods across states, and let $W\subseteq\mathbb{R}^{N}$ be the linear span of the vectors $\left(f^{\omega}(s_{1})\cdots f^{\omega}(s_{r})\right)_{\omega=1}^{N}$ over all finite strings; each such vector is the entrywise (coordinate-by-coordinate) product of $\bm{f}(s_{1}),\dots,\bm{f}(s_{r})$. $W$ is closed under entrywise products (the entrywise product of two string vectors is the vector of the concatenated string, and the property extends to linear combinations), contains the all-ones vector $\bm{1}$ (sum $\bm{f}(s)$ over $s\in S$ and use $\sum_{s}f^{\omega}(s)=1$), and separates states: for $\omega\neq\omega'$ there exists $s$ with $f^{\omega}(s)\neq f^{\omega'}(s)$, as no two signal structures coincide. Fix any $\omega_{0}$; for each $\omega\neq\omega_{0}$ pick $s(\omega)$ with $f^{\omega}(s(\omega))\neq f^{\omega_{0}}(s(\omega))$; then each vector
	\begin{equation*}
		\bm{g}(\omega)=\frac{\bm{f}(s(\omega))-f^{\omega}(s(\omega))\,\bm{1}}{f^{\omega_{0}}(s(\omega))-f^{\omega}(s(\omega))},\qquad\omega\neq\omega_{0},
	\end{equation*}
	lies in $W$ (a linear combination of $\bm{f}(s(\omega))\in W$ and $\bm{1}\in W$, with nonzero denominator by the choice of $s(\omega)$), and so does their entrywise product $\bm{v}$, with entries $v_{\omega'}=\prod_{\omega\neq\omega_{0}}g_{\omega'}(\omega)$, as $W$ is closed under entrywise products. Since $g_{\omega'}(\omega)=0$ for $\omega'=\omega$ and $g_{\omega_{0}}(\omega)=1$, the vector $\bm{v}$ vanishes at every $\omega'\neq\omega_{0}$ and equals $1$ at $\omega'=\omega_{0}$; hence $W$ contains every coordinate vector and $W=\mathbb{R}^{N}$. All string sums vanishing means $\bm{z}\perp W=\mathbb{R}^{N}$, i.e., $\bm{z}=\bm{0}$ — contradicting Step~1. Therefore, in Case~B as well, some admissible $(\varphi,\psi)$ satisfies Equation~\eqref{eq:newproofcriterion}: an entry signal $s$ and exit probabilities $\bm{t}$ with $\gamma_{s}(\bm{t})>0$, i.e., the constructed mechanism leads to a strict improvement.

	\emph{Step 5: uniformity in $k$.} Apply the construction with the fixed objects $(\tilde{m},\varphi,\psi,\varepsilon)$ chosen above to the $k$-th mechanism $(m_{1k},\mathscr{T}_{k},d)$. The derivation of Step~3 holds with $(\mu^{\omega}_{m,k},U_{\omega,k})$ in place of the limit objects, so the improvement $L(m_{1k},\mathscr{T}_{k},d)-L(m'_{1k},\mathscr{T}'_{k},d')$ equals the exact expression in Equation~\eqref{eq:newproofmaster} evaluated at the $k$-th quantities, which converges to $\Delta U\geq\varepsilon\gamma/2>0$ as $k\rightarrow\infty$. Setting $\bar{\epsilon}=\varepsilon\gamma/4$, there exists $K^{*}$ such that the improvement is at least $\bar{\epsilon}$ for all $k\geq K^{*}$, whence
	\begin{equation*}
		L^{*}_{M+1}\leq\liminf_{k\rightarrow\infty}L(m'_{1k},\mathscr{T}'_{k},d')\leq\lim_{k\rightarrow\infty}L(m_{1k},\mathscr{T}_{k},d)-\bar{\epsilon}=L^{*}_{M}-\bar{\epsilon}<L^{*}_{M}.
	\end{equation*}

\end{proof}

\begin{proof}[Proof of  Proposition~\ref{prop_smallworld} (i, convergence rate bound) and (ii)]
	I first define ``confirmatory signals'' for each state: A signal $s$ supports state $\omega$ with probability $G^\omega(s)$ and supports no state with probability $1-\sum_{\omega=1}^{N}G^\omega(s)$, where
	\begin{equation}\label{eq:signalsetdefinition}
		G^\omega(s)\propto\dfrac{f^{\omega}(s)}{\sqrt{\sum_{s'\in S} (f^{\omega}(s'))^{2}}}
	\end{equation}
	with common normalization such that $\sum_{\omega=1}^{N}G^\omega(s)\leq 1$ for all $s$.\footnote{In the mechanisms constructed in the proofs, the common normalization factor multiplies every transition probability between distinct memory states, which leaves the long-run distribution unchanged.} With some abuse of notations, I use $F^{\omega}(G^{\omega'}(s))$ to denote the probability of receiving a signal supporting state $\omega'$, or equivalently ``a signal $G^{\omega'}$'', in state $\omega$, i.e., $F^{\omega}(G^{\omega'})\equiv\sum_{s\in S} f^{\omega}(s)G^{\omega'}(s)$. Importantly, Equation~\ref{eq:signalsetdefinition} implies it is more likely that the DM receives a signal supporting a correct state than a signal supporting a wrong state:
	\begin{equation*}
		F^{\omega}(G^{\omega})=\dfrac{\sum_{s\in S} (f^{\omega}(s))^{2}}{\sqrt{\sum_{s\in S} (f^{\omega}(s))^{2}}}=\sqrt{\sum_{s\in S} (f^{\omega}(s))^{2}}
		>\dfrac{\sum_{s\in S} f^{\omega}(s)f^{\omega'}(s)}{\sqrt{\sum_{s\in S} (f^{\omega'}(s))^{2}}}=F^{\omega}(G^{\omega'}),
	\end{equation*}			
	implied by the Cauchy–Schwarz inequality.
	
	I now formally describe the simple updating mechanism described in the main text. Denote $\lambda=\floor*{(M-1)/N}$, relabel the memory states as $0,11,\cdots,1\lambda,$ $21,\cdots,2\lambda,\cdots,N\lambda$ and denote the unused memory states as $N\lambda+1,N\lambda+2,\cdots,M-1$. The decision rule is as follows:
	\begin{equation*}
		\begin{split}
			d(0)&= 1;\\
			d(\omega k)&=\omega \text{ for all $\omega=1,\dots,N$ and $k=1,\cdots,\lambda$};\\
			d(m)&=1 \text{ for all $m>N\lambda$}.
		\end{split}
	\end{equation*}

	The transition function between the memory states is defined below for some big enough $\delta>1$ such that 
	\begin{equation*}
		\dfrac{\delta F^{\omega}(G^{\omega'})}{\sum_{\omega''\neq\omega'} F^{\omega}(G^{\omega''})}>1\text{ for all $\omega,\omega'\in\Omega$.}
	\end{equation*}
	Suppose the DM receives some signal $s$, he follows the following transition rule:
		\begin{equation*}
		\mathscr{T}(0,s)=
		\begin{cases}
			\omega1 &\text{ with probability $G^\omega(s)$ for all $\omega=1,\cdots,N$};\\
			0 &\text{ with probability $1-\sum_{\omega=1}^{N} G^{\omega}(s)$}.
		\end{cases}
	\end{equation*}
	\begin{equation*}
		\mathscr{T}(\omega1,s)=
		\begin{cases}
			\omega2 &\text{ with probability $G^{\omega}(s)$;}\\
			0 & \text{ with probability $\sum_{j\in\Omega\setminus \lbrace \omega\rbrace}\frac{G^{j}(s)}{\delta}$;}\\
			\omega1 &\text{ with probability $1-\sum_{j\in\Omega\setminus \lbrace \omega\rbrace} \frac{G^{j}(s)}{\delta}-G^{\omega}(s)$.}
		\end{cases}
	\end{equation*}
	\begin{equation*}
		\mathscr{T}(\omega\lambda,s)=\begin{cases}
			\omega(\lambda-1) &\text{ with probability $\sum_{j\in\Omega\setminus \lbrace \omega\rbrace}\frac{G^{j}(s)}{\delta}$;}\\
			\omega\lambda &\text{ with probability $1-\sum_{j\in\Omega\setminus \lbrace \omega\rbrace}\frac{G^{j}(s)}{\delta}$.}
		\end{cases}
	\end{equation*}
	while for $k=2,3,\cdots,\lambda-1$,
	\begin{equation*}
		\mathscr{T}(\omega k,s)=\begin{cases}
			\omega(k+1) &\text{ with probability $G^{\omega}(s)$;}\\
			\omega(k-1) &\text{ with probability $\sum_{j\in\Omega\setminus \lbrace \omega\rbrace}\frac{G^{j}(s)}{\delta}$;}\\
			\omega k &\text{ with probability $1-\sum_{j\in\Omega\setminus \lbrace \omega\rbrace}\frac{G^{j}(s)}{\delta}-G^{\omega}(s)$.}
		\end{cases}
	\end{equation*}
	Finally, for $m>N\lambda$, $\mathscr{T}(m,s)=m$ for all $s$. By restricting the initial memory state to one of $0,11,12,\cdots,1\lambda,21,22,\cdots,2\lambda,\cdots,N\lambda$, the DM will never transition to memory states $m>N\lambda$. 
		
	Note that this updating mechanism does not depend on $p^\omega$ nor $\bm{u}^\omega$. Now I compute the long-run distribution $\bm{\mu}^{\omega}$ and the utility loss $L(m_1,\mathscr{T},d)$. Fix a state $\omega$, at the two extreme memory states in branch $\omega'$, i.e., memory states $\omega'\lambda$ and $\omega'(\lambda-1)$,
	\begin{equation*}
		\begin{split}
			\mu^{\omega}_{\omega'(\lambda-1)}F^{\omega}(G^{\omega'})&=\mu^{\omega}_{\omega'\lambda}\frac{1}{\delta}\sum_{\omega''\neq\omega'}F^{\omega}(G^{\omega''})\\
			\mu^{\omega}_{\omega'(\lambda-1)}&=\mu^{\omega}_{\omega'\lambda}\left[\frac{\delta F^{\omega}(G^{\omega'})}{\sum_{\omega''\neq\omega'}F^{\omega}(G^{\omega''})}\right]^{-1}
		\end{split}
	\end{equation*}
	for all $\omega'$. Next, at memory state $\omega'(\lambda-1)$,
	\begin{equation*}
		\begin{split}
			\mu^{\omega}_{\omega'\lambda}\frac{1}{\delta}\sum_{\omega''\neq\omega'}F^{\omega}(G^{\omega''})+\mu^{\omega}_{\omega'(\lambda-2)}F^{\omega}(G_{\omega'})&=\mu^{\omega}_{\omega'(\lambda-1)}\left[\frac{1}{\delta}\sum_{\omega''\neq\omega'}F^{\omega}(G^{\omega''})+F^{\omega}(G^{\omega'})\right]\\
			\mu^{\omega}_{\omega'(\lambda-2)}&=\mu^{\omega}_{\omega'(\lambda-1)}\left[\frac{\delta F^{\omega}(G^{\omega'})}{\sum_{\omega''\neq\omega'}F^{\omega}(G^{\omega''})}\right]^{-1}.
		\end{split}
	\end{equation*}
	Repeating the same procedures implies that for all $k=1,\cdots,\lambda$ and $\omega'=1,\cdots,N$
	\begin{equation}
		\mu^{\omega}_{\omega'k}=\mu^{\omega}_{\omega'\lambda}\left[\frac{\delta F^{\omega}(G^{\omega'})}{\sum_{\omega''\neq\omega'}F^{\omega}(G^{\omega''})}\right]^{-(\lambda-k)}, 
	\end{equation} 
	\begin{equation}
		\mu^{\omega}_{\omega'k}=\mu^{\omega}_{0}\left[\frac{\delta F^{\omega}(G^{\omega'})}{\sum_{\omega''\neq\omega'}F^{\omega}(G^{\omega''})}\right]^{k}.
	\end{equation}
	As $\sum_{\omega'=1}^{N}\sum_{k=1}^{\lambda}\mu^{\omega}_{\omega'k}+\mu^{\omega}_{0}=1$, and denote $\frac{\delta F^{\omega}(G^{\omega'})}{\sum_{\omega''\neq\omega'}F^{\omega}(G^{\omega''})}$ by $r^{\omega\omega'}$, we have
	\begin{equation*}
		\begin{split}
			\mu^{\omega}_{\omega\lambda}\sum_{k=1}^{\lambda}\left[r^{\omega\omega}\right]^{-(\lambda-k)}+\mu^{\omega}_{\omega\lambda}\left[r^{\omega\omega}\right]^{-\lambda}
			+\mu^{\omega}_{\omega\lambda}\left[r^{\omega\omega}\right]^{-\lambda}\sum_{\omega'\neq\omega}\sum_{k=1}^{\lambda}\left[r^{\omega\omega'}\right]^{k}=1\\
			\Rightarrow\quad  \mu_{\omega\lambda}^{\omega}=\frac{1}{\sum_{k=1}^{\lambda}\left[r^{\omega\omega}\right]^{-(\lambda-k)}+\left[r^{\omega\omega}\right]^{-\lambda}+\left[r^{\omega\omega}\right]^{-\lambda}\sum_{\omega'\neq\omega}\sum_{k=1}^{\lambda}\left[r^{\omega\omega'}\right]^{k}}
		\end{split}
	\end{equation*}
	and the probability of choosing actions in $\Omega\setminus\omega$ in state~$\omega$ is smaller than $\sum_{m\notin\lbrace\omega1,\cdots,\omega\lambda\rbrace}\mu_{m}^{\omega}$ which is as follows:
	\begin{align*}
		&\frac{\left[r^{\omega\omega}\right]^{-\lambda}+\left[r^{\omega\omega}\right]^{-\lambda}\sum_{\omega'\neq\omega}\sum_{k=1}^{\lambda}\left[r^{\omega\omega'}\right]^{k}}{\sum_{k=1}^{\lambda}\left[r^{\omega\omega}\right]^{-(\lambda-k)}+\left[r^{\omega\omega}\right]^{-\lambda}+\left[r^{\omega\omega}\right]^{-\lambda}\sum_{\omega'\neq\omega}\sum_{k=1}^{\lambda}\left[r^{\omega\omega'}\right]^{k}}\\
		<&\left[r^{\omega\omega}\right]^{-\lambda}+\left[r^{\omega\omega}\right]^{-\lambda}\sum_{\omega'\neq\omega}\sum_{k=1}^{\lambda}\left[r^{\omega\omega'}\right]^{k}\\
		=&\left[r^{\omega\omega}\right]^{-\lambda}+\sum_{\omega'\neq\omega}\frac{r^{\omega\omega'}}{r^{\omega\omega'}-1}\left[\left(\frac{r^{\omega\omega'}}{r^{\omega\omega}}\right)^{\lambda}-[r^{\omega\omega}]^{-\lambda}\right]\\
		<&\left[r^{\omega\omega}\right]^{-\lambda}+\sum_{\omega'\neq\omega}\frac{r^{\omega\omega'}}{r^{\omega\omega'}-1}\left[\left(\frac{r^{\omega\omega'}}{r^{\omega\omega}}\right)^{\lambda}\right]\stepcounter{equation}\tag{\theequation}\label{eq:upperboundonL}\\
		<&\left[r^{\omega\omega}\right]^{-\lambda}+(N-1)\frac{\max_{\omega'\neq\omega}r^{\omega\omega'}}{\max_{\omega'\neq\omega}r^{\omega\omega'}-1}\left[\max_{\omega'\neq\omega}\left(\frac{r^{\omega\omega'}}{r^{\omega\omega}}\right)^{\lambda}\right]\\
		<&\left[(N-1)\frac{\max_{\omega'\neq\omega}r^{\omega\omega'}}{\max_{\omega'\neq\omega}r^{\omega\omega'}-1}+1\right]\max\left\lbrace\left[r^{\omega\omega}\right]^{-\lambda},\max_{\omega'\neq\omega}\left(\frac{r^{\omega\omega'}}{r^{\omega\omega}}\right)^{\lambda}\right\rbrace
	\end{align*}
	The first inequality of Equation~\eqref{eq:upperboundonL} is implied by the fact that the denominator is strictly greater than $1$. Now, using Equation~\eqref{eq:upperboundonL} and denoting $K^{\omega}=\left[(N-1)\frac{\max_{\omega'\neq\omega}r^{\omega\omega'}}{\max_{\omega'\neq\omega}r^{\omega\omega'}-1}+1\right]$, we can compute the upper bound of the utility loss $L(m_1,\mathscr{T},d)$
	\begin{equation}
		\begin{split}
			L(m_1,\mathscr{T},d)&\leq \sum_{\omega=1}^{N}p^\omega\max_{a\neq\omega}\lbrace u(\omega,\omega)-u(a,\omega)\rbrace\left[\sum_{m\notin\lbrace\omega1,\cdots,\omega\lambda\rbrace}\mu_{m}^{\omega}\right]\\
			&<	\sum_{\omega=1}^{N}p^\omega\max_{a\neq\omega}\lbrace u(\omega,\omega)-u(a,\omega)\rbrace K^{\omega}\max\left\lbrace\left[r^{\omega\omega}\right]^{-\lambda},\max_{\omega'\neq\omega}\left(\frac{r^{\omega\omega'}}{r^{\omega\omega}}\right)^{\lambda}\right\rbrace\\
			&<\max_{\omega}\left[\max_{a\neq\omega}\lbrace u(\omega,\omega)-u(a,\omega)\rbrace K^{\omega}\max\left\lbrace\left[r^{\omega\omega}\right]^{-\lambda},\max_{\omega'\neq\omega}\left(\frac{r^{\omega\omega'}}{r^{\omega\omega}}\right)^{\lambda}\right\rbrace\right] 
		\end{split}
	\end{equation}
	As $[r^{\omega\omega}]^{-1}$ and $\max_{\omega'\neq\omega}\left(\frac{r^{\omega\omega'}}{r^{\omega\omega}}\right)$ are strictly smaller than $1$ for all $\omega$, the exponential rate bound in Proposition~\ref{prop_smallworld}(i) follows. Moreover, the right-hand side of Equation~\eqref{eq:upperboundonL} converges to $0$ as $\lambda$ goes to $\infty$. This proves Proposition~\ref{prop_smallworld}(ii).

\end{proof}

	\subsection{Proof of Proposition~\ref{prop_smallworldinfiniteN}}
\begin{proof}
	I first prove part (i), i.e., $h<1$. I then turn to big worlds ($h\geq 1$): I first prove Equation~\eqref{eq:limitignorancedisplay} of part (ii), then show that it implies the payoff consequence (whenever $L^{*}_{N}$ converges, $\lim_{N\rightarrow\infty}L^{*}_{N}=1-\overline{p}_{0}$; convergence holds under Assumption~\ref{assumption_fullsupportinfiniteplus}), and lastly verify the concluding statement of the proposition.

	\paragraph{1) Part (i): $h<1$.} I show that for all $\epsilon>0$, there exists a sequence of updating mechanisms whose utility losses are bounded above by $\epsilon$ in the limit; this gives $\lim_{N\rightarrow\infty}L^{*}_{N}=0$. Apply Assumption~\ref{assumption_identifiableinfinte} with $\varepsilon=\epsilon$: there exist a sequence of subsets $\widehat{\Omega}_{N}\subseteq\lbrace 1,\cdots,N\rbrace$ with $\liminf_{N\rightarrow\infty}\sum_{\omega\in\widehat{\Omega}_{N}}p_{N}^{\omega}\geq 1-\epsilon$ and a constant $\kappa>0$ (the liminf of the pairwise Cauchy--Schwarz distances within $\widehat{\Omega}_{N}$). By the definition of $\liminf$, for all sufficiently large $N\in\mathcal{N}$,
	\begin{equation}\label{eq:IDpairwise}
		\min_{\substack{\omega,\omega'\in\widehat{\Omega}_{N}\\ \omega'\neq\omega}}\left\lbrace-\log\dfrac{\sum_{s\in S_{N}} f_{N}^{\omega}(s)f_{N}^{\omega'}(s)}{\sqrt{\sum_{s\in S_{N}} (f_{N}^{\omega}(s))^{2}}\sqrt{\sum_{s\in S_{N}} (f_{N}^{\omega'}(s))^{2}}}\right\rbrace>\frac{\kappa}{2}.
	\end{equation}
	For each $N$, let $(G^{\omega}_{N})_{\omega=1}^{N}$ be the confirmatory-signal probabilities of Equation~\eqref{eq:signalsetdefinition}, applied to the $N$-th decision problem (i.e., to the signal structures $(f^{\omega}_{N})_{\omega=1}^{N}$ on $S_{N}$), and recall the notation $F^{\omega}_{N}(G^{\omega'}_{N})=\sum_{s\in S_{N}}f^{\omega}_{N}(s)G^{\omega'}_{N}(s)$. For a constant $\delta_{N}>1$, all states $\omega$, and $\omega'\in\widehat{\Omega}_{N}$, define the branch ratios
	\begin{equation*}
		r^{\omega\omega'}_{N}=\dfrac{\delta_{N}F^{\omega}_{N}(G^{\omega'}_{N})}{\sum_{\omega''\in\widehat{\Omega}_{N}\setminus\lbrace\omega'\rbrace}F^{\omega}_{N}(G^{\omega''}_{N})};
	\end{equation*}
	these are the quantities $r^{\omega\omega'}$ from the proof of Proposition~\ref{prop_smallworld}, computed for the updating mechanism that includes only the branches of the states in $\widehat{\Omega}_{N}$ (formally described below). Note that the normalizing constant in Equation~\eqref{eq:signalsetdefinition} cancels from every ratio used below. The following lemma collects the two properties of $(r^{\omega\omega'}_{N})$ used in the sequel.
	\begin{lemma}\label{lemmasmallworldinfoinfiniteas}
		Let $\widehat{\Omega}_{N}$ and $\kappa>0$ be given by Assumption~\ref{assumption_identifiableinfinte} as above, and set $\tilde{\kappa}:=\exp(\kappa/2)-1>0$. There exists a sequence of constants $\delta_{N}>1$ such that
		\begin{enumerate}[label=(\alph*)]
			\item $r^{\omega\omega'}_{N}>2$ for all states $\omega$, all $\omega'\in\widehat{\Omega}_{N}$, and all $N\in\mathcal{N}$;
			\item $\dfrac{r^{\omega\omega}_{N}}{r^{\omega\omega'}_{N}}>1+\tilde{\kappa}$ for all $\omega,\omega'\in\widehat{\Omega}_{N}$ with $\omega'\neq\omega$, for all sufficiently large $N\in\mathcal{N}$.
		\end{enumerate}
	\end{lemma}
	\begin{proof}
		(a) As $F_{N}^{\omega}(G_{N}^{\omega'})>0$ for all $\omega,\omega'$ (signal structures have full support on $S_{N}$ by Assumption~\ref{assumption_noperfectsignalinfinite}), for each $N$ there exists a big enough $\delta_{N}$ such that (a) holds. (b) does not restrict the choice of $\delta_{N}$ further, as $\delta_{N}$ cancels from the ratio: for $\omega,\omega'\in\widehat{\Omega}_{N}$ with $\omega'\neq\omega$,
		\begin{equation}\label{eq:smallworldinfinitelemma}
				\resizebox{0.9\hsize}{!}{$
			\dfrac{r^{\omega\omega}_{N}}{r^{\omega\omega'}_{N}}=\dfrac{F_{N}^{\omega}(G_{N}^{\omega})}{F_{N}^{\omega}(G_{N}^{\omega'})}\cdot\dfrac{\sum_{\omega''\in\widehat{\Omega}_{N}\setminus\lbrace\omega'\rbrace} F_{N}^{\omega}(G_{N}^{\omega''})}{\sum_{\omega''\in\widehat{\Omega}_{N}\setminus\lbrace\omega\rbrace}F_{N}^{\omega}(G_{N}^{\omega''})}
			\geq\dfrac{F_{N}^{\omega}(G_{N}^{\omega})}{F_{N}^{\omega}(G_{N}^{\omega'})}
			=\dfrac{\sqrt{\sum_{s\in S_{N}} (f_{N}^{\omega}(s))^{2}}\sqrt{\sum_{s\in S_{N}} (f_{N}^{\omega'}(s))^{2}}}{\sum_{s\in S_{N}} f_{N}^{\omega}(s)f_{N}^{\omega'}(s)},
			$}		\end{equation}
		where the inequality holds because $F_{N}^{\omega}(G_{N}^{\omega})\geq F_{N}^{\omega}(G_{N}^{\omega'})$ (the Cauchy--Schwarz comparison following Equation~\eqref{eq:signalsetdefinition}) implies $\sum_{\omega''\in\widehat{\Omega}_{N}\setminus\lbrace\omega'\rbrace} F_{N}^{\omega}(G_{N}^{\omega''})\geq\sum_{\omega''\in\widehat{\Omega}_{N}\setminus\lbrace\omega\rbrace} F_{N}^{\omega}(G_{N}^{\omega''})$, and the last equality follows from the definition of $G^{\omega}_{N}$. With Equation~\eqref{eq:IDpairwise}, implied by Assumption~\ref{assumption_identifiableinfinte}, the right-hand side of Equation~\eqref{eq:smallworldinfinitelemma} exceeds $\exp(\kappa/2)=1+\tilde{\kappa}$ for all sufficiently large $N\in\mathcal{N}$.
	\end{proof}
		Now consider the sequence of updating mechanisms described in the proof of Proposition~\ref{prop_smallworld}, constructed on the states in $\widehat{\Omega}_{N}$ only: there are only branches corresponding to actions in $\widehat{\Omega}_{N}$, each of length $\lambda_{N}=\floor*{(M_{N}-1)/\vert\widehat{\Omega}_{N}\vert}$, and there exists no memory state $m$ with $d(m)=\omega'$ for $\omega'\notin\widehat{\Omega}_{N}$. Since $\vert\widehat{\Omega}_{N}\vert\leq N=\Theta(M_{N}^{h})$ with $h<1$, there exists a constant $c>0$ such that $\lambda_{N}\geq cM_{N}^{1-h}$ for all large enough $N\in\mathcal{N}$; in particular $\lambda_{N}\rightarrow\infty$. The stationary distribution under state $\omega\in\widehat{\Omega}_{N}$ satisfies
	\begin{equation}
		\resizebox{0.9\hsize}{!}{$
		\mu^{\omega}_{N\omega\lambda_{N}}\sum_{i=1}^{\lambda_{N}}\left[r^{\omega\omega}_{N}\right]^{-(\lambda_{N}-i)}+\mu^{\omega}_{N\omega\lambda_{N}}\left[r^{\omega\omega}_{N}\right]^{-\lambda_{N}}
		+\mu^{\omega}_{N\omega\lambda_{N}}\left[r^{\omega\omega}_{N}\right]^{-\lambda_{N}}\sum_{\omega'\in \widehat{\Omega}_{N}\setminus\lbrace\omega\rbrace}\sum_{i=1}^{\lambda_{N}}\left[r^{\omega\omega'}_{N}\right]^{(\lambda_{N}-i)}=1$}
	\end{equation}
	where $\mu^{\omega}_{Nm}$ is the stationary probability of memory state $m$ under state $\omega$ in the $N$-th decision problem. Given Lemma~\ref{lemmasmallworldinfoinfiniteas}(a), $r^{\omega\omega}_{N}>2$ for all $\omega$ and $N$, and therefore $\lim_{N\rightarrow\infty}\mu^{\omega}_{N\omega\lambda_{N}}\left[r^{\omega\omega}_{N}\right]^{-\lambda_{N}}=0$. On the other hand, for all sufficiently large $N\in\mathcal{N}$,
	\begin{align*}
			\left[r^{\omega\omega}_{N}\right]^{-\lambda_{N}}\sum_{\omega'\in \widehat{\Omega}_{N}\setminus\lbrace\omega\rbrace}\sum_{i=1}^{\lambda_{N}}\left[r^{\omega\omega'}_{N}\right]^{(\lambda_{N}-i)}
			=&\left[r^{\omega\omega}_{N}\right]^{-\lambda_{N}}\sum_{\omega'\in \widehat{\Omega}_{N}\setminus\lbrace\omega\rbrace}\frac{\left[r^{\omega\omega'}_{N}\right]^{\lambda_{N}}-1}{r^{\omega\omega'}_{N}-1}\\
			\leq&\sum_{\omega'\in \widehat{\Omega}_{N}\setminus\lbrace\omega\rbrace}\left[\frac{r^{\omega\omega'}_{N}}{r^{\omega\omega}_{N}}\right]^{\lambda_{N}}\\
			\leq&\sum_{\omega'\in \widehat{\Omega}_{N}\setminus\lbrace\omega\rbrace}(1+\tilde{\kappa})^{-\lambda_{N}}
			\leq N(1+\tilde{\kappa})^{-\lambda_{N}}
	\end{align*}
	where the first inequality uses $r^{\omega\omega'}_{N}-1>1$ (Lemma~\ref{lemmasmallworldinfoinfiniteas}(a)), and the second uses Lemma~\ref{lemmasmallworldinfoinfiniteas}(b). Note that these bounds do not depend on $\omega$. Taking logarithms,
	\begin{equation*}
		\log\left[N(1+\tilde{\kappa})^{-\lambda_{N}}\right]\leq \log N-cM_{N}^{1-h}\log(1+\tilde{\kappa})=h\log M_{N}+O(1)-cM_{N}^{1-h}\log(1+\tilde{\kappa})\rightarrow-\infty,
	\end{equation*}
	since $M_{N}^{1-h}$ grows polynomially ($h<1$) while $\log M_{N}$ grows only logarithmically. Hence $\lim_{N\rightarrow\infty}N(1+\tilde{\kappa})^{-\lambda_{N}}=0$, which implies $\lim_{N\rightarrow\infty}\left[r^{\omega\omega}_{N}\right]^{-\lambda_{N}}\sum_{\omega'\in \widehat{\Omega}_{N}\setminus\lbrace\omega\rbrace}\sum_{i=1}^{\lambda_{N}}\left[r^{\omega\omega'}_{N}\right]^{(\lambda_{N}-i)}=0$ uniformly over $\omega\in\widehat{\Omega}_{N}$, and
	\begin{equation}
		\lim_{N\rightarrow\infty}\sum_{i=1}^{\lambda_{N}}\mu^{\omega}_{N\omega i}=\lim_{N\rightarrow\infty}\mu^{\omega}_{N\omega\lambda_{N}}\sum_{i=1}^{\lambda_{N}}\left[r^{\omega\omega}_{N}\right]^{-(\lambda_{N}-i)}=1,
	\end{equation}
	again uniformly over $\omega\in\widehat{\Omega}_{N}$. In state $\omega\in\widehat{\Omega}_{N}$, the DM thus chooses action~$\omega$ with asymptotic probability converging to $1$. As $\liminf_{N\rightarrow\infty}\sum_{\omega\in\widehat{\Omega}_{N}}p_{N}^{\omega}\geq 1-\epsilon$ and $u_{N}(\omega,\omega)=1$, the asymptotic utility loss of the proposed sequence of updating mechanisms is bounded above by $\epsilon$ in the limit. Since $\epsilon>0$ was arbitrary, $\lim_{N\rightarrow\infty}L^{*}_{N}=0$.

%
	\paragraph{2) Part (ii), $h\geq1$: proof of Equation~\eqref{eq:limitignorancedisplay}.}
The proof uses the following lemma.
\begin{lemma}\label{lemma_bigworldnonconvergence}
	Fix a sequence of decision problems satisfying Assumptions~\ref{assumption_fullsupportinfinite}, \ref{assumption_noperfectsignalinfinite}, and \ref{assumption_identifiableinfinte}, where $N=\Theta(M_{N}^{h})$ with $h\geq 1$. First, $\liminf_{N\rightarrow\infty}L^*_N > 0$. Moreover, for any sequence of updating mechanisms, there does not exist a sequence of subsets $\check{\Omega}^{*}_{N}\subseteq\lbrace 1,\ldots,N\rbrace$ with $\liminf_{N\rightarrow\infty}\vert\check{\Omega}^{*}_{N}\vert/M_{N}>0$ such that $\lim_{N\rightarrow\infty}\max_{m}\mu^{\omega}_{Nm}/\mu^{\omega'}_{Nm}=\infty$ for all $\omega,\omega'\in\check{\Omega}^{*}_{N}$ with $\omega\neq\omega'$.
\end{lemma}
\begin{proof}[Proof of Lemma~\ref{lemma_bigworldnonconvergence}]
	First, note that if in contrast $\liminf_{N\rightarrow \infty}L_{N}^{*}=0$, then, passing to a subsequence of $\mathcal{N}$ along which $L^{*}_{N}\rightarrow 0$ and denoting $M_N^{\omega}=\lbrace m: d_N(m)=\omega\rbrace$, we must have for some $\lbrace m_{1N},\mathscr{T}_N,d_N\rbrace_{N\in\mathcal{N}}$,
	\begin{equation*}
		\lim_{N\rightarrow \infty}\sum_{m\in M_N^{\omega}}\mu_{Nm}^{\omega}=1
	\end{equation*}
	for almost all $\omega$ measured in prior probability, i.e., there must exist a sequence of subsets of states $\check{\Omega}^{p}_{N}\subseteq\lbrace1,\cdots,N\rbrace$ where $\lim_{N\rightarrow\infty}\sum_{\omega\in\check{\Omega}^{p}_{N}}p_{N}^{\omega}=1$ and
	\begin{equation*}
		\lim_{N\rightarrow \infty}\sum_{m\in M_N^{\omega}}\mu_{Nm}^{\omega}=1
	\end{equation*}
	for all $\omega\in\check{\Omega}^{p}_{N}$. Additionally, Assumption~\ref{assumption_fullsupportinfinite} (applied along a convergent subsequence) implies that there must exist a sequence of subsets of states $\check{\Omega}_{N}\subseteq\lbrace1,\cdots,N\rbrace$ where $\lim_{N\rightarrow\infty}\frac{\vert\check{\Omega}_{N}\vert}{N}=1$ and
	\begin{equation*}
		\lim_{N\rightarrow \infty}\sum_{m\in M_N^{\omega}}\mu_{Nm}^{\omega}=1,
	\end{equation*}
	for all $\omega\in\check{\Omega}_{N}$. That is, the DM chooses the optimal action in almost all states, measured in both prior probability and in fraction w.r.t.~$N$. It implies that for all $\omega$ in $\check{\Omega}_{N}$, there must exist a set of memory states $\hat{M}_N^{\omega}\subseteq M_N^{\omega}$ such that
	\begin{equation}
		\begin{split}
			\lim_{N\rightarrow\infty}\sum_{m\in\hat{M}_{N}^{\omega}}\mu^{\omega}_{Nm}&=1;\\
			\lim_{N\rightarrow\infty}\sum_{m\in\hat{M}_{N}^{\omega}}\mu^{\omega'}_{Nm}&=0\text{ for all $\omega'\in\check{\Omega}_{N}\setminus\lbrace\omega\rbrace$;}\\
			\lim_{N\rightarrow\infty}\max_{m\in\hat{M}_{N}^{\omega}}\frac{\mu_{Nm}^{\omega}}{\mu_{Nm}^{\omega'}}&=\infty\text{ for all $\omega'\in\check{\Omega}_{N}\setminus\lbrace\omega\rbrace$}.\\
		\end{split}
		\label{eq:bigworldfirstequation}
	\end{equation}
	In the following I prove that for Equation~\eqref{eq:bigworldfirstequation} to hold, $\frac{M_{N}}{N}$ has to go to $\infty$. The same argument below also proves the second statement of the Lemma.
	
	First consider an irreducible automaton $\mathscr{T}_{N}$. Fix a $\omega'\in\check{\Omega}_{N}\setminus\lbrace\omega\rbrace$, without loss of generality rearrange the memory states such that $\frac{\mu^{\omega}_{Nm}}{\mu^{\omega'}_{Nm}}$ is weakly decreasing in $m$. As argued in the proof of Lemma~\ref{lemmma:appendixiff}, Lemma 2 of \cite{hellman1970learning} holds in the current setting, so that for all $m<M_{N}$,
	\begin{equation}
		\begin{split}
			\frac{\mu^{\omega}_{N(m+1)}}{\mu^{\omega'}_{N(m+1)}}\geq (\overline{l}_{N}^{\omega\omega'}\overline{l}_{N}^{\omega'\omega})^{-1}\frac{\mu^{\omega}_{Nm}}{\mu^{\omega'}_{Nm}}
			\geq\varsigma^{2}\frac{\mu^{\omega}_{Nm}}{\mu^{\omega'}_{Nm}}.
		\end{split}
		\label{eq:bigworldsecondequation}
	\end{equation}
	If $\max_{m}\frac{\mu^{\omega}_{Nm}}{\mu^{\omega'}_{Nm}}=\frac{\mu^{\omega}_{N1}}{\mu^{\omega'}_{N1}}>K$, Equation~\eqref{eq:bigworldsecondequation} implies that
	\begin{equation*}
		\frac{\mu^{\omega}_{Nm'}}{\mu^{\omega'}_{Nm'}}\geq\varsigma^{2(m'-1)}\frac{\mu^{\omega}_{N1}}{\mu^{\omega'}_{N1}}>\varsigma^{2(m'-1)}K.
	\end{equation*}
	As Equation~\eqref{eq:bigworldfirstequation} implies that there must exist some $m'$ with $\frac{\mu^{\omega}_{Nm'}}{\mu^{\omega'}_{Nm'}}\leq 1$, there must exist at least $\frac{\log K}{-2\log \varsigma}+1$ memory states with $\frac{\mu^{\omega}_{Nm}}{\mu^{\omega'}_{Nm}}\geq 1$. Repeating the same analysis for other $\omega''\in \check{\Omega}_{N}\setminus\lbrace\omega\rbrace$ implies that if $\max_m\frac{\mu^{\omega}_{Nm}}{\mu^{\omega''}_{Nm}}>K$ for all $\omega''\in \check{\Omega}_{N}\setminus\lbrace\omega'\rbrace$, there must exist at least $\frac{\log K}{-2\log \varsigma}+1$ memory states with a likelihood ratio $\min_{\omega'\in\check{\Omega}_{N}}\frac{\mu_{Nm}^{\omega}}{\mu_{Nm}^{\omega'}}\geq 1$ (i.e., $\frac{\mu_{Nm}^{\omega}}{\mu_{Nm}^{\omega'}}\geq 1$ for all $\omega'\in\check{\Omega}_{N}\setminus \lbrace\omega\rbrace$).

	With the same argument, if $\max_{m}\frac{\mu^{\omega''}_{Nm}}{\mu^{\omega'}_{Nm}}>K$ for all $\omega''\in\check{\Omega}_{N}$ and all $\omega'\in\check{\Omega}_{N}\setminus\lbrace\omega''\rbrace$,
	there must exist at least $\frac{\log K}{-2\log \varsigma}+1$ memory states with a likelihood ratio $\min_{\omega'\in\check{\Omega}_{N}}\frac{\mu_{Nm}^{\omega''}}{\mu_{Nm}^{\omega'}}\geq 1$ for all $\omega''\in\check{\Omega}_{N}$. Therefore, $\frac{M_{N}}{\vert\check{\Omega}_{N}\vert}\geq \frac{\log K}{-2\log \varsigma}+1$. This implies that $\frac{M_{N}}{\vert\check{\Omega}_{N}\vert}$ goes to $\infty$ as $K$ goes to $\infty$, which contradicts the fact that $\lim_{N\rightarrow\infty}\frac{M_{N}}{\vert \check{\Omega}_{N}\vert}=\lim_{N\rightarrow\infty}\frac{M_{N}}{N}<\infty$ in big worlds.

	For reducible automata, the same conclusion follows. Denote the recurrent communicating classes as $\mathscr{R}_{N1},\cdots,\mathscr{R}_{Nr}$ and the set of transient memory states as $\mathscr{R}_{N0}$. By Lemma~\ref{lemmma:appendixiff}, the global ratio decomposes via absorption probabilities: for $m\in\mathscr{R}_{Nj}$,
	\begin{equation*}
		\frac{\mu^{\omega}_{Nm}}{\mu^{\omega'}_{Nm}}\leq \varsigma^{-2}\cdot\max_{m\in\mathscr{R}_{N0}}\underbrace{\frac{\mu^{0,\omega}_{Nm}}{\mu^{0,\omega'}_{Nm}}}_{\text{within-transient ratio}}\cdot\underbrace{\frac{\mu^{j,\omega}_{Nm}}{\mu^{j,\omega'}_{Nm}}}_{\text{within-class-}\mathscr{R}_{Nj}\text{ ratio}}.
	\end{equation*}
	Hence if $\max_{m}\mu^{\omega}_{Nm}/\mu^{\omega'}_{Nm}>K$, then $\varsigma^{-2}\cdot a\cdot b>K$ where $a$ is the within-transient ratio and $b$ is the within-recurrent-class ratio. Equivalently $ab>\varsigma^{2}K$, so $\max(a,b)>\varsigma\sqrt{K}$. The irreducible-case counting argument applied to whichever class realizes the maximum yields a chain of $\frac{\log(\varsigma\sqrt{K})}{-2\log\varsigma}+1$ memory states with elevated ratios in that class. Repeating for all $\omega\in\check{\Omega}_{N}$ as in the irreducible case yields $\frac{M_{N}}{|\check{\Omega}_{N}|}\geq\frac{\log(\varsigma\sqrt{K})}{-2\log\varsigma}+1\rightarrow\infty$ as $K\rightarrow\infty$, contradicting the big-world condition.
\end{proof}

\begin{proof}[Proof of Equation~\eqref{eq:limitignorancedisplay}]
	Fix a sequence of updating mechanisms $\lbrace m_{1N},\mathscr{T}_{N},d_{N}\rbrace_{N\in\mathcal{N}}$ and suppose to the contrary that Equation~\eqref{eq:limitignorancedisplay} fails for some $\xi'>0$. Writing
	\begin{equation*}
		\tilde{\Omega}_{N}:=\left\lbrace \omega:\; \lim_{T\rightarrow\infty}\frac{1}{T}\sum_{t=1}^{T}\Pr\nolimits^{\omega'}(a_{t}=\omega)\geq\xi'\ \text{ for some state }\omega'\right\rbrace,
	\end{equation*}
	this means there exist $\theta>0$ and an infinite subset $\mathcal{N}'\subseteq\mathcal{N}$ such that $\vert\tilde{\Omega}_{N}\vert/M_{N}\geq\theta$ for all $N\in\mathcal{N}'$; the remainder of the proof works along $\mathcal{N}'$. For each $\omega\in\tilde{\Omega}_{N}$, fix a state $\omega'(\omega)$ attaining the defining inequality of $\tilde{\Omega}_{N}$. Denote  $M^{\omega}_{N}=\lbrace m: d_{N}(m)=\omega\rbrace$, we have, for all $\omega\in\tilde{\Omega}_{N}$ and all $N\in\mathcal{N}'$,
	\begin{equation*}
		\sum_{m\in M^{\omega}_{N}}\mu^{\omega'(\omega)}_{Nm}\geq\xi'.
	\end{equation*}
	Since $\sum_{\omega}\vert M^{\omega}_{N}\vert\leq M_{N}$, for any $C>0$ the number of $\omega$ with $\vert M^{\omega}_{N}\vert>C$ is at most $M_{N}/C$, i.e., there are at most $M_{N}/C$ states of the world with more than $C$ memory states allocated to their associated actions. Pick $C=2/\theta$. Then the subset $\tilde{\Omega'}_{N}:=\lbrace\omega\in\tilde{\Omega}_{N}:\vert M^{\omega}_{N}\vert\leq C\rbrace$ satisfies $\vert\tilde{\Omega'}_{N}\vert\geq\vert\tilde{\Omega}_{N}\vert-M_{N}/C\geq\frac{\theta}{2}M_{N}$. By the pigeonhole principle, for each $\omega\in\tilde{\Omega'}_{N}$ there exists a memory state $m(\omega)\in M^{\omega}_{N}$ with $\mu^{\omega'(\omega)}_{Nm(\omega)}\geq\xi'/C=:\xi''$; note that the $m(\omega)$ are distinct across $\omega$, as the sets $M^{\omega}_{N}$ are disjoint.

	Now denote the set of the states $\omega'(\omega)$ in which the DM picks some action $\omega\in\tilde{\Omega'}_{N}$ with long-run frequency at least $\xi'$ as $\tilde{\Omega''}_{N}$. First, $\vert\tilde{\Omega''}_{N}\vert\geq\xi''\vert\tilde{\Omega'}_{N}\vert\geq\frac{\xi''\theta}{2}M_{N}$: each $\omega''\in\tilde{\Omega''}_{N}$ can serve as $\omega'(\omega)$ for at most $1/\xi''$ many $\omega\in\tilde{\Omega'}_{N}$, since the associated memory states $m(\omega)$ are distinct and each carries $\mu^{\omega''}_{Nm(\omega)}\geq\xi''$, while $\sum_{m}\mu^{\omega''}_{Nm}=1$. Second, for all $\omega''\in\tilde{\Omega''}_{N}$ and any threshold $\eta>0$, the number of memory states with $\mu^{\omega''}_{Nm}\geq\eta$ is at most $1/\eta$ (again since $\sum_{m}\mu^{\omega''}_{Nm}=1$). To briefly summarize, we have a set of tuples $(\omega,\omega'(\omega),m(\omega))$, where the $m(\omega)$ are distinct across $\omega\in\tilde{\Omega'}_{N}$; both $\tilde{\Omega'}_N$ and $\tilde{\Omega''}_N$ have cardinality at least a positive fraction of $M_{N}$ along $\mathcal{N}'$ (as $\vert\tilde{\Omega'}_{N}\vert\geq \frac{\theta}{2}M_{N}$ and $\vert\tilde{\Omega''}_{N}\vert\geq\frac{\xi''\theta}{2}M_{N}$), and for each $\omega''\in \tilde{\Omega''}_N$, only a bounded number of memory states carry probability at least $\eta$.

	I now prove that for almost all $\omega\in\tilde{\Omega'}_{N}$ (measured in $/M_{N}$ fraction along $\mathcal{N}'$), the memory state $m(\omega)$ satisfies $\mu^{\omega'(\omega)}_{Nm(\omega)}\geq\xi''$ and $\lim_{N\rightarrow\infty}\mu^{\omega''}_{Nm(\omega)}=0$ for almost all $\omega''\in\tilde{\Omega''}_N\setminus\{\omega'(\omega)\}$ (in $/M_{N}$ fraction). By the previous step, each $\omega''\in\tilde{\Omega''}_N$ places non-vanishing limiting probability on only finitely many memory states, so the total number of pairs $(\omega'',m)$ with $\omega''\in\tilde{\Omega''}_N$ and $\lim_{N\to\infty}\mu^{\omega''}_{Nm}>0$ is at most a constant multiple of $\vert\tilde{\Omega''}_N\vert=\Theta(M_{N})$. Next, for $\omega\in\tilde{\Omega'}_N$, let $X(\omega)$ denote the number of states $\omega''\in\tilde{\Omega''}_N$ whose long-run probability at $\omega$'s own memory state $m(\omega)$ does not vanish. A pair $(\omega'',m)$ contributes to $X(\omega)$ only if $m=m(\omega)$; since the $m(\omega)$ are distinct across $\omega$, each pair is counted in at most one $X(\omega)$, so $\sum_{\omega\in\tilde{\Omega'}_N}X(\omega)=O(M_{N})$. Finally, by Markov's inequality, for any $\eta>0$ the number of $\omega\in\tilde{\Omega'}_N$ with $X(\omega)>\eta M_{N}$ is at most $\sum_{\omega}X(\omega)/(\eta M_{N})=O(1/\eta)$ --- a vanishing fraction of $\vert\tilde{\Omega'}_N\vert=\Theta(M_{N})$. For all remaining $\omega$ --- almost all of $\tilde{\Omega'}_N$ --- the memory state $m(\omega)$ carries non-vanishing probability for at most $\eta M_{N}$ of the states in $\tilde{\Omega''}_N$; as $\eta>0$ can be taken arbitrarily small, the claim follows. This implies that there exists some sequence of subsets of states $\tilde{\tilde{\Omega}}_{N}$ with $\liminf_{N\rightarrow\infty}\frac{\vert\tilde{\tilde{\Omega}}_{N}\vert}{M_{N}}>0$ (along $\mathcal{N}'$) such that
		\begin{equation*}
			\lim_{N\rightarrow\infty}\max_{m}\frac{\mu^{\omega'}_{Nm}}{\mu^{\omega''}_{Nm}}=\infty
		\end{equation*}
		for all $\omega'\in\tilde{\tilde{\Omega}}_{N}$ and all $\omega''\in\tilde{\tilde{\Omega}}_{N}\setminus\lbrace\omega'\rbrace$ (as the numerator is at least $\xi''$ at $m(\omega)$ but the denominator vanishes as $N\rightarrow\infty$). The sequence $\tilde{\tilde{\Omega}}_{N}$ thus has a non-vanishing fraction w.r.t.~$M_{N}$ and exhibits unbounded likelihood ratios for all pairs within --- precisely ruled out by Lemma~\ref{lemma_bigworldnonconvergence}. The result thus follows.
\end{proof}

\paragraph{3) Equation~\eqref{eq:limitignorancedisplay} implies the payoff consequence.} I establish an unconditional upper bound $\limsup_{N\rightarrow\infty}L^{*}_{N}\leq 1-\overline{p}_{0}$; a lower-bound accounting valid for arbitrary mechanisms; the equality $\lim_{N\rightarrow\infty}L^{*}_{N}=1-\overline{p}_{0}$ whenever $L^{*}_{N}$ converges; and unconditional convergence to $1$ under Assumption~\ref{assumption_fullsupportinfiniteplus}.

\emph{Upper bound: $\limsup_{N\rightarrow\infty}L^*_N\leq 1-\overline{p}_0$.} Fix any $\eta>0$. By the definition of $\overline{p}_{0}$, there exists a sequence of subsets $\Omega^{f}_{N}\subseteq\lbrace 1,\ldots,N\rbrace$ with $\vert\Omega^{f}_{N}\vert/M_{N}\rightarrow 0$ and $\sum_{\omega\in\Omega^{f}_{N}}p^{\omega}_{N}$ converging to a limit at least $\overline{p}_{0}-\eta$. Apply the Part-1 construction with $\Omega^{f}_{N}\cap\widehat{\Omega}_{N}$ (where $\widehat{\Omega}_{N}$ is given by Assumption~\ref{assumption_identifiableinfinte} applied with $\varepsilon=\eta$, so that pairwise identifiability holds within the branch set) in place of $\widehat{\Omega}_{N}$: since $\lambda_{N}=\floor*{(M_{N}-1)/\vert\Omega^{f}_{N}\cap\widehat{\Omega}_{N}\vert}\rightarrow\infty$, the same argument yields long-run utility converging to $1$ in every state $\omega\in\Omega^{f}_{N}\cap\widehat{\Omega}_{N}$ (and the DM allocates no memory states to other $\omega$, contributing $0$ utility there). Therefore
\begin{equation*}
	\limsup_{N\rightarrow\infty}L^{*}_{N}\leq 1-\liminf_{N\rightarrow\infty}\sum_{\omega\in\Omega^{f}_{N}\cap\widehat{\Omega}_{N}}p^{\omega}_{N}\leq 1-(\overline{p}_{0}-2\eta).
\end{equation*}
Letting $\eta\rightarrow 0$ gives $\limsup_{N\rightarrow\infty}L^{*}_{N}\leq 1-\overline{p}_{0}$.


\emph{Lower-bound.} Fix any sequence of updating mechanisms and any $\xi'\in(0,1)$, and let $\Omega^{NI}_{N}(\xi')=\lbrace\omega:\lim_{T\rightarrow\infty}\frac{1}{T}\sum_{t=1}^{T}\Pr\nolimits^{\omega'}(a_{t}=\omega)\geq\xi'\text{ for some }\omega'\rbrace$ denote the set of actions taken with long-run frequency at least $\xi'$ in some state; by Equation~\eqref{eq:limitignorancedisplay}, $\lim_{N\rightarrow\infty}\vert\Omega^{NI}_{N}(\xi')\vert/M_{N}=0$. In every state $\omega\notin\Omega^{NI}_{N}(\xi')$, the DM picks action~$\omega$ with long-run frequency below $\xi'$ and thus earns long-run utility below $\xi'$ (recall $u_{N}(\omega,\omega)=1$ and $u_{N}(\omega',\omega)=0$ for $\omega'\neq\omega$), so the loss $L_{N}$ of the $N$-th mechanism satisfies
\begin{equation}\label{eq:lowerboundaccounting}
1-L_{N}\,\leq\,\sum_{\omega\in\Omega^{NI}_{N}(\xi')}p^{\omega}_{N}+\xi'.
\end{equation}

\emph{The equality for convergent $L^{*}_{N}$.} Suppose $L^{*}_{N}$ converges to a limit $L_{\infty}$; I show $\overline{p}_{0}\geq 1-L_{\infty}$, which combined with the upper bound yields $L_{\infty}=1-\overline{p}_{0}$. For each $N$, apply Equation~\eqref{eq:lowerboundaccounting} to an $\epsilon_{N}$-optimal mechanism, where $\epsilon_{N}\rightarrow 0$: since $L_{N}\leq L^{*}_{N}+\epsilon_{N}$,
\begin{equation*}
\liminf_{N\rightarrow\infty}\sum_{\omega\in\Omega^{NI}_{N}(\xi')}p^{\omega}_{N}\,\geq\,1-L_{\infty}-\xi'.
\end{equation*}
Conversely, I also show that every subsequential limit of $\sum_{\omega\in\Omega^{NI}_{N}(\xi')}p^{\omega}_{N}$ is at most $1-L_{\infty}$; together with the equation above, $\sum_{\omega\in\Omega^{NI}_{N}(\xi')}p^{\omega}_{N}$ therefore eventually lies in $[1-L_{\infty}-2\xi',\,1-L_{\infty}+\xi']$ (the extra $\xi'$ converts the liminf/limsup bounds into bounds holding for all large $N$). Indeed, if $\sum_{\omega\in\Omega^{NI}_{N}(\xi')}p^{\omega}_{N}$ converges to $c$ along an infinite subsequence $\mathcal{N}'$, then applying the upper-bound construction with $\Omega^{NI}_{N}(\xi')$ in place of $\Omega^{f}_{N}$ along $\mathcal{N}'$ gives $L^{*}_{N}\leq 1-c+2\eta+o(1)$ along $\mathcal{N}'$ for every $\eta>0$, and $L^{*}_{N}\rightarrow L_{\infty}$ forces $c\leq 1-L_{\infty}$ (letting $\eta\rightarrow 0$). A standard diagonalization over $\xi'_{k}\downarrow 0$ then yields a single sequence of subsets with vanishing fraction w.r.t.~$M_{N}$ and prior mass converging to $1-L_{\infty}$. It is therefore admissible in Equation~\eqref{eq:p0bar}, whence $\overline{p}_{0}\geq 1-L_{\infty}$.

\emph{Convergence under Assumption~\ref{assumption_fullsupportinfiniteplus}.} Fix any sequence of updating mechanisms and $\xi'\in(0,1)$. Since $M_{N}\leq O(N)$ when $h\geq 1$, $\vert\Omega^{NI}_{N}(\xi')\vert/N\rightarrow 0$ as well; by Assumption~\ref{assumption_fullsupportinfiniteplus}, every subsequential limit of $\sum_{\omega\in\Omega^{NI}_{N}(\xi')}p^{\omega}_{N}$ is $0$, and hence $\lim_{N\rightarrow\infty}\sum_{\omega\in\Omega^{NI}_{N}(\xi')}p^{\omega}_{N}=0$. Equation~\eqref{eq:lowerboundaccounting} then gives $\liminf_{N\rightarrow\infty}L_{N}\geq 1-\xi'$ for every sequence of mechanisms; applying this to $\epsilon_{N}$-optimal mechanisms with $\epsilon_{N}\rightarrow 0$ and letting $\xi'\rightarrow 0$ yields $\liminf_{N\rightarrow\infty}L^{*}_{N}\geq 1$. As $L^{*}_{N}\leq 1$, $L^{*}_{N}$ converges and $\lim_{N\rightarrow\infty}L^{*}_{N}=1$.

\paragraph{4) The concluding statement.} If $h<1$, part (i) gives $L^{*}_{N}\rightarrow 0$; hence for any $N_{0}\in\mathcal{N}$ with $L^{*}_{N_{0}}>0$, we have $L^{*}_{N}<L^{*}_{N_{0}}$ for all large enough $N\in\mathcal{N}$. If $h\geq 1$ and Assumption~\ref{assumption_fullsupportinfiniteplus} holds, part (ii) gives $L^{*}_{N}\rightarrow 1$; hence for any $N_{0}\in\mathcal{N}$ with $L^{*}_{N_{0}}<1$, we have $L^{*}_{N}>L^{*}_{N_{0}}$ for all large enough $N\in\mathcal{N}$.

\end{proof}

\subsection{Proof of Corollary~\ref{prop_bigworldignorance2} and Proposition~\ref{coro_bigworldignorance2example}}

\begin{proof}[Proof of Corollary~\ref{prop_bigworldignorance2} (ii)]
	First, an ignorant updating mechanism induces utility loss weakly greater than $\min_{\omega\in\Omega}\min_{a\neq\omega}[p^{\omega}(u(\omega,\omega)-u(a,\omega))]>0$. On the other hand, as shown in Proposition~\ref{prop_smallworld}, $L_{M}^{*}$ converges to $0$ as $M\rightarrow\infty$. The monotonicity of $L^*_M$ implies there exists some big enough $\bar{M}$ such that for $M\geq\bar{M}$, $L_{M}^{*}\leq L_{\bar{M}}^{*}< \min_{\omega\in\Omega}\min_{a\neq\omega}[p^{\omega}(u(\omega,\omega)-u(a,\omega))]$. The result follows. 
\end{proof}

\begin{proof}[Proof of Corollary~\ref{prop_bigworldignorance2} (i) and Proposition~\ref{coro_bigworldignorance2example}]
	First note that Proposition~\ref{coro_bigworldignorance2example} implies Corollary~\ref{prop_bigworldignorance2} (i). I first prove Proposition~\ref{coro_bigworldignorance2example}(i). Consider an updating mechanism $(m_1,\mathscr{T},d)$ such that $\mu_{m}^{\tilde{\omega}}>0$ for some $\tilde{\omega}$ and $m$ where $d(m)=\omega$. Now I prove that there exists a different updating mechanism that yields higher asymptotic utility if $p^{\omega}$ is small enough.

	Now using Equation~\eqref{eq:profignorance}, denote $u^{\omega}=u(\omega,\omega)$, we have
	\begin{align}
		\min_{m}\frac{\mu^{\omega'}_{m}}{\mu^{\omega}_{m}}&\geq \varsigma^{2(M-1)}\nonumber\\
		\frac{u^{\omega'}p^{\omega'}\mu^{\omega'}_{m}}{u^{\omega}p^{\omega}\mu^{\omega}_{m}}&\geq\varsigma^{2(M-1)}\frac{u^{\omega'}p^{\omega'}}{u^{\omega}p^{\omega}} \text{ for all $m$.}\nonumber
	\end{align}
    
	Now, denote $\overline{u}=\max_{\omega\in\Omega}\min_{a\neq\omega}[u(\omega,\omega)-u(a,\omega)]$ and $\underline{u}=\min_{\omega\in\Omega}\min_{a\neq\omega}[u(\omega,\omega)-u(a,\omega)]$, as $\min\max_{\omega'}\frac{u^{\omega'}p^{\omega'}}{u^{\omega}p^{\omega}}=\frac{\underline{u}}{\overline{u}}\frac{\frac{1-p^{\omega}}{N-1}}{p^{\omega}}$, there exists some $\omega'$ such that
	\begin{equation}\label{eq:eqprofignroance2}
		\frac{u^{\omega'}p^{\omega'}\mu^{\omega'}_{m}}{u^{\omega}p^{\omega}\mu^{\omega}_{m}}\geq\varsigma^{2(M-1)}\min\max_{\omega''}\frac{u^{\omega''}p^{\omega''}}{u^{\omega}p^{\omega}}\geq\varsigma^{2(M-1)}\frac{\underline{u}}{\overline{u}}\frac{\frac{1-p^{\omega}}{N-1}}{p^{\omega}}\text{ for all $m\in M$.}
	\end{equation}
	When $p^\omega<\frac{1}{2}\varsigma^{2(M-1)}\frac{\underline{u}}{\overline{u}}\frac{1-p^{\omega}}{N-1}$, or equivalently, $\frac{p^{\omega}}{1-p^{\omega}}<\frac{1}{2}\varsigma^{2(M-1)}\frac{\underline{u}}{\overline{u}}\frac{1}{N-1}$, Equation~\eqref{eq:eqprofignroance2} implies that 	$u^{\omega'}p^{\omega'}\mu^{\omega'}_{m}>2u^{\omega}p^{\omega}\mu^{\omega}_{m}$, and the DM is better off choosing $\omega'$ than choosing $\omega$ for all $m\in M$. As $u^{\omega'}p^{\omega'}\mu^{\omega'}_{m}>u^{\omega}p^{\omega}\mu^{\omega}_{m}$, it violates that $d(m)$ is optimal at memory state $m$. This proves Proposition~\ref{coro_bigworldignorance2example}(i)(a). For Proposition~\ref{coro_bigworldignorance2example}(i)(b), the improvement in utility by choosing $\omega'$ rather than choosing $\omega$ in $m$ is equal to
	\begin{equation*}
		u^{\omega'}p^{\omega'}\mu^{\omega'}_{m}-u^{\omega}p^{\omega}\mu^{\omega}_{m}>u^{\omega}p^{\omega}\mu^{\omega}_{m}.
	\end{equation*}
	The updating mechanism $(m_1,\mathscr{T},d)$ is therefore not $\epsilon$-optimal for $\epsilon\leq u^{\omega}p^{\omega}\mu^{\omega}_{m}$. Equivalently, if $(m_1,\mathscr{T},d)$ is $\epsilon$-optimal, then $\mu^{\omega}_{m}<\epsilon/(u^{\omega}p^{\omega})$ for every memory state $m$ with $d(m)=\omega$. Summing over all such $m$ (of which there are at most $M$), the long-run probability that the DM picks action~$\omega$ under state~$\omega$ satisfies
	\begin{equation*}
		\sum_{m:\,d(m)=\omega}\mu^{\omega}_{m}<\frac{M\epsilon}{u^{\omega}p^{\omega}}.
	\end{equation*}
	Applying the ratio bound $\mu^{\omega''}_{m}\leq\varsigma^{-2(M-1)}\mu^{\omega}_{m}$ (Lemma~\ref{lemmma:appendixiff}), the long-run probability of picking action~$\omega$ under any state~$\omega''$ is bounded by $M\epsilon\varsigma^{-2(M-1)}/(u^{\omega}p^{\omega})=O(\epsilon)$, which vanishes as $\epsilon\rightarrow 0$. This proves Proposition~\ref{coro_bigworldignorance2example}(i)(b). Similar argument proves Proposition~\ref{coro_bigworldignorance2example}(ii)(a) and (ii)(b).

	I now prove Proposition~\ref{coro_bigworldignorance2example}(iii)(a) and (iii)(b). Without loss of generality, assume $u^{\omega'}p^{\omega'}\geq u^{\omega}p^{\omega}$. Again, consider an updating mechanism $\mathscr{T}$ such that $\mu_{m}^{\tilde{\omega}}>0$ for some $\tilde{\omega}$ and $m$ where $d(m)=\omega$. Similar to the proof of the first part of Proposition~\ref{coro_bigworldignorance2example},
	\begin{equation}
		\begin{split}
			\min_{m}\frac{\mu^{\omega'}_{m}}{\mu^{\omega}_{m}}&\geq (\overline{l}^{\omega\omega'}\overline{l}^{\omega'\omega})^{-(M-1)}\max_{m}\frac{\mu^{\omega'}_{m}}{\mu^{\omega}_{m}}\\
			\min_{m}\frac{\mu^{\omega'}_{m}}{\mu^{\omega}_{m}}&\geq \overline{F}^{-(M-1)}\max_{m}\frac{\mu^{\omega'}_{m}}{\mu^{\omega}_{m}}\\
			\min_{m}\frac{\mu^{\omega'}_{m}}{\mu^{\omega}_{m}}&\geq \overline{F}^{-(M-1)}\\
			\frac{u^{\omega'}p^{\omega'}\mu^{\omega'}_{m}}{u^{\omega}p^{\omega}\mu^{\omega}_{m}}&\geq\overline{F}^{-(M-1)}\frac{u^{\omega'}p^{\omega'}}{u^{\omega}p^{\omega}} \text{ for all $m$.}
		\end{split}
		\label{eq:eqprofignroance3}
	\end{equation}
	As $u^{\omega'}p^{\omega'}\geq u^{\omega}p^{\omega}$, for $\overline{F}<\frac{1}{2}\left[ \frac{u^{\omega}p^{\omega}}{u^{\omega'}p^{\omega'}}\right]^{\frac{1}{M-1}}$, Equation~\eqref{eq:eqprofignroance3} implies that $u^{\omega'}p^{\omega'}\mu^{\omega'}_{m}>2u^{\omega}p^{\omega}\mu^{\omega}_{m}$. With similar argument in the proof of Proposition~\ref{coro_bigworldignorance2example}(i)(a) and (i)(b), the result follows.
	\end{proof}

\subsection{Proof of Corollary~\ref{coro_ignorancesmallworldinfinteN}}

\begin{proof}
	Fix $N\in\mathcal{N}$ and let $I_{N}\subseteq\lbrace 1,\cdots,N\rbrace$ denote the set of states $\xi$-ignored by the $N$-th mechanism. In each state $\omega\in I_{N}$, the long-run frequency of the correct action~$\omega$ is below $\xi$ (take $\omega'=\omega$ in Definition~\ref{def:limitignorance}); since $u_{N}(\omega,\omega)=1$ and $u_{N}(\omega',\omega)=0$ for $\omega'\neq\omega$, the utility loss of the $N$-th mechanism satisfies $L_{N}\geq(1-\xi)\sum_{\omega\in I_{N}}p^{\omega}_{N}$. By $\epsilon$-optimality at the limit and Proposition~\ref{prop_smallworldinfiniteN}(i), $\limsup_{N\rightarrow\infty}L_{N}\leq\lim_{N\rightarrow\infty}L^{*}_{N}+\epsilon=\epsilon$. Combining, $\limsup_{N\rightarrow\infty}\sum_{\omega\in I_{N}}p^{\omega}_{N}\leq\frac{\epsilon}{1-\xi}$.
\end{proof}

	\section{Discounted future utility}\label{section:wilson}
		In this section, I consider the alternative formulation where future utility is discounted. More specifically, consider the following updating mechanism
		\begin{equation*}
			(m^{*D}_1(\rho),\mathscr{T}^{*D}(\rho),d^{*D}(\rho))=\argmax_{m_1,\mathscr{T},d} E\left[(1-\rho)\sum_{t=1}^{\infty}\rho^t u(a_t,\omega)\right],
		\end{equation*}
		for some discount factor $\rho<1$. We have the following result.
		\begin{proposition}\label{prop_discountfactor}
			For all $\epsilon>0$, there exists some $\bar{\rho}<1$ such that $L(m^{*D}_1(\rho),\mathscr{T}^{*D}(\rho),d^{*D}(\rho))$ $<L^*_M+\epsilon$ for all $\rho>\bar{\rho}$.
		\end{proposition}

	\begin{proof}
		Suppose to the contrary that $\lim_{\rho\rightarrow 1}L(m^{*D}_1(\rho),\mathscr{T}^{*D}(\rho),d^{*D}(\rho))>L^*_M+\epsilon$ for some $\epsilon>0$. It implies that there exists some updating mechanism $(m_1,\mathscr{T},d)$ that yields strictly higher asymptotic utility than $(m^{*D}_1(\rho),\mathscr{T}^{*D}(\rho),d^{*D}(\rho))$ for all $\rho$ close to 1. By a standard Abelian-type result \citep{hardy1949divergent} --- for any bounded sequence $\{u_{t}\}$ with a well-defined long-run average $\lim_{T\rightarrow\infty}\frac{1}{T}\sum_{t=1}^{T}u_{t}$, the discounted average $(1-\rho)\sum_{t=1}^{\infty}\rho^{t}u_{t}$ converges to the same value as $\rho\rightarrow 1$ --- $(m_1,\mathscr{T},d)$ also yields strictly higher discounted sum of utility than $(m^{*D}_1(\rho),\mathscr{T}^{*D}(\rho),d^{*D}(\rho))$ for all $\rho$ close to $1$. This contradicts that $(m^{*D}_1(\rho),\mathscr{T}^{*D}(\rho),d^{*D}(\rho))$ is the discounted-utility maximizer, proving the result.
	\end{proof}

		In other words, the updating mechanism that maximizes the discounted sum of utility, $(m^{*D}_1(\rho),\mathscr{T}^{*D}(\rho),d^{*D}(\rho))$, is $\epsilon$-optimal for any $\epsilon>0$ given $\rho$ close enough to $1$. This implies that the main results extend to the discounted case as $\rho\rightarrow 1$. To see why, consider the results on ignorance (Corollary~\ref{prop_bigworldignorance2} and Proposition~\ref{prop_smallworldinfiniteN}(ii)): $\epsilon$-optimal mechanisms with sufficiently small $\epsilon$ must ignore some states in some big-world environments (Corollary~\ref{prop_bigworldignorance2}) and ignore almost all states in any big-world environment at the limit (Proposition~\ref{prop_smallworldinfiniteN}(ii)). Since the discount-optimal mechanism is $\epsilon$-optimal for small enough $\epsilon$ when $\rho$ is close to $1$, it must also ignore some states. Conversely, consider the results on non-ignorance (Corollary~\ref{prop_bigworldignorance2}(ii) and Corollary~\ref{coro_ignorancesmallworldinfinteN}): in the small world, every $\epsilon$-optimal mechanism with sufficiently small $\epsilon$ ignores no state (fixed $N$), and $\xi$-ignores only states of vanishing total prior probability (at the limit, for all $\xi\in(0,1)$). Since the discount-optimal mechanism falls in this class for $\rho$ close to $1$, the same conclusions apply to it.

\end{document}


\onehalfspacing
	\begin{center}
		{\LARGE \textbf{{\LARGE \textbf{Online Appendix of ``Learning in a Small/Big World"}}}}\\
		\vspace{5mm}
		\textit{Benson Tsz Kin Leung\footnote{Hong Kong Baptist University. Email: \href{mailto:btkleung@hkbu.edu.hk}{\texttt{btkleung@hkbu.edu.hk}}, \href{mailto:ltkbenson@gmail.com}{\texttt{ltkbenson@gmail.com}}. }}\\
		\textit{\today}\\
	\end{center}

\onehalfspacing

\renewcommand{\thefootnote}{\arabic{footnote}}
\setcounter{footnote}{0}

\renewcommand\thesection{\Alph{section}}
\setcounter{section}{2}

\renewcommand\theequation{\thesection.\arabic{equation}}
\numberwithin{equation}{section}
\setcounter{equation}{0}
\renewcommand\theproposition{\thesection.\arabic{proposition}}
\numberwithin{proposition}{section}
\setcounter{equation}{0}
\renewcommand\thelemma{\thesection.\arabic{lemma}}
\numberwithin{lemma}{section}
\renewcommand\thecorollary{\thesection.\arabic{corollary}}
\numberwithin{corollary}{section}

\renewcommand\thefigure{\thesection.\arabic{figure}}
\setcounter{section}{0}

	\newtheorem{innercustomgeneric}{\customgenericname}
\providecommand{\customgenericname}{}
\newcommand{\newcustomtheorem}[2]{%
	\newenvironment{#1}[1]
	{%
		\renewcommand\customgenericname{#2}%
		\renewcommand\theinnercustomgeneric{##1}%
		\innercustomgeneric
	}
	{\endinnercustomgeneric}
}
\newcustomtheorem{assumption}{Assumption}

\begin{center}
	\Large \textbf{For online publication only}
\end{center}

\section{Switching between different automatons}\label{section:switch}
This section illustrates how the setup in this paper encompasses learning mechanisms that involve switching between automatons with size smaller than $M$. 

I first argue that assuming that the DM could switch between multiple learning mechanisms with $M$ memory states implicitly implies that he has a larger memory capacity than $M$. Suppose that a DM starts with $(\mathscr{T},d)$ at memory state $m_{1}$ and switches to $(\mathscr{T}',d')$ and $(\mathscr{T}'',d'')$ once he transit to memory state $m_{1}$ and $m_{2}$ respectively. As $(\mathscr{T},d)\neq(\mathscr{T}',d')\neq(\mathscr{T}'',d'')$, when the DM receives a signal $s$ at memory state $m$ and decides to which memory state he transit to, or when he decides which action he takes at memory state $m$, he has to remember whether he has once transited to memory state $m_{1}$ and $m_{2}$. In other words, the DM has to track not only his current memory states, but also has to memorize the (incomplete) history of his previous memory states. Switching between multiple automatons thus implicitly implies a larger memory capacity.

Now, I illustrate through an example that a $M$ memory states automaton could be designed to involve switching between automatons with smaller sizes. The example is illustrated in Figure~\ref{fig:switch}. In this example, the DM starts at memory state $3$ with a learning mechanism $(\mathscr{T},d)$ that involves 5 memory states $1$ to $5$. Once he transits to memory state $1$, he switches to another learning mechanism $(\mathscr{T}',d')$ that involves 5 memory states $1$, $10, 11, 12, 13$. On the other hand, once he transits to memory state $5$, he switches to the learning mechanism $(\mathscr{T}',d')$ that involves 5 memory states $5, 6, 7, 8, 9$. Thus, the proposed 13 memory states automaton can be interpreted as a mechanism that involves switching between three 5 memory states automatons.

\tikzstyle{cloud} = [draw, align=left, circle,fill=cyan!20, node distance=3cm,
minimum height=1em]
\tikzstyle{cloud1} = [draw, align=left, circle,fill=black, node distance=1cm,
minimum height=0.01cm]
\tikzstyle{cloud2} = [draw, align=left, circle,fill=red, node distance=1cm,
minimum height=0.01cm]
\tikzset{
	dot/.style = {circle, fill, minimum size=#1,
		inner sep=0pt, outer sep=0pt},
	dot/.default = 3pt 
}

\begin{figure}
	\centering
	\begin{tikzpicture}[every text node part/.style={align=center}]
		\node[cloud] (m1) {\footnotesize 1};
		\node[cloud, right = 3em of m1] (m2) {\footnotesize 2};
		\node[cloud, right = 3em of m2] (m3) {\footnotesize 3};
		\node[above = 3em of m3, align=center] (m0) {Initial \\memory state};
		\node[cloud, right = 3em of m3] (m4) {\footnotesize 4};
		\node[cloud, right = 3em of m4] (m5) {\footnotesize 5};
		\node[cloud, fill=red!60, above right= 2em and 2em of m5] (m6) {\footnotesize 6};
		\node[cloud, fill=red!60, above right= 2em and 2em of m6] (m7) {\footnotesize 7};
		\node[cloud, fill=red!60, below right= 2em and 2em of m5] (m8) {\footnotesize 8};
		\node[cloud, fill=red!60, below right= 2em and 2em of m8] (m9) {\footnotesize 9};
		\node[cloud, fill=green!40, above left= 2em and 2em of m1] (m10) {\footnotesize 10};
		\node[cloud, fill=green!40, above left= 2em and 2em of m10] (m11) {\footnotesize 11};
		\node[cloud, fill=green!40, below left= 2em and 2em of m1] (m12) {\footnotesize 12};
		\node[cloud, fill=green!40, below left= 2em and 2em of m12] (m13) {\footnotesize 13};
		\node[below left = 2em and 0.05em of m1] (1below) {};
		\node[below right = 2em and 0.05em of m5] (5below) {};
		\draw [decorate,decoration={brace,mirror,amplitude=5pt},xshift=0pt,yshift=0pt]
		(1below) -- (5below) node [black,midway,yshift=-0.6cm]
		{\footnotesize $(\mathscr{T},d)$};
		\node[below right = 1em and 0.05em of m9] (9below) {};
		\node[below left = 7.5em and 0.05em of m5] (5below') {};
		\draw [decorate,decoration={brace,mirror,amplitude=5pt},xshift=0pt,yshift=0pt]
		(5below') -- (9below) node [black,midway,yshift=-0.6cm]
		{\footnotesize $(\mathscr{T}'',d'')$};
		\node[below right = 8em and 0.05em of m1] (1below') {};
		\node[below left = 1em and 0.05em of m13] (13below) {};
		\draw [decorate,decoration={brace,mirror,amplitude=5pt},xshift=0pt,yshift=0pt]
		(13below) -- (1below') node [black,midway,yshift=-0.6cm]
		{\footnotesize $(\mathscr{T}',d')$};
		\draw[-Latex]
		(m0) edge (m3)
		(m2) edge[bend right] node [below, midway] {} (m3)
		(m3) edge[bend right] node [below, midway] {} (m4)
		(m4) edge[bend right] node [above, midway] {} (m3)
		(m3) edge[bend right] node [above, midway] {} (m2)
		(m2) edge[bend right] node [above, midway] {} (m1)
		(m4) edge[bend left] node [above, midway] {} (m5)
		(m1) edge[bend left] node [above, midway] {} (m10)
		(m10) edge[bend left] node [above, midway] {} (m11)
		(m10) edge[bend left] node [above, midway] {} (m1)
		(m11) edge[bend left] node [above, midway] {} (m10)
		(m1) edge[bend left] node [above, midway] {} (m12)
		(m12) edge[bend left] node [above, midway] {} (m13)
		(m12) edge[bend left] node [above, midway] {} (m1)
		(m13) edge[bend left] node [above, midway] {} (m12)
		(m5) edge[bend left] node [above, midway] {} (m6)
		(m6) edge[bend left] node [above, midway] {} (m7)
		(m5) edge[bend left] node [above, midway] {} (m8)
		(m8) edge[bend left] node [above, midway] {} (m9)
		(m6) edge[bend left] node [above, midway] {} (m5)
		(m7) edge[bend left] node [above, midway] {} (m6)
		(m8) edge[bend left] node [above, midway] {} (m5)
		(m9) edge[bend left] node [above, midway] {} (m8);
	\end{tikzpicture}
	\caption{An example of a 13 memory states automaton that involves switching between three 5 memory states automatons. The DM starts at memory state $3$ with $(\mathscr{T},d)$, and switches to $(\mathscr{T}',d')$ and $(\mathscr{T}'',d'')$ once he transits to memory state $1$ and $5$, respectively.}
	\label{fig:switch}
\end{figure}


\section{Numerical illustration: naive memory versus the designed mechanism}\label{section:naivememory}

This section presents the numerical illustration referenced in Section~5.1 of the main text: a family of decision problems in which a naive (truncated-Bayesian) memory remains bounded away from the Bayesian benchmark as $M\rightarrow\infty$, while the designed updating mechanism converges to it.

\paragraph{Setup.} There are $N=2$ states with uniform prior and $u(a,\omega)=1$ if $a=\omega$ and $0$ otherwise. Signals are composite, $s=(b,x)$: an informative component $b\in\lbrace1,2\rbrace$ with $\Pr[b=\omega\mid\omega]=q=0.6$, and a noise component $x$ drawn uniformly from $\lbrace1,\cdots,K\rbrace$ independently of the state, so that $\vert S\vert=2K$ and $f^{\omega}(b,x)=\varphi^{\omega}(b)/K$ with $\varphi^{\omega}(\omega)=q$. The informativeness of the signals is $\mathscr{I}=q/(1-q)=1.5$, regardless of $K$.

\paragraph{The two memories.} The \emph{naive} memory recalls the last $T$ signals verbatim and updates as a Bayesian on that truncated history. It is an automaton whose memory states are the $\vert S\vert^{T}$ possible signal strings, so a budget of $M$ memory states supports the recall horizon $T(M)=\floor{\log M/\log\vert S\vert}$. Given $T$ recalled signals, the optimal decision is the majority of the informative components (ties split at random), whose error is
\begin{equation*}
	e(T)=\Pr\left[\mathrm{Bin}(T,q)<\tfrac{T}{2}\right]+\tfrac{1}{2}\Pr\left[\mathrm{Bin}(T,q)=\tfrac{T}{2}\right],\qquad e(0)=\tfrac{1}{2}.
\end{equation*}
The \emph{designed} mechanism is the chain of Examples~1 and~2 of the main text driven by the informative component $b$ --- with the stochastic labeling of the main text, the noise component cancels from every labeling ratio, so the construction ignores it automatically. By \cite{hellman1970learning}, the chain is optimal for $N=2$, with asymptotic utility loss $1/(1+\mathscr{I}^{M-1})$ (in the limit $\delta\rightarrow0$), \emph{independently of} $\vert S\vert$.

\paragraph{Comparison.} Table~\ref{tab:naivecomparison} reports the asymptotic utility loss (the Bayesian benchmark is $0$) as $M$ grows, for two sequences of signal spaces: $\vert S\vert=2M$ ($K=M$) and $\vert S\vert=2\ceil{\sqrt{M}}$ ($K=\ceil{\sqrt{M}}$). In the first, the naive memory cannot store even one full signal ($T(M)=0$ for all $M$) and learns nothing; in the second, its horizon is stuck at $T(M)=1$ --- one signal --- for every $M$, so its loss stays at $1-q=0.4$. In both, the naive memory remains bounded away from the Bayesian benchmark while the designed mechanism converges to it exponentially fast.

\begin{table}[h]
\centering
\begin{tabular}{l|ccccc}
	 & $M=4$ & $M=8$ & $M=16$ & $M=32$ & $M=64$ \\ \hline
	Designed mechanism (any $\vert S\vert$) & $0.229$ & $0.055$ & $2.3\times10^{-3}$ & $3.5\times10^{-6}$ & $8.1\times10^{-12}$ \\
	Naive, $\vert S\vert=2M$ & $0.5$ & $0.5$ & $0.5$ & $0.5$ & $0.5$ \\
	Naive, $\vert S\vert=2\ceil{\sqrt{M}}$ & $0.4$ & $0.4$ & $0.4$ & $0.4$ & $0.4$ \\
\end{tabular}
\caption{Asymptotic utility loss of the naive (truncated-Bayesian) memory and of the designed chain mechanism, with $q=0.6$ ($\mathscr{I}=1.5$). The naive memory's recall horizon is $T(M)=\floor{\log M/\log\vert S\vert}$: it equals $0$ for all $M$ when $\vert S\vert=2M$, and $1$ for all $M$ when $\vert S\vert=2\ceil{\sqrt{M}}$.}
\label{tab:naivecomparison}
\end{table}

Two remarks. First, even with a \emph{fixed} signal space (say $\vert S\vert=4$), the naive memory converges only polynomially: its horizon is $T(M)=\floor{\log_{2}M/2}$ and the majority error decays as $\exp(-cT)$ with $c=-\log\left(2\sqrt{q(1-q)}\right)\approx 0.02$, i.e., as $M^{-0.015}$. Second, a cleverer use of the $M$ memory states --- for instance, recording only the informative component of each signal --- is precisely what the designed mechanism does; the naive memory, by definition, treats every signal string as distinct regardless of its decision relevance, which is what makes it wasteful.

\section{Example of a sequence of signal structures that satisfies Assumptions~\ref{assumption_noperfectsignalinfinite} and~\ref{assumption_identifiableinfinte}}\label{section:signalstructure}

In this section, I provide an example of a sequence of signal structures that satisfies Assumptions~\ref{assumption_noperfectsignalinfinite} and~\ref{assumption_identifiableinfinte}.\footnote{The signal structures are described by densities on $S=[0,1)$ for convenience. Since each $f^{\omega}_{N}$ is constant on the dyadic intervals $[\frac{i}{2^{N}},\frac{i+1}{2^{N}})$, the example is equivalent to one with the finite signal space $\lbrace 0,\cdots,2^{N}-1\rbrace$ in which signal $i$ carries the mass $f^{\omega}_{N}(i)/2^{N}$, matching the finite-signal setting of the main text; the Cauchy--Schwarz distances are unchanged.} Consider the following sequence of signal structures $\lbrace(f_{N}^{\omega})_{\omega=1}^{N}\rbrace_{N=1}^{\infty}$ with $S=[0,1)$:
	\begin{equation*}
		f_{N}^{\omega}(s)=\begin{cases}
			\frac{2}{3}\text{ for $s\in [\frac{i}{2^{\omega}},\frac{i+1}{2^{\omega}})$ where $i=0,2,\cdots, 2^{\omega}-2$};\\
			\frac{4}{3}\text{ for $s\in [\frac{i}{2^{\omega}},\frac{i+1}{2^{\omega}})$ where $i=1,3,\cdots, 2^{\omega}-1$.}
		\end{cases}
	\end{equation*}
	The signal structures are illustrated in Figure~\ref{fig:appendixb}. First, they satisfy Equation~\ref{assumption_noperfectsignalinfinite} as $\frac{f_{N}^{\omega}(s)}{f_{N}^{\omega'}(s)}\geq\frac{1}{2}$ for all $\omega,\omega'\in\Omega$, $N$ and $s$. Second,  for any $N$ and given any $\omega$ and $\omega'\neq\omega$, the Cauchy-Schwarz distance is equal to:
	\begin{equation*}
		\begin{split}
			-\log\dfrac{\int f_{N}^{\omega}(s)f_{N}^{\omega'}(s)ds}{\sqrt{\int (f_{N}^{\omega}(s))^{2}ds}\sqrt{\int (f_{N}^{\omega'}(s))^{2}ds}}=&-\log\frac{\frac{1}{4}(\frac{2}{3})^{2}+\frac{1}{2}\frac{2}{3}\frac{4}{3}+\frac{1}{4}(\frac{4}{3})^{2}}{\frac{1}{2}(\frac{2}{3})^{2}+\frac{1}{2}(\frac{4}{3})^{2}}\\
			=&-\log\frac{1}{10/9}>1+\kappa
		\end{split}
	\end{equation*}
	where $\kappa<\log\frac{10}{9}-1$ and thus satisfies Assumption~\ref{assumption_identifiableinfinte}.
	
	\begin{figure}
		\centering
		\begin{tikzpicture}
			\draw[->] (0,0) node[left] {$\omega=1$} --(10,0) node[right] {$s$};
			\draw (1,0.1)--(1,-0.1) node[below] {$0$};
			\draw (9,0.1)--(9,-0.1) node[below] {$1$};
			\draw (5,0.1)--(5,-0.1) node[below] {$\frac{1}{2}$};
			\node[above] at (3,0.1) {$f^{1}_{N}=\frac{2}{3}$};
			\node[above] at (7,0.1) {$f^{1}_{N}=\frac{4}{3}$};
			\draw[->] (0,-2) node[left] {$\omega=2$} --(10,-2) node[right] {$s$};
			\draw (1,-1.9)--(1,-2.1) node[below] {$0$};
			\draw (9,-1.9)--(9,-2.1) node[below] {$1$};
			\draw (5,-1.9)--(5,-2.1) node[below] {$\frac{2}{2^{2}}$};
			\draw (3,-1.9)--(3,-2.1) node[below] {$\frac{1}{2^{2}}$};
			\draw (7,-1.9)--(7,-2.1) node[below] {$\frac{3}{2^{2}}$};
			\node[above] at (2,-1.9) {$f^{2}_{N}=\frac{2}{3}$};
			\node[above] at (4,-1.9) {$f^{2}_{N}=\frac{4}{3}$};
			\node[above] at (6,-1.9) {$f^{2}_{N}=\frac{2}{3}$};
			\node[above] at (8,-1.9) {$f^{2}_{N}=\frac{4}{3}$};
			\draw[->] (0,-4) node[left] {$\omega=3$} --(10,-4) node[right] {$s$};
			\draw (1,-3.9)--(1,-4.1) node[below] {$0$};
			\draw (9,-3.9)--(9,-4.1) node[below] {$1$};
			\draw (2,-3.9)--(2,-4.1) node[below] {$\frac{1}{2^{3}}$};
			\draw (3,-3.9)--(3,-4.1) node[below] {$\frac{2}{2^{3}}$};
			\draw (4,-3.9)--(4,-4.1) node[below] {$\frac{3}{2^{3}}$};
			\draw (5,-3.9)--(5,-4.1) node[below] {$\frac{4}{2^{3}}$};
			\draw (6,-3.9)--(6,-4.1) node[below] {$\frac{5}{2^{3}}$};
			\draw (7,-3.9)--(7,-4.1) node[below] {$\frac{6}{2^{3}}$};
			\draw (8,-3.9)--(8,-4.1) node[below] {$\frac{7}{2^{3}}$};
			\node[above] at (1.5,-3.9) {\tiny $f^{3}_{N}=\frac{2}{3}$};
			\node[above] at (2.5,-3.9) {\tiny $f^{3}_{N}=\frac{4}{3}$};
			\node[above] at (3.5,-3.9) {\tiny $f^{3}_{N}=\frac{2}{3}$};
			\node[above] at (4.5,-3.9) {\tiny $f^{3}_{N}=\frac{4}{3}$};
			\node[above] at (5.5,-3.9) {\tiny $f^{3}_{N}=\frac{2}{3}$};
			\node[above] at (6.5,-3.9) {\tiny $f^{3}_{N}=\frac{4}{3}$};
			\node[above] at (7.5,-3.9) {\tiny $f^{3}_{N}=\frac{2}{3}$};
			\node[above] at (8.5,-3.9) {\tiny $f^{3}_{N}=\frac{4}{3}$};
			\node at (5,-5.5) {$\vdots$};
		\end{tikzpicture}
		\caption{Signal structures that satisfies Equation~\ref{assumption_noperfectsignalinfinite} and Assumption~\ref{assumption_identifiableinfinte} as $N\rightarrow\infty$. The signal structures comprise low and high density alternatively in $2^{\omega}$ equal-sized intervals.}
		\label{fig:appendixb}
	\end{figure}

\section{$N,M\rightarrow\infty$: moved to main text}\label{section:infinite}
The analysis of the limit $N,M\rightarrow\infty$ --- the formal setup, assumptions, and the results on learning and ignorance, with their proofs --- appears in Sections~\ref{sec:NMlimit_small} and~\ref{sec:NMlimit_ignorance} of the main paper and the corresponding subsections of the main paper's Appendix. The limit results on disagreement appear in Section~\ref{section:disagreementonline} below.

\section{Further numerical comparison for the symmetric environment}\label{section:symmetricsupp}

\begin{figure}
		\centering
		\includegraphics[width=25em]{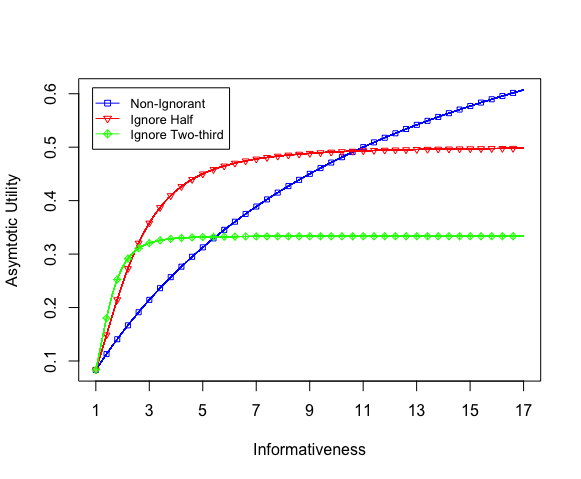}
		\caption{Consider $N=M=12$. The y-axis is the asymptotic utility of the ignorant mechanism ignoring half of the states illustrated in Figure~\ref{fig:symmetricignorestates1}, a symmetric ignorant mechanism that ignores $2/3$ of the states, and the non-ignorant mechanism illustrated in Figure~\ref{fig:symmetricallstates1}, while the x-axis is $\frac{f^{\omega}(s^\omega)}{f^{\omega}(s^{\omega'})}$. As the informativeness $\frac{f^{\omega}(s^\omega)}{f^{\omega}(s^{\omega'})}$ decreases, it becomes more beneficial to ignore more states.} 
		\label{fig:compareignoreor23nonignorant}
	\end{figure}

This section extends the numerical comparison of Section~5.2.2 of the main text by adding a symmetric ignorant mechanism that ignores $2/3$ of the states of the world. Figure~\ref{fig:compareignoreor23nonignorant} shows that when the signals are very noisy, i.e., when $\frac{f^{\omega}(s^\omega)}{f^{\omega}(s^{\omega'})}$ is close to $1$, ignoring $2/3$ of the states outperforms the mechanism that ignores half of the states, which in turn outperforms the non-ignorant mechanism: as learning becomes more difficult, the DM optimally concentrates his memory on fewer states.

\section{Persistent disagreement: formal results}\label{section:disagreementonline}

This section contains the formal definitions and statements of the disagreement results summarized in Section~5.3 of the main text; all proofs appear in Section~\ref{section:onlineproofs}. The result on disagreement driven by cognitive ability, Corollary~\ref{prop_disagreementbasedonM}, is stated in the main text; its proof also appears in Section~\ref{section:onlineproofs}.

	Lastly, I turn to the question of whether disagreement could persist asymptotically in small and big worlds. Consider two individuals $A$ and $B$ who have different utility matrices $(\bm{u}^{\omega}_{A})_{\omega=1}^{N}$ and $(\bm{u}^{\omega}_{B})_{\omega=1}^{N}$, and/or different prior beliefs $(p^{\omega}_{A})_{\omega=1}^{N}$ and $(p^{\omega}_{B})_{\omega=1}^{N}$, and/or different abilities of information acquisition captured by $(f^{\omega}_{A})_{\omega=1}^{N}$ and $(f^{\omega}_{B})_{\omega=1}^{N}$.\footnote{For example, individual $A$ could receive noisier signals than individual $B$, i.e., $f^{\omega}_{A}=\gamma+(1-\gamma)f^{\omega}_{B}$ for some $\gamma\in(0,1)$; or individual $A$ could have different learning advantages in identifying some states better but other states worse than individual $B$, i.e., $\sup_{s}f^{\omega}_{A}(s)/f^{\omega'}_{A}(s)>\sup_{s}f^{\omega}_{B}(s)/f^{\omega'}_{B}(s)$ but $\sup_{s}f^{\omega''}_{A}(s)/f^{\omega'''}_{A}(s)<\sup_{s}f^{\omega''}_{B}(s)/f^{\omega'''}_{B}(s)$ for some $\omega,\omega',\omega'',\omega'''$.} Their updating mechanisms, $(m_{A1},\mathscr{T}_A,d_A)$ and $(m_{B1},\mathscr{T}_B,d_B)$, induce a (random) sequence of actions over time. To define disagreement, I sample one action from each individual~A and~B and define the ``disagreement between $(m_{A1},\mathscr{T}_A,d_A)$ and $(m_{B1},\mathscr{T}_B,d_B)$ in state $\omega$'' as the probability that the two sampled actions are different.\footnote{For example, if both individuals alternate between action~1 and~2, their disagreement is $\frac{1}{2}$. An alternative definition is to measure the proportion of time $t$ where $a^A_t\neq a^B_t$: If both individuals alternate between action~1 and~2, their disagreement is $0$ if they both start with the same action and is~$1$ otherwise. Corollary~\ref{coro_bigworlddisagreement} (i) and the limit result of (ii), i.e., disagreement approaches $0$ as $M$ goes to infinity, hold with this alternative definition.} For a given $\epsilon$, the disagreement between individual~A and~B in state~$\omega$ is the supremum of disagreement between all pairs of $\epsilon$-optimal updating mechanisms in state $\omega$. Lastly, the two individuals almost always disagree (resp. agree) with each other for a given $\epsilon$ if their disagreement equals $1$ (resp. $0$) in all states.

	\begin{corollary}\label{coro_bigworlddisagreement}
		In big worlds, disagreement can persist; in small worlds, individuals cannot disagree forever.
		\begin{enumerate}[label=(\roman*)]
			\item For all $N>2$, $M_A$ and $M_B$, there exists some  $(\bm{u}^{\omega}_{A},p^{\omega}_{A},f^{\omega}_{A})_{\omega=1}^{N}$ and $(\bm{u}^{\omega}_{B},p^{\omega}_{B},f^{\omega}_{B})_{\omega=1}^{N}$ such that individuals~A and~B almost always disagree when both adopt $\epsilon$-optimal updating mechanisms, when $\epsilon\rightarrow 0$.
			\item In contrast, take $(\bm{u}^{\omega}_{A},p^{\omega}_{A},f^{\omega}_{A})_{\omega=1}^{N}$ and $(\bm{u}^{\omega}_{B},p^{\omega}_{B},f^{\omega}_{B})_{\omega=1}^{N}$, there exist $K_1, K_2>0$ and $r<1$ such that when both individuals adopt $\epsilon$-optimal updating mechanisms, their disagreement is bounded above by $\left(K_1 r^{\floor{\frac{\min\lbrace M_A,M_B\rbrace-1}{N}}}+K_2\epsilon\right)$ in all states.
		\end{enumerate}
	\end{corollary}

	(i) shows that in general, there exist decision environments such that disagreement almost always happens and is persistent. However, as implied by (ii), when the world gets smaller ($\frac{N}{M}$ goes to $0$), such set of decision environments with disagreement vanishes: different individuals with different prior beliefs and/or information acquisition abilities who adopt (almost) optimal updating mechanisms are bound to agree with each other.

\subsection*{The limit $N,M\rightarrow\infty$}

	The following results extend Corollary~\ref{coro_bigworlddisagreement} to the limit $N,M\rightarrow\infty$. In small worlds, Proposition~\ref{prop_smallworldinfiniteN} implies that individuals are bound to agree:

	\begin{corollary}\label{coro_agreesmallworld1infiniteN}
    	Under the same conditions as Proposition~\ref{prop_smallworldinfiniteN} for individuals $i=A,B$ (each with their own memory sequence $M_{iN}$ satisfying $N=\Theta(M_{iN}^{h_i})$ with $h_i<1$), suppose the two individuals have the same limit support, in the sense that their priors agree on which events vanish: for every sequence of subsets $\tilde{\Omega}_{N}\subseteq\lbrace 1,\cdots,N\rbrace$, $\lim_{N\rightarrow\infty}\sum_{\omega\in\tilde{\Omega}_{N}}p^{\omega}_{NA}=0$ if and only if $\lim_{N\rightarrow\infty}\sum_{\omega\in\tilde{\Omega}_{N}}p^{\omega}_{NB}=0$. Then individuals~A and~B must almost always agree with each other (measured in prior probability $p^{\omega}_{NA}$ or $p^{\omega}_{NB}$) at the limit as $\epsilon\rightarrow 0$.
    \end{corollary}
	The intuition is direct. In a small world, the optimal loss vanishes ($L^*_N\rightarrow 0$), so each individual's $\epsilon$-optimal mechanism picks the correct action in (almost measured in prior probability) every true state at the limit. The same-limit-support condition ensures that both individuals consider the same set of states as a priori plausible: without it, an individual with vanishing prior on a state would not bother to learn that state (the contribution to loss is also vanishing) and would almost never play the corresponding action, while the other individual (with non-vanishing prior) would, generating trivial disagreement. With shared support, both individuals optimally pick the same correct action in every true state, and disagreement vanishes. The corollary thus establishes the substantive content of the small-world result: persistent disagreement does not occur when complexity is low compared to cognitive ability.

	In big worlds, however, disagreement is possible.

	\begin{corollary}\label{coro_bigworlddisagreementinfiniteN}
		Fix sequences of decision problems $\left\lbrace (\bm{u}^{\omega}_{Ni},p^{\omega}_{Ni},f^{\omega}_{Ni})_{\omega=1}^{N},M_{iN}\right\rbrace_{N\in\mathcal{N}}$ for $i=A,B$, satisfying Assumptions~\ref{assumption_fullsupportinfinite}, \ref{assumption_noperfectsignalinfinite}, and \ref{assumption_identifiableinfinte}, where $N=\Theta(M_{iN}^{h_i})$ with $h_i\geq 1$. Suppose the two individuals have the same limit support, in the sense of Corollary~\ref{coro_agreesmallworld1infiniteN}.\footnote{The same-limit-support restriction only strengthens the existence claim: it rules out the trivial route to disagreement in which the two individuals simply care about different states. The example constructed in the proof is in fact stronger still --- the two individuals share identical priors --- so the disagreement is driven by the scarcity of memory alone.}There exist such sequences of decision problems for which, for all $\epsilon>0$, there exist sequences of $\epsilon$-optimal updating mechanisms for individuals A and B under which they disagree in all states.
	\end{corollary}

	The intuition is as follows. By Proposition~\ref{prop_smallworldinfiniteN}(ii), each individual in a big world is forced to focus on a vanishing fraction of states --- almost all states are ignored. The driving force of the result is most starkly visible under \ref{assumption_fullsupportinfiniteplus}, where any updating mechanism is $\epsilon$-optimal in the limit. This freedom explains why Corollary~\ref{coro_bigworlddisagreementinfiniteN} is slightly weaker than Corollary~\ref{coro_bigworlddisagreement}(i), in that it asserts \emph{some} (rather than all) $\epsilon$-optimal mechanisms lead to disagreement: a large set of mechanisms are $\epsilon$-optimal in big worlds, leaving ample room for divergent sets of actions chosen by the individuals. Nonetheless, combining Corollaries~\ref{coro_agreesmallworld1infiniteN} and~\ref{coro_bigworlddisagreementinfiniteN} confirms the same qualitative picture as in the baseline model with fixed $N$: disagreement does not arise in small worlds but can arise in big worlds.

\section{Robustness of the results in small worlds to updating mistakes}\label{section:robustness}
Below, I show that the main implications in small worlds, i.e., the asymptotic distribution of actions is close to Bayesian and that disagreement does not persist, hold even when individuals make ``updating mistakes". In other words, the results are robust to individuals' limited ability to design and follow an ``optimal" updating mechanism.

Consider two individuals $A$ and $B$. Individual $A$ adopts the simple updating mechanism described in the proof of Proposition~\ref{prop_smallworld} while individual $B$ ``attempts" to adopt the same updating mechanism but makes local mistakes as he randomly transits to neighboring memory states with some probability $\gamma\in(0,1)$. Formally, the transition rule of individual $B$, denoted as $\mathscr{T}'(m,s)$, is as follows:
\begin{equation*}
	\mathscr{T}'(0,s)=\begin{cases}
		\mathscr{T}(0,s) &\text{ with probability $1-\gamma$;}\\
		ji &\text{ with probability $\frac{\gamma}{N}$ for all $j=1,\cdots,N$.}
	\end{cases}
\end{equation*}
\begin{equation*}
	\mathscr{T}'(i1,s)=\begin{cases}
		\mathscr{T}(i1,s) &\text{ with probability $1-\gamma$;}\\
		i2 &\text{ with probability $\frac{\gamma}{2}$;}\\
		0&\text{ with probability $\frac{\gamma}{2}$.}
	\end{cases}
\end{equation*}
\begin{equation*}
	\mathscr{T}'(i\lambda,s)=
	\begin{cases}
		\mathscr{T}(i\lambda,s)&\text{ with probability $1-\gamma$;}\\
		i(\lambda-1) &\text{ with probability $\gamma$.}
	\end{cases}
\end{equation*}
while for $k=2,3,\cdots,\lambda-1$,
\begin{equation*}
	\mathscr{T}'(ik,s)=\begin{cases}
		\mathscr{T}(ik,s) &\text{ with probability $1-\gamma$;}\\
		i(k-1) &\text{ with probability $\frac{\gamma}{2}$;}\\
		i(k+1)&\text{ with probability $\frac{\gamma}{2}$.}
	\end{cases}
\end{equation*}
where $\mathscr{T}(m,s)$ is defined in the proof of Proposition~\ref{prop_smallworld}. The initial memory states and decision rule of individual B stay the same as individual A. Such updating mistakes could be induced by memory imperfections, i.e., the DM's memory state is subject to local fluctuations, or imperfect perceptions on signals, e.g., the DM may mistakenly perceive any signal as a signal that supports state~$\omega$. 

\begin{proposition}\label{prop_smallworldmistakes}
	Consider individual $A$ who adopts the updating mechanism $\mathscr{T}$ and individual $B$ who makes local mistakes with some probability $\gamma\in(0,1)$, characterized above by $\mathscr{T}'(m,s)$. Fix $N$ and $(\bm{u}_B^{\omega}, p_B^{\omega}, f_B^{\omega})_{\omega=1}^{N}$, for all $\gamma\in(0,1)$,
	\begin{equation*}
		\lim_{k\rightarrow\infty} L_k(m_1,\mathscr{T}',d)=0.
	\end{equation*}
	Moreover, individual $A$ and $B$ are bound to agree in small worlds. 
\end{proposition}

\section{Managerial implications}\label{section:managerial}

This section discusses the managerial implications of the results, moved here from the main text for brevity.

\paragraph{Encourage Strategic Ignorance and Focused Inquiry}
One of the important insights of this paper is that ignorance could be an optimal strategy, especially when cognitive resources are limited. Moreover, perhaps counterintuitively, as shown in Proposition~\ref{prop:symmetricignorance}, strategic ignorance is optimal especially when uncertainty is high or when information is noisy. This result complements a recent study \cite{yoon2021working}, which found that consuming excessive news information during the COVID-19 pandemic decreases employees' goal progress and creativity due to anxiety. The results of this paper demonstrate that even without psychological channels, reacting to excessive information can compromise the quality of inference and decision-making. For example, as Alan Mulally took over as CEO of Ford in 2006, he streamlined Ford's product lines, prioritized core brands, and sold off premier automobile brands like Jaguar and Land Rover. He inherited 97 models and reduced Ford's product lineup to as few as 20 models. While Ford might lose out on a potential boom in the premier automobile market, the streamlining of product lines allowed Ford to conserve resources and return to profitability by 2010, outpacing competitors who pursued broader, less focused strategies (\cite{edmondson2021turnaround}). Alan Mulally was included in the 2009 Time 100 list due to his successful strategies at Ford.\footnote{See \url{https://content.time.com/time/specials/packages/article/0,28804,1894410_1893837_1894182,00.html}.} Similarly, Steve Jobs has emphasized Apple's corporate culture of ``saying no to hundreds of good ideas''.\footnote{See \url{https://www.forbes.com/sites/bernhardschroeder/2022/10/31/entrepreneurs-and-small-business-owners-steve-jobs-advice-on-saying-no-more-often-to-achieve-your-goals/}.}

\paragraph{Summarizing Information to Save Cognitive Resources and Reduce Ignorance}
On the other hand, the results of this paper also shed light on how to discourage ignorance if one aims to do so. Despite the optimality of ignorance, a manager might want to discourage ignorance in an employee if the employee has a different utility matrix or disagrees on what possibilities (states of the world) are more probable. As shown in Proposition~\ref{coro_bigworldignorance2example} and Proposition~\ref{prop:symmetricignorance}, individuals tend to ignore possibilities when information is noisy. Thus, conversely, managers can use AI to summarize information before employees process it. Different from a Bayesian setting and models with entropy-based rational inattention, summarizing multiple signals into one helps individuals process information with limited cognitive resources. 

Thanks to the recent advancements in LLM technologies, summarizing information has never been easier. For example, Bloomberg integrated BloombergGPT to summarize earnings calls, news, and filings. Analysts using the tool can easily process multiple signals simultaneously, which reduces their cognitive load when processing information and reduces their incentive to ignore possibilities.\footnote{See \cite{wu2023bloomberggpt} and \url{https://www.forbes.com/sites/jamielsheikh/2023/04/05/the-chatgpt-of-finance-is-here-bloomberg-is-combining-ai-and-fintech/}.} Similarly, in a recent Microsoft study, they reported that users of their Copilot system are more efficient in pulling together information from multiple sources, which reduces employees' incentive to ignore information, as suggested by the results of this paper.\footnote{See \url{https://www.microsoft.com/en-us/worklab/work-trend-index/copilots-earliest-users-teach-us-about-generative-ai-at-work}.}

\section{Omitted Proofs in the main text and Proof of the results in online appendix}\label{section:onlineproofs}

\subsection{Proof of Proposition~\ref{prop:symmetricignorance} (stated in the main text)}
\begin{proof}
    As mentioned in the main text, I consider a case where $N=M$ and states of the world are a priori uniformly distributed, i.e., $p^{\omega}=\frac{1}{N}$ for all $\omega=\lbrace 1,\cdots, N\rbrace$. Moreover, without loss of generality I assume $u(\omega,\omega)=1>u(\omega',\omega)=0$ for all $\omega$ and $\omega'\neq\omega$. Signals are distributed in $S=\lbrace s^{1},\cdots.s^{N}\rbrace$ and $s^{\omega}$ is a signal that supports state $\omega$. The signal distribution follows:
		\begin{equation*}
			F^{\omega}(s^{\omega})=\mathscr{I}F^{\omega}(s^{\omega'}) \text{ for all $\omega$ and $\omega'\neq\omega$ where $\mathscr{I}>1$.}
		\end{equation*} 
		Under all states of the world, it is $\mathscr{I}$ times more likely to receive a signal that supports the true state than a signal that supports one of the other states.
		
	First, note that if the DM ignores no state, he allocates one memory state to each action. Without loss of generality, assume he takes action~$\omega$ in memory state $\omega$, i.e., $d(m)=m$ for all $m=1,\cdots, M$. I prove the Proposition in three steps: 1) to characterize the infimum utility loss of asymmetric non-ignorant updating mechanisms, it is without loss to assume that the DM transits to memory state $m$ only if he receives a signal $s^m$, i.e., a signal supporting state $m$; 2) the asymptotic utility of an optimal symmetric non-ignorant mechanism equals $\frac{\mathscr{I}}{\mathscr{I}+N-1}$; 3) a ``simple ignorant updating mechanism'' that ignores half of the states outperforms the optimal symmetric non-ignorant mechanism when $N$ is large enough or when $\mathscr{I}$ is small enough.

    To prove 1), note that under symmetry and $d(m)=m$, the transition function is characterized by five parameters:
    \begin{itemize}
        \item $\alpha=\Pr[\mathscr{T}(i,s^{i})=i]$ (stay on signal supporting the current action);
        \item $\beta=\Pr[\mathscr{T}(i,s^{k})=i]$ for $k\neq i$ (stay on non-matching signal);
        \item $\gamma=\Pr[\mathscr{T}(i,s^{i})=j]$ for $j\neq i$ (transit on own-state signal);
        \item $\delta=\Pr[\mathscr{T}(i,s^{j})=j]$ for $j\neq i$ (transit on the target's matching signal);
        \item $\eta=\Pr[\mathscr{T}(i,s^{k})=j]$ for $i,j,k$ all distinct (transit on a cross signal),
    \end{itemize}
    with $\alpha+(N-1)\gamma=1$ and $\beta+\delta+(N-2)\eta=1$. Part~1's claim --- transitions only on matching signals --- corresponds to $\gamma=\eta=0$.

    By symmetry, the long-run distribution under any state $\omega$ takes the form $\mu^{\omega}_{\omega}=\mu^{*}$ and $\mu^{\omega}_{m}=\nu$ for $m\neq\omega$, with $\mu^{*}+(N-1)\nu=1$ and $\mu^{*}$ the same across $\omega$. In the following I focus on state~1 and characterize $\mu^*$. Let $A:=q^{1}_{11}$ denote the self-transition probability of memory state~$1$ in state~$1$, and $B:=q^{1}_{j1}$ for $j\neq 1$ the transition probability from any other memory state into memory state~$1$ in state~$1$ (well-defined because, by symmetry, $q^{1}_{j1}$ is the same for all $j\neq 1$). Using $F^{1}(s^{1})=\mathscr{I}/(\mathscr{I}+N-1)$ and $F^{1}(s^{k})=1/(\mathscr{I}+N-1)$ for $k\neq 1$,
    \begin{align*}
        A&=F^{1}(s^{1})\,\alpha+\sum_{k\neq 1}F^{1}(s^{k})\,\beta=\frac{\mathscr{I}\alpha+(N-1)\beta}{\mathscr{I}+N-1},\\
        B&=F^{1}(s^{1})\,\delta+F^{1}(s^{j})\,\gamma+\sum_{k\neq 1,k\neq j}F^{1}(s^{k})\,\eta=\frac{\mathscr{I}\delta+\gamma+(N-2)\eta}{\mathscr{I}+N-1}.
    \end{align*}
    As the long-run probabilities of transiting in and out of memory state 1 are equal, we have $\mu^{*}(1-A)=(N-1)\nu B$. Combined with $\nu=(1-\mu^{*})/(N-1)$ gives
    \begin{equation*}
        \mu^{*}=\frac{B}{1-A+B}.
    \end{equation*}
    Maximizing $\mu^{*}$ is equivalent to minimizing $(1-A)/B$. Substituting the normalizations $\alpha=1-(N-1)\gamma$ and $\beta=1-\delta-(N-2)\eta$ into the formula for $A$ above gives $1-A=(N-1)[\mathscr{I}\gamma+\delta+(N-2)\eta]/(\mathscr{I}+N-1)$, and therefore
    \begin{equation*}
        \frac{1-A}{B}=(N-1)\,\frac{\mathscr{I}\gamma+\delta+(N-2)\eta}{\mathscr{I}\delta+\gamma+(N-2)\eta}.
    \end{equation*}
   Differentiating the ratio with respect to $\gamma$ gives
    \begin{equation*}
        \frac{\partial}{\partial\gamma}\frac{\mathscr{I}\gamma+\delta+(N-2)\eta}{\mathscr{I}\delta+\gamma+(N-2)\eta}=\frac{(\mathscr{I}-1)\big[(\mathscr{I}+1)\delta+(N-2)\eta\big]}{(\mathscr{I}\delta+\gamma+(N-2)\eta)^{2}}\geq 0,
    \end{equation*}
    so the ratio is minimized at $\gamma=0$. Differentiating with respect to $\eta$,
    \begin{equation*}
        \frac{\partial}{\partial\eta}\frac{\mathscr{I}\gamma+\delta+(N-2)\eta}{\mathscr{I}\delta+\gamma+(N-2)\eta}=\frac{(N-2)(\mathscr{I}-1)(\delta-\gamma)}{(\mathscr{I}\delta+\gamma+(N-2)\eta)^{2}}.
    \end{equation*}
    Given $\gamma=0$ and $\delta>0$, this is non-negative, so the ratio is further minimized at $\eta=0$. At $\gamma=\eta=0$ (with any $\delta>0$),
    \begin{equation*}
        \frac{1-A}{B}=\frac{(N-1)\delta}{\mathscr{I}\delta}=\frac{N-1}{\mathscr{I}},\qquad\mu^{*}=\frac{\mathscr{I}}{\mathscr{I}+N-1}.
    \end{equation*}
     Hence (one of) the optimal symmetric non-ignorant mechanism has $\gamma=\eta=0$, i.e., the DM transits to a different memory state $j$ only upon receiving the matching signal $s^{j}$. This proves Part~1 within the symmetric class; the maximum value $\mu^{*}=\mathscr{I}/(\mathscr{I}+N-1)$ also delivers Part~2 directly.


%
		I now prove 3). Consider an ignorant mechanism that follows the similar idea of the non-ignorant updating mechanism illustrated in Figure~\ref{fig:symmetricallstates1}, but ignores half of the states of the world. For simplicity, assume that $N$ is even. The ignorant mechanism is illustrated in Figure~\ref{fig:symmetricignorestates1}, with an example of $N=4$. By ignoring half of the states, the DM allocates two memory states to each action that he does not ignore. Without loss of generality, assume that the DM takes action~$\omega$ in memory states $2\omega-1$ and $2\omega$ for $\omega\leq\frac{N}{2}$, and ignores all actions $\omega'>\frac{N}{2}$. In the ``more confident'' memory state $2\omega$, upon receiving a signal supporting state $\omega'\neq\omega$ where $\omega'\leq \frac{N}{2}$, the DM transits to state $2\omega-1$ with probability $\delta<1$ and stays in his current memory state otherwise. In the ``less confident'' memory state $2\omega-1$, upon receiving a signal that supports state $\omega$, he transits to the ``more confident'' memory state $2\omega$; upon receiving a signal that supports state $\omega'\neq\omega$ where $\omega'\leq \frac{N}{2}$, the DM transits to state $2\omega'-1$ with probability $\delta<1$ and stays in his current memory state otherwise.

		Formally, the transition function is as follows:
		\begin{small}
			\begin{alignat*}{2}
				\mathscr{T}(2\omega,s^{\omega})&= 2\omega  &&\text{ for all $\omega\leq\frac{N}{2}$,}\\
				\mathscr{T}(2\omega,s^{\omega'})&=\begin{cases}
					2\omega-1 \text{ with probability $\delta$}\\
					2\omega  \text{ with probability $1-\delta$}
				\end{cases}
				&&\text{ for all $\omega\leq\frac{N}{2}$, $\omega'\neq\omega$ and $\omega'\leq\frac{N}{2}$,}\\
				\mathscr{T}(2\omega-1,s^{\omega})&= 2\omega  &&\text{ for all $\omega\leq\frac{N}{2}$,}\\
				\mathscr{T}(2\omega-1,s^{\omega'})&=\begin{cases}
					2\omega'-1 \text{ with probability $\delta$}\\
					2\omega  \text{ with probability $1-\delta$}
				\end{cases} &&\text{ for all $\omega\leq\frac{N}{2}$, $\omega'\neq\omega$ and $\omega'\leq\frac{N}{2}$,}\\
				\mathscr{T}(m,s^{\omega'})&= m &&\text{ for all $\omega>\frac{N}{2}$.}
			\end{alignat*}
		\end{small}
		Suppose the true state is $1$, In the stationary distribution, we must have
		\begin{equation}
			\begin{split}
				\delta\mu^{1}_{2\omega}\sum_{\omega'\neq\omega,\omega'\leq\frac{N}{2}}F^{1}(s^{\omega'})&=\mu^{1}_{2\omega-1}F^{1}(s^{\omega})\text{ for all $\omega\leq\frac{N}{2}$,}\\
				\delta\mu^{1}_{2\omega-1}F^{1}(s^{\omega'})&=\delta\mu^{1}_{2\omega'-1}F^{1}(s^{\omega})\text{ for all $\omega,\omega'\leq\frac{N}{2}$ and $\omega\neq\omega'$.}
			\end{split}
			\label{eq:ignorehalfstate}
		\end{equation}
		From the first Equation of Equation~\eqref{eq:ignorehalfstate}, we know that when $\delta$ is close to $0$, $\mu^{1}_{2\omega-1}$ is close to $0$ for all $\omega\leq\frac{N}{2}$. Moreover, we have
		\begin{equation*}
			\begin{split}
				\mu^{1}_{2}&=\frac{F^{1}(s^{1})}{\delta\sum_{\omega'\neq1,\omega'\leq\frac{N}{2}}F^{1}(s^{\omega'})}\times\frac{\delta F^{1}(s^{1})}{\delta F^{1}(s^{\omega})}\times \frac{\delta\sum_{\omega'\neq\omega,\omega'\leq\frac{N}{2}}F^{1}(s^{\omega'})}{F^{1}(s^{\omega})}\times\mu^{1}_{2\omega}\\
				&=\frac{\mathscr{I}}{\frac{N}{2}-1}\times\mathscr{I}\times\left(\frac{N}{2}-2+\mathscr{I}\right)\times\mu^{1}_{2\omega}\\
				&=\frac{\mathscr{I}^{2}(N+2\mathscr{I}-4)}{N-2}\mu^{1}_{2\omega}
			\end{split}
		\end{equation*} 
		for all $\omega\neq1$ and $\omega\leq\frac{N}{2}$. We have
		\begin{equation*}
			\begin{split}
				\mu^{1}_{2}+\sum_{\omega\neq 1,\omega\leq\frac{N}{2}}\mu^{1}_{2\omega}&=1\\
				\mu^{1}_{2}+\frac{N-2}{\mathscr{I}^{2}(N+2\mathscr{I}-4)}\left(\frac{N}{2}-1\right)\mu^{1}_{2}&=1\\
				\mu^{1}_{2}&=\frac{2\mathscr{I}^{2}(N+2\mathscr{I}-4)}{2\mathscr{I}^{2}(N+2\mathscr{I}-4)+(N-2)^{2}}.
			\end{split}
		\end{equation*}
		By repeating the same computation for all $\omega\leq\frac{N}{2}$, the asymptotic utility equals:
		\begin{equation}
			\sum_{\omega=1}^{\frac{N}{2}}\frac{2\mathscr{I}^{2}(N+2\mathscr{I}-4)}{2\mathscr{I}^{2}(N+2\mathscr{I}-4)+(N-2)^{2}}\times\frac{1}{N}=\frac{\mathscr{I}^{2}(N+2\mathscr{I}-4)}{2\mathscr{I}^{2}(N+2\mathscr{I}-4)+(N-2)^{2}}
		\end{equation}
		which is larger than the utility of an optimal ignorant mechanism if and only if:
		\begin{equation*}
			\frac{\mathscr{I}^{2}(N+2\mathscr{I}-4)}{2\mathscr{I}^{2}(N+2\mathscr{I}-4)+(N-2)^{2}}\geq \frac{\mathscr{I}}{\mathscr{I}+N-1}
		\end{equation*}
		or equivalently:
		\begin{equation*}
			\mathscr{I}(N+2\mathscr{I}-4)(\mathscr{I}+N-1)-2\mathscr{I}^2(N+2\mathscr{I}-4)-(N-2)^2\geq 0
		\end{equation*}
		Simple algebra shows that the Inequality holds for $\mathscr{I}\in[1,\frac{1}{4}\left(\sqrt{9N^2-32N+32}+N\right)]$, where the segment increases in $N$. Setting $\bar{\mathscr{I}}(N)=\frac{1}{4}\left(\sqrt{9N^2-32N+32}+N\right)$, one has $\bar{\mathscr{I}}(4)=\frac{1}{4}(\sqrt{48}+4)>1$, $\bar{\mathscr{I}}(N)$ increasing in $N$, and $\bar{\mathscr{I}}(N)\geq N/2\rightarrow\infty$, which gives the statement (i) and (ii) in the proposition.

\end{proof}

\subsection{Proof of Corollary~\ref{coro_bigworlddisagreement}}
	\begin{proof}
		First, (i) is directly implied by Proposition~\ref{coro_bigworldignorance2example}. If $p_{A}^{\omega}$ is small enough for all $\omega\in N_{A}\subset N$ and $p_{B}^{\omega}$ is small enough for all $\omega\in N_{B}= N\setminus N_{A}$, individual $A$ (almost) never picks action $\omega$ for all $\omega \in N_{A}$ and individual $B$ (almost) never picks action $\omega$ for all $\omega\in N\setminus N_{A}$. Therefore, they must disagree with each other.
		
		To prove (ii), note that by Proposition~\ref{prop_smallworld}, we know that if an individual adopts an $\epsilon$-optimal updating mechanism, his utility loss is bounded above by $K r^{\floor{\frac{M-1}{N}}}+\epsilon$ for some $K>0$ and $r<1$. Thus, for all $\omega$, the probability that the individual chooses a ``wrong'' action $\omega'\neq\omega$ is strictly smaller than $\left[\frac{K r^{\floor{\frac{M-1}{N}}}+\epsilon}{\min_{\omega}p^{\omega}}\right]$. For all $\omega$, the probability that both individuals choose action~$\omega$ and thus agree with each other is greater than
		\begin{equation*}
			\begin{split}
				&\left[1-\left[\frac{K_A r_A^{\floor{\frac{M_A-1}{N}}}+\epsilon}{\min_{\omega}p^{\omega}_A}\right]\right]\left[1-\left[\frac{K_B r_B^{\floor{\frac{M_B-1}{N}}}+\epsilon}{\min_{\omega}p^{\omega}_B}\right]\right]\\
				>&1-\left[\frac{1}{\min_{\omega}p^{\omega}_A}+\frac{1}{\min_{\omega}p^{\omega}_B}\right](\max\lbrace K_A,K_B\rbrace (\max\lbrace r_A,r_B\rbrace)^{\floor{\frac{\min\lbrace M_A,M_B\rbrace-1}{N}}}+\epsilon)
			\end{split}
		\end{equation*}
		for all $\epsilon$-optimal updating mechanisms of individual A and B. The result follows.
		
	\end{proof}

\subsection{Proof of Corollary~\ref{prop_disagreementbasedonM} (stated in the main text)}
			\begin{proof}
				I prove the Corollary using the following simple example. Consider a setting with $N=3$ and two individuals, $A$ and $B$, who share the same prior belief and the same objective signal structure:
			\begin{equation}
				\begin{split}
					p^{1}&=\frac{1}{3}+2\nu\\
					p^{2}=p^{3}&=\frac{1}{3}-\nu
				\end{split}
			\end{equation}
			\begin{equation}
				\begin{split}
					\sup_{s}\frac{f^{1}(s)}{f^{n}(s)}&=\sup_{s}\frac{f^{n}(s)}{f^{1}(s)}=\sqrt{1+\tau} \text{ for $n=2,3$}\\
					\sup_{s}\frac{f^{2}(s)}{f^{3}(s)}&=\sup_{s}\frac{f^{3}(s)}{f^{2}(s)}=\sqrt{1+\Psi}\text{ where $\Psi>\tau$.}
				\end{split}
			\end{equation}
			with $1+\tau\geq\frac{\frac{1}{3}+2\nu}{\frac{1}{3}-\nu}$.\footnote{It ensures that if $M\geq 2$, the DM never chooses action~$1$ with probability $1$ and he can achieve a strictly lower utility loss compared to the benchmark of no information.} Moreover, assume that $u(1,1)=u(2,2)=u(3,3)=1$ and $u(\omega,\omega')=0$ for all $\omega\neq\omega'$. The only difference the two individuals have is their levels of cognitive ability, where $M_{A}=1$ and $M_{B}=2$.
			
			First, as $M=1$ for individual $A$, his action is constant in all periods for all signal realizations and thus $d(1)=1$. Individual~$A$ always takes action~$1$. Now I characterize the $\epsilon$-optimal updating mechanism of individual $B$. With some abuse of notations, denote $L_{2}^{*}(nn')$ as optimal utility loss where the DM chooses action $n$ in memory state~$1$ and action~$n'$ in memory state~$2$. We have, obviously,
			\begin{equation*}
			\begin{split}
			L_{2}^{*}(11)&=\frac{2}{3}-2\nu,\\
			L_{2}^{*}(22)=L_{2}^{*}(33)&=\frac{2}{3}+\nu.
			\end{split}	
			\end{equation*}
			Moreover, as argued in the proof of Corollary~\ref{prop_bigworldignorance2} (i), suppose $d(1)=n$ and $d(2)=n'\neq n$, we have
			\begin{equation*}
				 \frac{\mu^{n}_{1}}{1-\mu^{n'}_{2}}/\frac{1-\mu^{n}_{1}}{\mu^{n'}_{2}}=\frac{\mu^{n}_{1}}{\mu^{n'}_{1}}/\frac{\mu^{n}_{2}}{\mu^{n'}_{2}}\leq\sup_{s}\frac{f^{n}(s)}{f^{n'}(s)}\sup_{s}\frac{f^{n'}(s)}{f^{n}(s)}.
			\end{equation*}
			The upper bound of $L^*(nn')$ is given by the following minimization problem:
			\begin{equation*}
				\begin{split}
					&\min_{\mu^{n}_{1},\mu^{n'}_{2}} 1-p^n-p^{n'}+p^{n}(1-\mu^{n}_{1})+p^{n'}(1-\mu^{n'}_{2})\\
					&\text{ given  $\frac{\mu^{n}_{1}}{1-\mu^{n'}_{2}}/\frac{1-\mu^{n}_{1}}{\mu^{n'}_{2}}\leq\sup_{s}\frac{f^{n}(s)}{f^{n'}(s)}\sup_{s}\frac{f^{n'}(s)}{f^{n}(s)}$,}
				\end{split}
			\end{equation*}
			\cite{hellman1970learning} shows that the upper bound is tight with the updating mechanism where $\mathscr{T}(1,s')=2$ if and only if $\frac{f^{n'}(s')}{f^{n}(s')}$ is close to $\sup_{s}\frac{f^{n'}(s)}{f^{n}(s)}$, and $\mathscr{T}(2,s')=1$ if and only if $\frac{f^{n}(s')}{f^{n'}(s')}$ is close to $\sup_{s}\frac{f^{n}(s)}{f^{n'}(s)}$. Therefore,
			\begin{equation*}
			\begin{split}
			L_{2}^{*}(12)=L^{*}_{2}(13)&=\frac{1}{3}-\nu+\frac{2\sqrt{(1+\tau)\left(\frac{1}{3}+2\nu\right)\left(\frac{1}{3}-\nu\right)}-\left(\frac{2}{3}-\nu\right)}{\tau},\\
			L^{*}_{2}(23)&=\frac{1}{3}+2\nu+\frac{2\left(\frac{1}{3}-\nu\right)\sqrt{1+\Psi}-\left(\frac{2}{3}-2\nu\right)}{\Psi}.
			\end{split}
			\end{equation*}
			where $1+\tau\geq\frac{\frac{1}{3}+2\nu}{\frac{1}{3}-\nu}$ implies $L^{*}_{2}(22)=L^{*}_{2}(33)>L^{*}_{2}(11)\geq L^{*}_{2}(12)=L^{*}_{2}(13)$. In the following, I prove $L^{*}_{2}(12)> L^{*}_{2}(23)$ if $\nu$ is small enough which implies that $a_{t}^{B}=2$ or $3$ for all $t$ in all $\epsilon$-optimal updating mechanism when $\epsilon$ is smaller than $L^{*}_{2}(12)-L^{*}_{2}(23)$. Now, $L^{*}_{2}(12)> L^{*}_{2}(23)$ if and only if
			\begin{equation*}
		L^{*}_{2}(12)-L^{*}_{2}(23)=3\nu+\frac{2(\frac{1}{3}-\nu)\sqrt{1+\Psi}}{\Psi}-\frac{\frac{2}{3}-2\nu}{\Psi}-\frac{2\sqrt{(1+\tau)(\frac{1}{3}+2\nu)(\frac{1}{3}-\nu)}}{\tau}+\frac{\frac{2}{3}+\nu}{\tau}<0.
			\end{equation*}
			When $\nu=0$,
			\begin{equation*}
			L^{*}_{2}(12)-L^{*}_{2}(23)=\frac{2}{3}\left(\frac{\sqrt{1+\Psi}}{\Psi}-\frac{\sqrt{1+\tau}}{\tau}\right)-\frac{2}{3}\left(\frac{1}{\Psi}-\frac{1}{\tau}\right).
			\end{equation*}
			As both $\frac{\sqrt{1+x}}{x}$ and $\frac{1}{x}$ decreases in $x$,  $\Psi>\tau$ implies that $L^{*}_{2}(12)-L^{*}_{2}(23)<0$ when $\nu=0$. 
			The result follows by continuity.

		\end{proof}

\subsection{Proof of Corollary~\ref{coro_agreesmallworld1infiniteN}}
\begin{proof}
	With some abuse of notation, denote $\pmb{\mathscr{T}^{\epsilon}_{Ni}}$ as the set of $\epsilon$-optimal updating mechanisms of individual $i=A,B$ in the $N$-th decision problem. By Proposition~\ref{prop_smallworldinfiniteN}(i), $\lim_{N\rightarrow\infty}L^*_N=0$ for $h<1$. Moreover, by Assumption~\ref{assumption_fullsupportinfinite}, any set of states with non-vanishing fraction w.r.t.~$N$ carries non-vanishing prior probability. Together these imply that the prior probability on states where the long-run probability of choosing the correct action tends to 1 must itself tend to 1; formally,
	\begin{equation*}
		\lim_{N\rightarrow\infty}\lim_{\epsilon\rightarrow 0}\sup_{(m_{1Ni},\mathscr{T}_{Ni},d_{Ni})\in\pmb{\mathscr{T}^{\epsilon}_{Ni}}}\sum_{\omega}\left[\mathbbm{1}\left\lbrace\lim_{T\rightarrow \infty}E_{\omega,\mathscr{T}_{Ni}}\frac{\sum_{t=1}^{T}\mathbbm{1}_{a^i_t=\omega}}{T}=1\right\rbrace p^{\omega}_{Ni}\right]=1
	\end{equation*}
	which is equivalent to
	\begin{equation*}
		\lim_{N\rightarrow\infty}\lim_{\epsilon\rightarrow 0}\sup_{(m_{1Ni},\mathscr{T}_{Ni},d_{Ni})\in\pmb{\mathscr{T}^{\epsilon}_{Ni}}}\sum_{\omega}\left[\mathbbm{1}\left\lbrace\lim_{T\rightarrow \infty}E_{\omega,\mathscr{T}_{Ni}}\frac{\sum_{t=1}^{T}\mathbbm{1}_{a^i_t=\omega}}{T}<1\right\rbrace p^{\omega}_{Ni}\right]=0,
	\end{equation*}
	i.e., individual $i$ would only take sub-optimal actions in probability $0$ events measured by $(p_{Ni}^{\omega})_{\omega=1}^{N}$. By the same-limit-support assumption,
	\begin{equation*}
		\lim_{N\rightarrow\infty}\sum_{\omega\in \tilde{\Omega}_{N}} p^{\omega}_{NA}=0 \text{ if and only if  }\lim_{N\rightarrow\infty}\sum_{\omega\in \tilde{\Omega}_{N}} p^{\omega}_{NB}=0
	\end{equation*}
	for all sequences of subsets $\tilde{\Omega}_{N}\subseteq\lbrace1,\cdots,N\rbrace$. Thus,
		\begin{equation*}
		\lim_{N\rightarrow\infty}\lim_{\epsilon\rightarrow 0}\sup_{(m_{1Ni},\mathscr{T}_{Ni},d_{Ni})\in\pmb{\mathscr{T}^{\epsilon}_{Ni}}}\sum_{\omega}\left[\mathbbm{1}\left\lbrace\lim_{T\rightarrow \infty}E_{\omega,\mathscr{T}_{Ni}}\frac{\sum_{t=1}^{T}\mathbbm{1}_{a^i_t=\omega}}{T}<1\right\rbrace p^{\omega}_{NI}\right]=0\text{ for all $i,I\in\lbrace A,B\rbrace$},
	\end{equation*}
	i.e., even when individual $i$'s mechanism is evaluated under the other individual's prior, the prior mass on suboptimal outcomes vanishes in the limit. Bounding disagreement by a union bound over the two individuals' suboptimal events,
	\begin{equation*}
		\begin{split}
			&\lim_{N\rightarrow\infty}\lim_{\epsilon\rightarrow 0}\sup_{\substack{(m_{1NA},\mathscr{T}_{NA},d_{NA})\in\pmb{\mathscr{T}^{\epsilon}_{NA}}\\ (m_{1NB},\mathscr{T}_{NB},d_{NB})\in\pmb{\mathscr{T}^{\epsilon}_{NB}}}}\sum_{\omega}\left[\mathbbm{1}\left\lbrace\lim_{T\rightarrow \infty}E_{\omega,\mathscr{T}_{NA},\mathscr{T}_{NB}}\frac{\sum_{t=1}^{T}\mathbbm{1}_{a^A_t=a^{B}_t=\omega}}{T}<1\right\rbrace p^{\omega}_{Ni}\right]\\
			\leq &\lim_{N\rightarrow\infty}\lim_{\epsilon\rightarrow 0}\sup_{\substack{(m_{1NA},\mathscr{T}_{NA},d_{NA})\\ \in\pmb{\mathscr{T}^{\epsilon}_{NA}}}}\sum_{\omega}\left[\mathbbm{1}\left\lbrace\lim_{T\rightarrow \infty}E_{\omega,\mathscr{T}_{NA}}\frac{\sum_{t=1}^{T}\mathbbm{1}_{a^A_t=\omega}}{T}<1\right\rbrace p^{\omega}_{Ni}\right]\\
			&+\lim_{N\rightarrow\infty}\lim_{\epsilon\rightarrow 0}\sup_{\substack{(m_{1NB},\mathscr{T}_{NB},d_{NB})\\ \in\pmb{\mathscr{T}^{\epsilon}_{NB}}}}\sum_{\omega}\left[\mathbbm{1}\left\lbrace\lim_{T\rightarrow \infty}E_{\omega,\mathscr{T}_{NB}}\frac{\sum_{t=1}^{T}\mathbbm{1}_{a^B_t=\omega}}{T}<1\right\rbrace p^{\omega}_{Ni}\right]\\
			=&0
		\end{split}
	\end{equation*}
	for $i=A,B$. The result follows.
\end{proof}

\subsection{Proof of Corollary~\ref{coro_bigworlddisagreementinfiniteN}}
\begin{proof}
	Consider an example where $\lim_{N\rightarrow\infty}p_{N}^{\omega}=0$ for all $\omega$ (so that a set of states has non-vanishing prior probability if and only if it has a non-vanishing fraction w.r.t.~$N$). As shown in Proposition~\ref{prop_smallworldinfiniteN}(ii), all sequences of updating mechanisms must ignore almost all actions as $N$ goes to infinity. Thus, $\lim_{N\rightarrow \infty}L_{N}^{*}=\lim_{N\rightarrow\infty}\sum_{\omega}u_{N}(\omega,\omega)p_{N}^{\omega}$ and in the limit, every updating mechanism is $\epsilon$-optimal. Thus, if individual $A$ adopts an updating mechanism with $d_N(m)=1$ for all $m$ and $N$, while individual $B$ adopts an updating mechanism with $d_N(m)=2$ for all $m$ and all $N\geq 2$, then they must disagree with each other under all $\omega$ even though they adopt a sequence of updating mechanisms that is $\epsilon$-optimal in the limit.
\end{proof}

\subsection{Proof of Proposition~\ref{prop_smallworldmistakes}}
\begin{proof}
	The proof follows closely the proof of Proposition~\ref{prop_smallworld}. For individual $B$, fixing the state $\omega$, in the stationary probability distribution, we have at the two extreme memory states in branch $\omega'$,
	\begin{equation*}
		\begin{split}
			\mu^{\omega}_{\omega'\lambda}\left[\frac{1-\gamma}{\delta}\sum_{\omega''\neq\omega'}F^{\omega}(G^{\omega''})+\gamma\right]&=\mu^{\omega}_{\omega'(\lambda-1)}\left[(1-\gamma)F^{\omega}(G^{\omega'})+\frac{\gamma}{2}\right]\\
			\mu^{\omega}_{\omega'(\lambda-1)}&=\mu^{\omega}_{\omega'\lambda}\left[\frac{(1-\gamma)F^{\omega}(G^{\omega'})+\frac{\gamma}{2}}{\frac{1-\gamma}{\delta}\sum_{\omega''\neq\omega'}F^{\omega}(G^{\omega''})+\gamma}\right]^{-1}
		\end{split}
	\end{equation*}
	for all $\omega'$. Similarly, at memory state $\omega'(\lambda-1)$,
	\begin{equation*}
		\resizebox{1\hsize}{!}{$
			\begin{split}
				\mu^{\omega}_{\omega'\lambda}\left[\frac{1-\gamma}{\delta}\sum_{\omega''\neq\omega'}F^{\omega}(G^{\omega''})+\gamma\right]+\mu^{\omega}_{\omega'(\lambda-2)}\left[(1-\gamma)F^{\omega}(G^{\omega'})+\frac{\gamma}{2}\right]&=\mu^{\omega}_{\omega'(\lambda-1)}\left[\frac{1-\gamma}{\delta}\sum_{\omega''\neq\omega'}F^{\omega}(G^{\omega''})+(1-\gamma)F^{\omega}(G^{\omega'})+\gamma\right]\\
				\mu^{\omega}_{\omega'(\lambda-2)}\left[(1-\gamma)F^{\omega}(G^{\omega'})+\frac{\gamma}{2}\right]&=\mu^{\omega}_{\omega'(\lambda-1)}\left[\frac{1-\gamma}{\delta}\sum_{\omega''\neq\omega'}F^{\omega}(G^{\omega''})+\frac{\gamma}{2}\right]\\
				\mu^{\omega}_{\omega'(\lambda-2)}&=\mu^{\omega}_{\omega'(\lambda-1)}\left[\frac{(1-\gamma)F^{\omega}(G^{\omega'})+\frac{\gamma}{2}}{\frac{1-\gamma}{\delta}\sum_{\omega''\neq\omega'}F^{\omega}(G^{\omega''})+\frac{\gamma}{2}}\right]^{-1}
			\end{split}
			$}
	\end{equation*}
	Repeating the same procedures implies that for all $k=1,\cdots,\lambda-1$
	\begin{equation}
		\mu^{\omega}_{\omega'k}=\mu^{\omega}_{\omega'\lambda}\left[\frac{(1-\gamma)F^{\omega}(G^{\omega'})+\frac{\gamma}{2}}{\frac{1-\gamma}{\delta}\sum_{\omega''\neq\omega'}F^{\omega}(G^{\omega''})+\gamma}\right]^{-1}\left[\frac{(1-\gamma)F^{\omega}(G^{\omega'})+\frac{\gamma}{2}}{\frac{1-\gamma}{\delta}\sum_{\omega''\neq\omega'}F^{\omega}(G^{\omega''})+\frac{\gamma}{2}}\right]^{-(\lambda-k-1)}	
	\end{equation}
	and 
	\begin{multline}
		\resizebox{1\hsize}{!}{$
			\mu^{\omega}_{0}=\mu^{\omega}_{\omega'\lambda}\left[\frac{(1-\gamma)F^{\omega}(G^{\omega'})+\frac{\gamma}{2}}{\frac{1-\gamma}{\delta}\sum_{\omega''\neq\omega'}F^{\omega}(G^{\omega''})+\gamma}\right]^{-1}\left[\frac{(1-\gamma)F^{\omega}(G^{\omega'})+\frac{\gamma}{2}}{\frac{1-\gamma}{\delta}\sum_{\omega''\neq\omega'}F^{\omega}(G^{\omega''})+\frac{\gamma}{2}}\right]^{-(\lambda-2)}	\left[\frac{(1-\gamma)F^{\omega}(G^{\omega'})+\frac{\gamma}{N}}{\frac{1-\gamma}{\delta}\sum_{\omega''\neq\omega'}F^{\omega}(G^{\omega''})+\frac{\gamma}{2}}\right]
			$}
	\end{multline}
	Note that for all $\gamma\in(0,1)$
	\begin{equation*}
		\frac{(1-\gamma)F^{\omega}(G^{\omega})+\frac{\gamma}{2}}{\frac{1-\gamma}{\delta}\sum_{\omega''\neq\omega}F^{\omega}(G^{\omega''})+\frac{\gamma}{2}}>1.
	\end{equation*}
	\begin{equation*}
		\frac{(1-\gamma)F^{\omega}(G^{\omega})+\frac{\gamma}{2}}{\frac{1-\gamma}{\delta}\sum_{\omega''\neq\omega}F^{\omega}(G^{\omega''})+\frac{\gamma}{2}}>\frac{(1-\gamma)F^{\omega}(G^{\omega'})+\frac{\gamma}{2}}{\frac{1-\gamma}{\delta}\sum_{\omega''\neq\omega}F^{\omega}(G^{\omega''})+\frac{\gamma}{2}}.
	\end{equation*}
	Then the result follows from the same steps in the proof of Proposition~\ref{prop_smallworld}.
\end{proof}

\bibliographystyle{econ}
\bibliography{reference_big}